%% file: main.tex
\documentclass[twocolumn]{aastex631}
\usepackage{amsmath}
\usepackage{graphicx}

\usepackage{natbib}
\usepackage[scaled]{helvet}
\usepackage{epsfig}
\usepackage{url}
\bibpunct{(}{)}{;}{a}{}{,}
\newcommand{\kms}{\,km\,s$^{-1}$}
\usepackage{textcomp}
\usepackage{mathpazo}
\usepackage{xspace}
\usepackage{subfigure}
\usepackage{hyperref}
\usepackage{apjfonts}
\usepackage{mathrsfs}
\usepackage{verbatim}
\usepackage{color}
\usepackage[normalem]{ulem}

    {\clearpage\pagebreak[4]\global\pdfpageattr\expandafter{\the\pdfpageattr/Rotate 90}}%
    {\clearpage\pagebreak[4]\global\pdfpageattr\expandafter{\the\pdfpageattr/Rotate 0}}%

\newcommand{\bjdtdb}{\ensuremath{\rm {BJD_{TDB}}}}
\newcommand{\feh}{\ensuremath{\left[{\rm Fe}/{\rm H}\right]}}

\newcommand{\teff}{\ensuremath{T_{\rm eff}}\xspace}

\newcommand{\ecosw}{\ensuremath{e\cos{\omega_*}}}
\newcommand{\esinw}{\ensuremath{e\sin{\omega_*}}}
\newcommand{\msun}{\ensuremath{\,M_\Sun}}
\newcommand{\rsun}{\ensuremath{\,R_\Sun}}
\newcommand{\lsun}{\ensuremath{\,L_\Sun}}
\newcommand{\mj}{\ensuremath{\,M_{\rm J}}}
\newcommand{\rj}{\ensuremath{\,R_{\rm J}}}

\newcommand{\re}{\ensuremath{\,R_{\rm \Earth}}\xspace}
\newcommand{\me}{\ensuremath{\,M_{\rm \Earth}}\xspace}
\newcommand{\fave}{\langle F \rangle}
\newcommand{\fluxcgs}{10$^9$ erg s$^{-1}$ cm$^{-2}$}

\newcommand{\be}{\begin{equation}}
\newcommand{\ee}{\end{equation}}

\newcommand{\tess}{{\it TESS}}
\newcommand\exofast{$\texttt{EXOFASTv2}$\xspace}

\begin{document}

\title{11 New Transiting Brown Dwarfs and Very Low Mass Stars from \textit{TESS}}

\input{authors.tex}
\input{affiliations.tex}

\shorttitle{11 New Transiting Companions from \textit{TESS}}
\shortauthors{Vowell et al.}

\begin{abstract}

We present the discovery of 11 new transiting brown dwarfs and low-mass M-dwarfs from NASA's \tess\ mission: TOI-2844, TOI-3122, TOI-3577, TOI-3755, TOI-4462, TOI-4635, TOI-4737, TOI-4759, TOI-5240, TOI-5467, and TOI-5882. They consist of 5 brown dwarf companions and 6 very low-mass stellar companions ranging in mass from 25 \mj\ to 128 \mj. We used a combination of photometric time-series, spectroscopic, and high-resolution imaging follow-up as a part of the \tess\ Follow-up Observing Program (TFOP) to characterize each system. With over 50 transiting brown dwarfs confirmed, we now have a large enough sample to directly test different formation and evolutionary scenarios. We provide a renewed perspective on the transiting brown dwarf desert and its role in differentiating between planetary and stellar formation mechanisms. Our analysis of the eccentricity distribution for the transiting brown dwarf sample does not support previous claims of a transition between planetary and stellar formation at $\sim$42 \mj. We also contribute a first look into the metallicity distribution of transiting companions in the range $7 - 150$ \mj, showing that this does not support a $\sim$42 \mj\ transition too. Finally, we also detect a significant lithium absorption feature in one of the brown dwarf hosts (TOI-5882). However, we determine that the host star is likely old based on rotation, kinematic, and photometric measurements. We therefore claim that TOI-5882 may be a candidate for planetary engulfment.

\end{abstract}

\section{Introduction}
Since the launch of NASA's Transiting Exoplanet Survey Satellite (\tess) in 2018 \citep{Ricker:2015}, the number of brown dwarfs (BDs) known to transit their host stars has increased rapidly from just 16 systems to $>50$. These BDs, which are defined as objects within the mass range of 13 - 80 \mj, fuse only deuterium in their cores. This differentiates them from planets, which undergo no fusion, and stars, which ignite hydrogen fusion. However, the deuterium and hydrogen burning limits have been shown to be less clear than this definition would imply. \citet{Spiegel2011} showed that the lower limit varies from 11 - 16 \mj, while \citet{Baraffe2003} showed that hydrogen fusion can ignite between 75 - 80 \mj. The spread in both of these estimates can be explained by variation in the chemical composition and formation conditions of the BD. While these definitions provide insight into the physical processes taking place in BD interiors, they offer little insight into how they form. 

Reframing our perspective on BDs into one motivated by formation and evolution has long been advocated for by some members of the BD community \citep{Chabrier2014, Burrows:2014, Carmichael2021}, where objects would be distinguished based on whether they form through a planet-like or a star-like formation mechanism. BDs forming like planets would undergo a core accretion pathway \citep{Pollack:1996}, commonly referred to as a "bottom-up" approach. The star-like BDs on the other hand would form via direct gravitational collapse, or "top-down", which can happen either within the circumstellar disk or at the core scale \citep{Adams1989, Bate2012, Kratter2016}. Differentiating between these two formation pathways remains challenging, since it is unclear under what conditions each mechanism dominates, and whether there are any observable parameters that could distinguish them. Fortunately, in the era of \tess, we have begun to accumulate transiting BDs en masse, allowing us to pursue the question of BD formation from a different perspective. This budding population of transiting BDs is particularly enticing for studying BD formation because it provides a complementary, and in many cases, more complete understanding of the BD compared to the previously studied objects, which have primarily been discovered via direct imaging or radial velocity (RV) techniques. The transiting population serves as a complementary dataset to these other populations because transits provide a model independent measurement of BD radii, a property which often can only be otherwise inferred with evolutionary models based on the observed spectrum and luminosity. This measurement is vital because BDs tend to contract with age, while also decreasing in size as mass increases \citep{Baraffe2003, SaumonMarley2008, Burrows2001, Phillips2020}. Thus there exists a degeneracy between mass, radius, and age for BDs making it difficult to test the substellar models with observed systems unless all three variables can be measured. These transiting systems provide direct, independent measurements on two of these degenerate parameters, and in cases where the host star's age can be precisely determined, all three \citep[e.g.][]{Gillen2017, Nowak2017, David2019, Vowell2023}.

This rapidly growing population of transiting BDs also allows us to revisit the longstanding idea of the so-called "brown dwarf desert". Prior work has shown a dearth of brown dwarfs orbiting main sequence host stars with semi-major axes $<5$ au. \citep{Marcy1997, Latham1998}. \citet{Ma&Ge2014} refined our understanding of the brown dwarf desert by investigating the population of all published brown dwarfs discovered with the RV method at the time. Here they found that the "driest land" of the desert lies between $35 < m\,\sin{i} < 55$ \mj\ and with period $P < 100$ days). The authors attribute this feature to being a result of different formation mechanisms dominating in different mass regimes. Namely, that stellar binary formation is responsible for the systems with BD companions $>42$\mj\ while formation in the protoplanetary disk explains the systems with BDs $<42$\mj. However, the sample in this study with period $P < 100$ days was quite small at only 25 brown dwarfs. Furthermore, by virtue of being a RV study, it was restricted to only probing $m\,\sin{i}$ rather than the BD mass directly, unable to break the $\sin{i}$ degeneracy in most cases, a complication that the transiting population does not have. 

As this population of transiting brown dwarfs expanded in the era of space-based transit surveys, several new discoveries noted an "oasis" forming in the desert \citep{Carmichael2020, Subjak2020, Henderson2024a} with new transiting systems beginning to populate the driest region of the desert noted by \citet{Ma&Ge2014}. With the new discoveries presented in this work, the transiting BD population now exceeds 50 systems, more than double the size of the population \citet{Ma&Ge2014} had access to, opening the door for a reevaluation of brown dwarf desert from a new perspective. Hence, in this paper we present the discovery of 11 new transiting companions from NASA's \tess\ mission. 6 of these systems are BDs, with 3 lying within the \citet{Ma&Ge2014} defined "driest" region of the BD desert. We confirmed the remaining 6 non-BD companions as very low-mass stars $<150$ \mj. In \S\ref{sec:obs} of this manuscript, we present all the observations collected for each system in this work. \S\ref{sec:analysis} details our analysis of each system using \exofast \citep{Eastman:2013, Eastman:2019}. In \S\ref{sec:discussion} we provide a discussion on how these new systems fit into the population as a whole with a renewed perspective on the BD desert. We also discuss a detection of lithium in the host star of the BD companions presented here (TOI-5882). Finally, we present our conclusions in \S\ref{sec:conclusions}.

\section{Observations}
\label{sec:obs}
In the following subsections we present all observations collected and analyzed for each target in this sample. To briefly summarize, each target has a suite of observations that serve to characterize the host star and/or companion and rule out false-positive scenarios. Generally, these observations include archival multiband observations from various ground-based missions, time-series photometry from both space and ground-based telescopes, spectroscopy, and high resolution imaging. See Table \ref{tab:lit} for the relevant results of the archival data associated with each system.

\begin{deluxetable*}{l>{\centering}cccccc}
\tablecaption{Literature and Measured Properties \label{tab:lit}}
\tabletypesize{\scriptsize}
\input{lit_table_4}
\end{deluxetable*}
\addtocounter{table}{-1}

\begin{deluxetable*}{l>{\centering}cccccc}
\tablecaption{\textit{(Continued)} \label{tab:lit}}
\tabletypesize{\scriptsize}
\input{lit_table_5}
\end{deluxetable*}
\addtocounter{table}{-1}

\begin{deluxetable*}{l>{\centering}cccccc}
\tablecaption{\textit{(Continued)} \label{tab:lit}}
\tabletypesize{\scriptsize}
\input{lit_table_6}
\footnotesize{ \textbf{\textsc{NOTES:}}
 The uncertainties of the photometric measurements have a systematic floor applied that is usually larger than the reported catalog errors. \\
 $\ddagger$ Right Ascension and Declination are in epoch J2000. The coordinates come from Vizier where the Gaia RA and Dec have been precessed and corrected to J2000 from epoch J2016.\\
 Sources: (1) \cite{GaiaDR3}; (2) \S \ref{subsec:TRES}; (3) \cite{Stassun2019}; (4) \cite{Cutri:2003, Skrutskie:2006}; (5) \cite{Wright:2010, Cutri:2012} \\
}
\end{deluxetable*}

\subsection{\tess\ Photometry}
Each system presented here initially showed signs of an orbiting companion via transits detected by \tess. \tess\ has a mosaic of four CCD cameras each with a $24^{\circ}$x$24^{\circ}$ field of view, and a pixel size of 21". In combination, this makes the \tess\ field of view $24^{\circ}$x$96^{\circ}$ for each sector, which is observed for approximately 27 days before moving to a new sector of sky. \tess\ observes at a 2-second, and in the \tess\ prime mission, the data were processed into 2-minute stacks for select stars, with the rest of the field being processed at 30-minute cadence. This prime mission observed $>$80\% of the entire sky with the largest gaps in coverage occurring near the ecliptic plane. As \tess\ transitioned to its first, and now second, extended missions it continues to observe even more of the ecliptic plane. In this second extended mission, most preselected targets are now processed at 120-seconds while a smaller number are processed at 20-second cadence. Full-frame images are processed at 200-seconds.

The systems presented here were observed by \tess\ between Sectors $6-76$ with cadences ranging from 30-minutes in the prime mission to as low as 2-minutes in the extended mission. The \tess\ data were originally downloaded and reduced using both the \tess\ Science Processing Operations Center (SPOC) Pipeline \citep{Jenkins:2016} and the MIT Quick-Look Pipeline \citep[QLP;][]{Huang2020a, Huang2020b, Kunimoto2021}. The initial detection of a transit-like signal was discovered and vetted by the faint-star QLP search \citep{Kunimoto2022} for 10 out the 12 systems presented in this paper. The remaining two, TOI-4462 and TOI-4635, were initially detected by the QLP and SPOC pipelines respectively, and then vetted by the \tess\ Science Office. The diagnostic tests described in \citet{Twicken:2018} were used to evaluate whether the transit-like signal is indeed Keplerian. Upon passing, each system was designated as a \tess\ Object of Interest \citep[TOI;][]{Guerrero:2021}. While both the QLP and SPOC pipelines correct for contamination by known nearby stars, we choose to use the SPOC light curves with the shortest cadence in our analysis wherever possible for consistency. See Table \ref{tab:tess} for full details on the sectors, cadence, and pipeline used for each source. It should be noted that all QLP light curves are processed from the full frame images. The TESS-SPOC \citep{TESSSPOC2020} light curves are produced by the SPOC on a best-effort basis under the leadership of Doug Caldwell, the PI, and delivered as high-level science products to the Mikulski Archive for Space Telescopes (MAST) rather than as part of the official mission data products produced by the SPOC.

We downloaded the individual light curves from the MAST using the {\it lightkurve}\footnote{\url{https://github.com/lightkurve/lightkurve}} package \citep{Lightkurve2018}. We then removed any long-term variability (both stellar and instumental) by fitting a spline to the flux and dividing the light curve by the best fitting spline model. We used the {\it Keplerspline}\footnote{\url{https://github.com/avanderburg/keplerspline}} package for this process as described in \citep{Vanderburg:2014}. We remove most of the out-of-transit data, electing to keep just one transit duration of baseline on either side of the transit.

\input{tess_table}

\subsection{Ground-Based Time-Series Photometry}
In order to confirm that the signal observed by \tess\ is on-target not originating from a nearby eclipsing binary that is blended with the target star, we gathered ground-based, time-series photometry of each system as the companion transited its host star. Since \tess\ has a relatively large pixel scale (21\arcsec\ per pixel), the shallow eclipses we measured, which are consistent with roughly 1 $R_J$, can be easily mimicked when a different, nearby eclipsing binary happens to fall on the same photometric aperture as the target star. The much deeper eclipses of the nearby eclipsing binary become diluted by the target star to mimic a much shallower event. Seeing-limited ground-based telescopes can have a much higher angular resolution than \tess\, typically 1-2\arcsec\ and therefore can confirm that the signal is on-target, thereby ruling out nearby eclipsing binaries at all but the closest separations. It also has the benefit of observing in multiple wavelengths to confirm that the transit-like signal is achromatic. This is helpful because the eclipse depth of an eclipsing binary is nearly always wavelength dependent since the occulting body cannot be treated as a non-luminous sphere. We also note here, that while 6 of the companions presented here are low-mass M-dwarfs, and hence are eclipsing binaries themselves, they are so low in mass that we can still treat them as black spheres since they contribute negligibly to the overall flux of the system \citep{Stevens2018}. This process not only rules out the nearby eclipsing binary false positive, but also serves to refine the ephemerides of systems in which these data are able to extend the photometric baseline. 

The observations for these systems were collected through the \tess\ Follow-up Observing Program \citep[TFOP;][]{Collins:2018} from various observatories as shown in Table \ref{tab:followup}. The Las Cumbres Observatory Global Telescope (LCOGT; \citealt{Brown:2013}) was responsible for 14 light curves from the following sites: McDonald Observatory (McD), Teide Observatory (TEID), South African Astronomical Observatory (SAAO), and Cerro Tololo Inter-American Observatory (CTIO). The remaining light curves were contributed by the following facilities: Calar Alto Observatory, Brierfield Observatory, the Telescopio Carlos S\'anchez (TCS) at Teide Observatory, Grand-Pra (GdP) Observatory, Thacher Observatory \citep{Swift2022}, KeplerCam at the Fred Lawrence Whipple Observatory (FLWO), and the Acton Sky Portal. 

All data sets, except for the observations of TOI-3577 from MuSCAT2, were reduced and their light curves extracted using \texttt{AstroImageJ} \citep[AIJ;][]{Collins:2017}. To do this, we use AIJ's multi-aperture photometry tool using at least five similarly bright comparison stars. We use AIJ's built-in transit fitting tool to assess the quality of the data and determine detrending parameters. Generally, we only detrend against the parameters that strongly correlate with the apparent brightness of the companion stars as they change over the course of the night. We choose to adopt detrending only when the Bayesian Information Criterion of AIJ's transit-only fit significantly favors the detrended model. The detrending parameters used for each light curve can be found in Table \ref{tab:followup}. See \S D in the appendix of \citet{Collins:2017} for a detailed description of each detrending parameter. Finally, we normalized the data to the out-of-transit baseline and incorporated each light curve (with detrending) into our global fitting process (see \S\ref{sec:analysis}).

Our follow-up observations of TOI-3577 were taken by MuSCAT2 on the TCS from Teide Observatory in Tenerife, Spain \citep{Narita2019}. MuSCAT2 is a multi-band imager with four cameras, each with a field of view of $7.4 ^{\prime} \times 7.4 ^{\prime} $. This set-up allows for simultaneous observation in multiple bands, which in our case, were the $g'$, $r'$, $i'$, and $z_s$ bands. These data were reduced by the dedicated MuSCAT2 pipeline \citep{Parviainen2019}, and incorporated into our global fit.

\input{followup_table}

\subsection{Spectroscopy}
\label{subsec:TRES}
We collected spectroscopic observations for each system to measure the mass and eccentricity of their companions while also further ruling out the false-positive scenario of nearby eclipsing binaries. While several of the systems presented here have companions above the hydrogen-burning boundary, and thus are eclipsing binaries themselves, none of them have companions that are bright enough to be detected photometrically or spectroscopically. Hence, they are all single-lined spectroscopic binaries. Any potential nearby eclipsing binaries (both bound and unbound) that cannot be ruled out by ground-based photometry can be ruled out by spectroscopy within the angular diameter of the fiber. These are ruled out by the fact that the companions presented here are significantly more massive than their giant planet counterparts. The Doppler motion of the host stars' spectral lines is too large to be mimicked by a nearby eclipsing binary without resolving a secondary set of spectral features. 

We obtained spectroscopic measurements for each system in this sample via the Tillinghast Reflector \'Echelle Spectrograph (TRES) on the 1.5-meter Tillinghast Reflector telescope at the Fred Lawrence Whipple Observatory on Mt. Hopkins, Arizona. The TRES instrument is a fiber-fed, \'echelle spectrograph with a resolving power of 44,000. We reduced the spectra according to \citet{Buchhave:2010} and analyzed each observation with the Stellar Parameter Classification (SPC) tool \citep{Buchhave:2012} in order to measure the metallicity, effective temperature, surface gravity, and projected rotational velocity of the star. We incorporated the average metallicity for each system into our analysis as a Gaussian prior in our global fits (see \S \ref{sec:analysis}). We did not incorporate the effective temperature or surface gravity measurements from SPC as priors because these quantities are better constrained by the fit itself. This is due to the fact that \exofast\ simultaneously models the spectral energy distribution, companion's transit, and stellar evolutionary models \citep{Eastman2023}

Finally, we derived the radial velocities according to the methods described in \citet{Quinn:2012}, except that we do not cross-correlate against a template spectrum. Instead, we create a high S/N, median-combined observed spectrum that we cross-correlate with each individual spectrum. See Table \ref{tab:rv} for a sample radial velocity point for each system (the full table of radial velocities is available in machine-readable form in the online journal).

\subsubsection{SOPHIE Spectroscopy}
We complement the TRES data with SOPHIE observations of TOI-5882. SOPHIE is a stabilized \'echelle spectrograph dedicated to high-precision radial-velocity
measurements at the 1.93-m telescope of the Observatoire
de Haute-Provence, France \citep{Perruchot2008, Bouchy2009, Bouchy2013}. We used its high-resolution mode (resolving power $R=75\,000$) and the fast readout of its~CCD.

Removing a few observations with low accuracy, we have a dataset of 21 SOPHIE measurements of TOI-5882 secured from December 2022 to July 2024. Exposure times ranged between 5 and 37~minutes, allowing signal-to-noise ratios between 14 and 40 to be reached per pixel at 550\,nm.

The radial velocities were extracted with the SOPHIE pipeline, as presented by \citet{Bouchy2009}
and refined by \citet{Heidari2024, Heidari2025}. That
procedure includes corrections for bad pixels, cosmic rays,
and charge transfer inefficiency of the CCD, as well as sky
background and instrumental drifts. It derives cross
correlation functions (CCF) from a numerical mask, then fit
the CCFs by Gaussians to derive the radial velocities \citep{Baranne1996, Pepe2002}. One sample measurement is
reported in Table \ref{tab:rv}.

\input{rv_table}

\subsection{High Resolution Imaging}
\label{sec:ao}
\input{hri_table}
While ground-based transits rule out nearby eclipsing binaries at most scales, if another source is close enough to the target star, it may be blended both in \tess\ and from the ground. Therefore, to verify that there is no contamination at these very small separations, and in order to detect any potentially bright companions, we utilized high-resolution imaging. We employed both Adaptive Optics (AO) and speckle imaging instruments to obtain our high resolution images for these systems. 

We used the ShARCS and PHARO instruments for AO imaging. The ShARCS instrument is on the Shane 3.0-meter telescope located at Lick Observatory \citep{Kupke2012, Gavel2014, McGurk2014}. The PHARO instrument is on the Palomar Hale 5-meter telescope at Palomar Observatory \citep{Hayward2001}. For our speckle observations we used the following telescopes and instruments: HRCam on the Southern Astrophysical Research (SOAR) 4.1-meter telescope at CTIO \citep{Tokovinin:2018, Ziegler2020}, the NN-EXPLORE Exoplanet Stellar Speckle Imager \citep[NESSI;][]{Scott2018} on the WIYN 3.5-meter telescope at Kitt Peak Observatory, the Speckle Polarimeter on the 2.5-meter telescope at the Caucasian Mountain Observatory of the Sternberg Astronomical Institute (SAI) at Lomonosov Moscow State University, and the Zorro instrument on the Gemini-South 8-meter telescope. The Speckle Polarimeter used an Andor iXon 897 Electron Multiplying CCD for the observation of TOI-3755 \citep{Safonov2017}. All other observations from this instrument used a Hamamatsu ORCA-quest CMOS detector \citep{Strakhov2023}. See Table \ref{tab:hri} for a summary of each observation including the dates each system was observed, filters used, contrast achieved, and whether a nearby companion was detected. 

\begin{figure*}
    \centering 
    \includegraphics[trim = 0 0 0 0,width=0.48\linewidth]{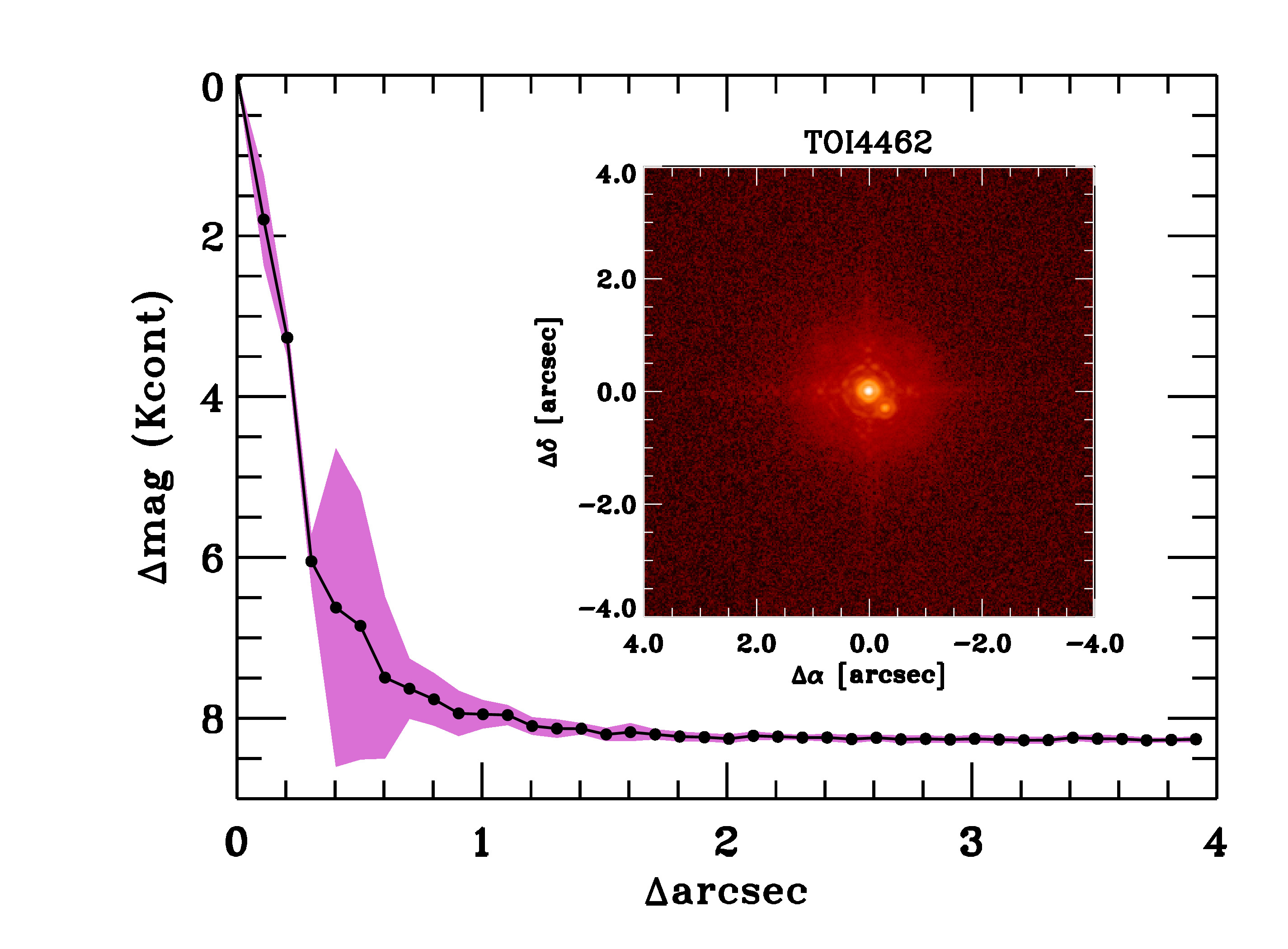}
    \includegraphics[trim = 0 0 0 0,width=0.48\linewidth]{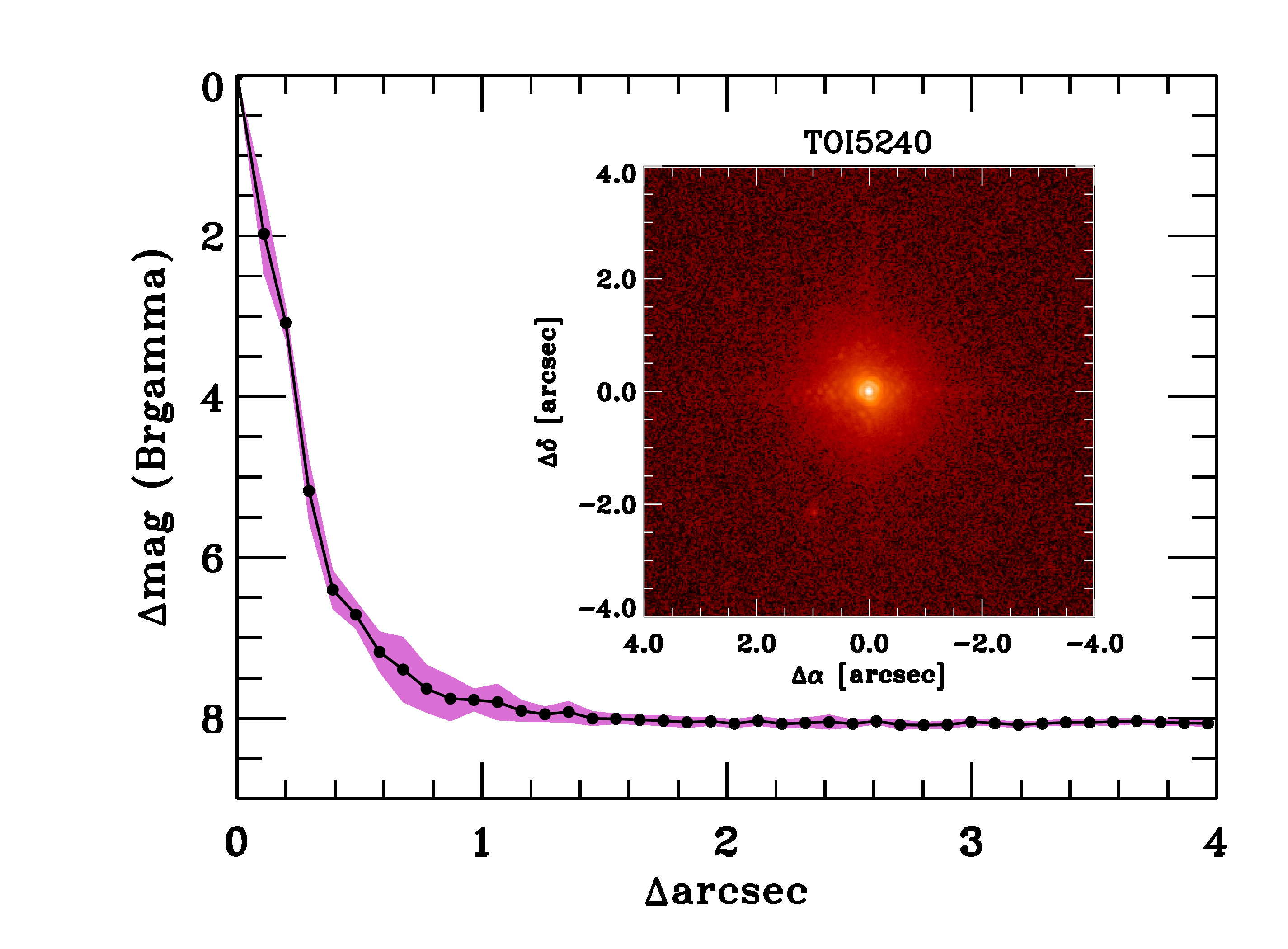}
    \caption{The adaptive optics image and contrast curves for TOI-4462 and TOI-5240 taken by PHARO on the 5.0 m Palomar telescope. (Left) TOI-4462 in the $K_{cont}$ filter with a bright companion clearly seen at a separation of 0.4$\arcsec$. (Right) TOI-5240 in the Br$\gamma$ filter with a faint companion at a separation of 2.4$\arcsec$.}
    \label{fig:AO}
\end{figure*}

Our high resolution imaging runs resulted in the detection of only two nearby companions, one in the TOI-4462 system and another in the TOI-5240 system (see Figure \ref{fig:AO}). The companion to TOI-5240 was detected only by PHARO and is 2.4$\arcsec$ away at a position angle of 156 degrees. It is 6.67 magnitudes dimmer in the Br$\gamma$ filter, contributing only 0.1\% of the total flux of the unresolved system. Even if this companion is a perfectly edge-on, equal mass eclipsing binary, the eclipse depths would be an order of magnitude smaller than the observed transit depths. Since it contributes a negligible amount of light to the overall flux of the system, we chose to neglect this companion in our analysis \citep[eg.][]{Mugrauer2020, Mugrauer2021}.

The companion to TOI-4462 A was resolved by both SAI and PHARO at 0.4$\arcsec$ separation and a position angle of 225 degrees. It is approximately 2.6 magnitudes dimmer in the $H_{cont}$ and $K_{cont}$ filters, too bright to neglect in our analysis (see \S\ref{sec:analysis}). However, we remain confident that the Keplerian signals detected in both our photometry and spectroscopy can only be attributed to the brighter, primary star. The transits observed by \tess\, KeplerCam, and LCO-TEID show no evidence of chromaticity, and the spectral line profiles show no evidence of a secondary set of spectral lines that would produce an apparent RV shift. Since TRES observed TOI-4462 with a 2.3$\arcsec$ fiber (larger than the companion's separation), the host star spectra were blended with light from the companion. However, this faint companion would only affect the measured RVs at the 10s of m/s level \citep{Buchhave:2011}, significantly smaller than 10000 m/s semi-amplitude we measure for the TOI-4462 system. Hence we are confident that the signal we detect is due to an unresolved transiting companion around the brighter primary star TOI-4462 A. 

\section{Analysis}
\label{sec:analysis}
We analyzed each system using \exofast\footnote{\url{https://github.com/jdeast/EXOFASTv2}} \citep{Eastman:2019}, a publicly available exoplanet fitting suite. \exofast\ is a Differential Evolution Markov Chain Monte Carlo (MCMC) code which globally fits both the star and the companion simultaneously, ensuring a self-consistent set of parameters for the entire system. In each fit, we generate a Spectral Energy Distribution (SED) model for the host star using MESA Isochrones and Stellar Tracks \citep[MIST;][]{Paxton:2011, Paxton:2013} in order to fit the host star, while the companion is fit with a standard Keplerian model. Our SED model is fit to broadband archival photometry which we collected from Gaia Data Release 3 \citep[DR3;][]{GaiaDR3}, 2MASS \citep{Cutri:2003,Skrutskie:2006}, and WISE \citep{Wright:2010, Cutri:2012}. The Keplerian model for the companion is fit to the \tess\, and ground based transits as well as the RV data from TRES and SOPHIE. For a more detailed explanation of the modeling process, see \citet{Eastman:2019}. 

The fit was generally set up in the same way for each system, except for TOI-4462, which required special consideration due to the presence of a bright nearby companion which we discuss in \S\ref{sec:toi4462}. For the other 11 systems, we first compiled the archival photometry for each target in the Gaia $G$, $Bp$, $Rp$, 2MASS $J$, $H$, $Ks$, and WISE $W1$, $W2$, and $W3$ bands to construct the SED. We then placed a set of priors on each system based on previous observations, the first of which was a Gaussian prior on the parallax from Gaia DR3 with the \citet{Lindegren2021B} correction applied. The parallax uncertainty was added in quadrature with 0.01 to account for any remaining systematic residuals. We also placed a Gaussian prior on the host star metallicity centered on the average value of the TRES-derived metallicity with a prior width of twice the standard deviation. Additionally, we place an upper limit on the \textit{V}-band extinction along the line-of-site using the dust maps from \citet{Schlegel:1998} and \citet{SchlaflyFinkbeiner2011}. 

In addition to the priors described above, we also fit for a dilution term in each system to account for unresolved contaminants.To do this, we placed a prior of 0\% $\pm$ 10\% of the contamination ratio reported by the \tess\ Input Catalog \citep[TIC;][]{Stassun:2018_TIC, Stassun2019}. While the QLP and SPOC light curves are both already corrected for known contaminants, we still chose to fit for a dilution term as a conservative assumption that the correction applied had a precision of at most 10\%. We did this because the contamination ratio reported by the TIC is only an estimate that does not account for the actual point spread functions, as the CCD location and camera were unknown until after the launch of \tess. We also provided each fit with starting points on several parameters from the TIC. Specifically, we adopted the TIC-derived values for the host star's mass, radius, and effective temperature as well as the companion's orbital period, time of conjunction, and radius. We retrieved these values from the \tess\ mission catalog on ExoFOP\footnote{\url{https://exofop.ipac.caltech.edu/tess/}} \citep{exofopdoi}. We performed a preliminary fit with \exofast\ on each system which included fitting a linear term to the radial velocities in order to account for a long-term drift due to unseen outer companions. In every case except for TOI-4737, this resulted in a slope consistent with zero within 1-sigma, and we subsequently fixed the slope to zero in all subsequent fits for these systems. For TOI-4737, we continued to fit for this long-term trend, and in the final iteration of these fits which we publish here, we found a slope of $-1.63 \pm 0.23$ m\,s$^{-1}$\,day$^{-1}$. Each system's final fit was run to the adopted convergence criteria suggested by \citet{Eastman:2019} of at least 1000 independent draws and a Gelman-Rubin statistic $< 1.01$. See Table \ref{tab:medians} for the priors used, and the median values determined from our analysis.

\begin{figure*}[p]
    \centering
    \includegraphics[width=\textwidth,height=0.8\textheight,keepaspectratio]{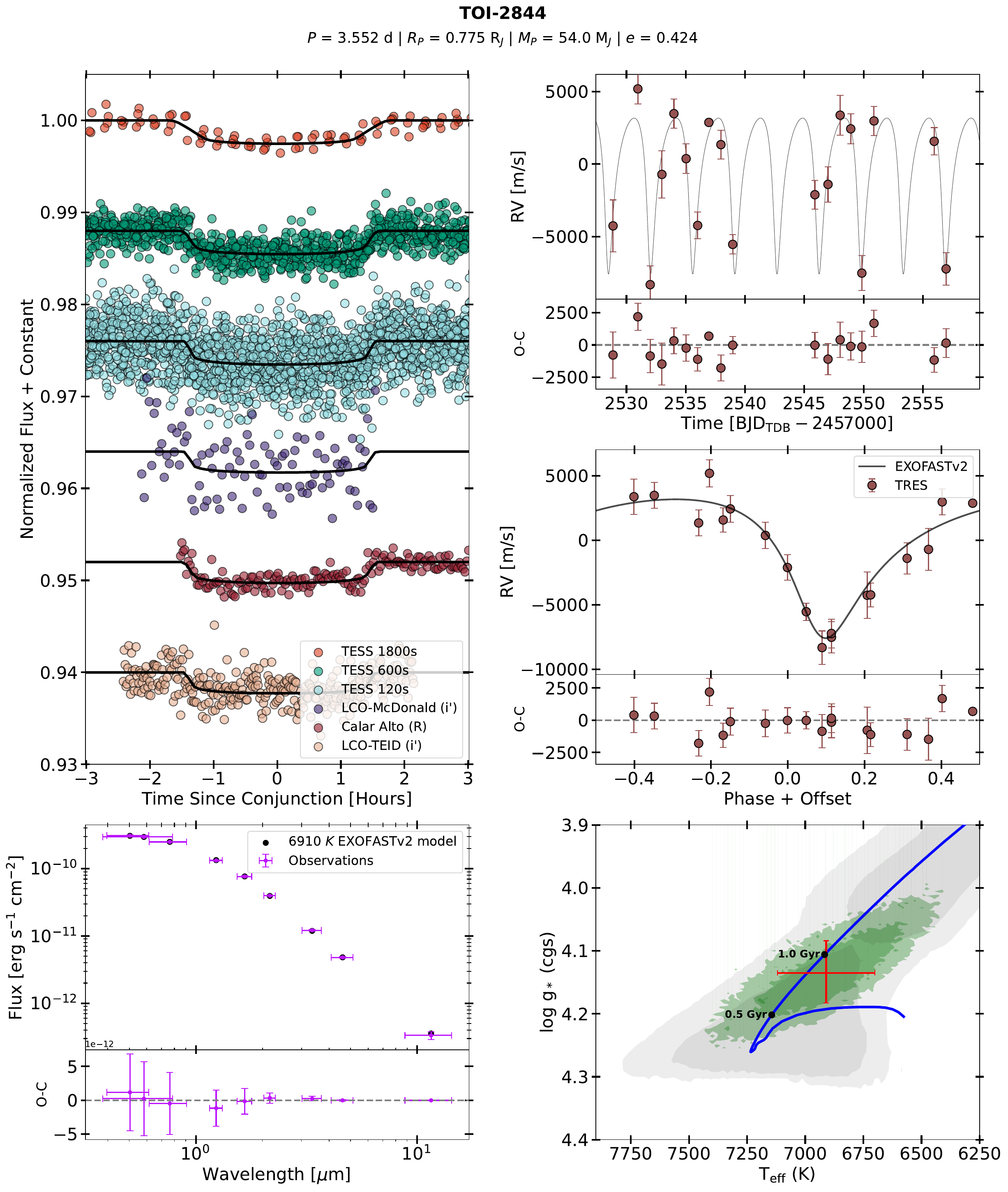}
    \caption{\tess , Follow-up and archival observations of TOI-2844 compared to the the \exofast\ results. \textbf{Upper left}: Unbinned \tess\ and follow-up ground-based transits, phase-folded and shown in comparison to the best fit \exofast\ model with an arbitrary normalized flux offset. Multiple \tess\ sectors in the same cadence are stacked on top of each other. \textbf{Bottom left}: The spectral energy distribution of the target star compared to the best-fit \exofast\ model. Residuals are shown on a linear scale, using the same units as the primary y-axis. \textbf{Upper right}: RV observations versus time, including any significant long-term trend. The residuals are shown in the subpanel below in the same units. \textbf{Middle right}: RV observations phase-folded using the best-fit ephemeris from the \exofast\ global fit. The phase is shifted so that the transit occurs at Phase + Offset = 0. The residuals are shown in the subpanel below in the same units. \textbf{Bottom right}: The evolutionary track and current evolutionary stage of the primary star according to the best-fit MESA Isochrones and Stellar Tracks (MIST) model. The blue line indicates the best-fit MIST track, while the gray shaded contours show the 1$\sigma$ and 2$\sigma$ constraints on the star’s current $T_{\rm eff}$ and $\log{g}$ from the MIST isochrone alone. The green contours represent the 1$\sigma$ and 2$\sigma$ constraints on the star’s $T_{\rm eff}$ and $\log{g}$ from the \exofast\ global fit, combining constraints from observations of the star and planet. The red cross indicates the median and 68\% confidence interval reported in Table \ref{tab:medians}.}
    \label{fig:2844}
\end{figure*}

\begin{figure*}[]
    \includegraphics[width=\linewidth]{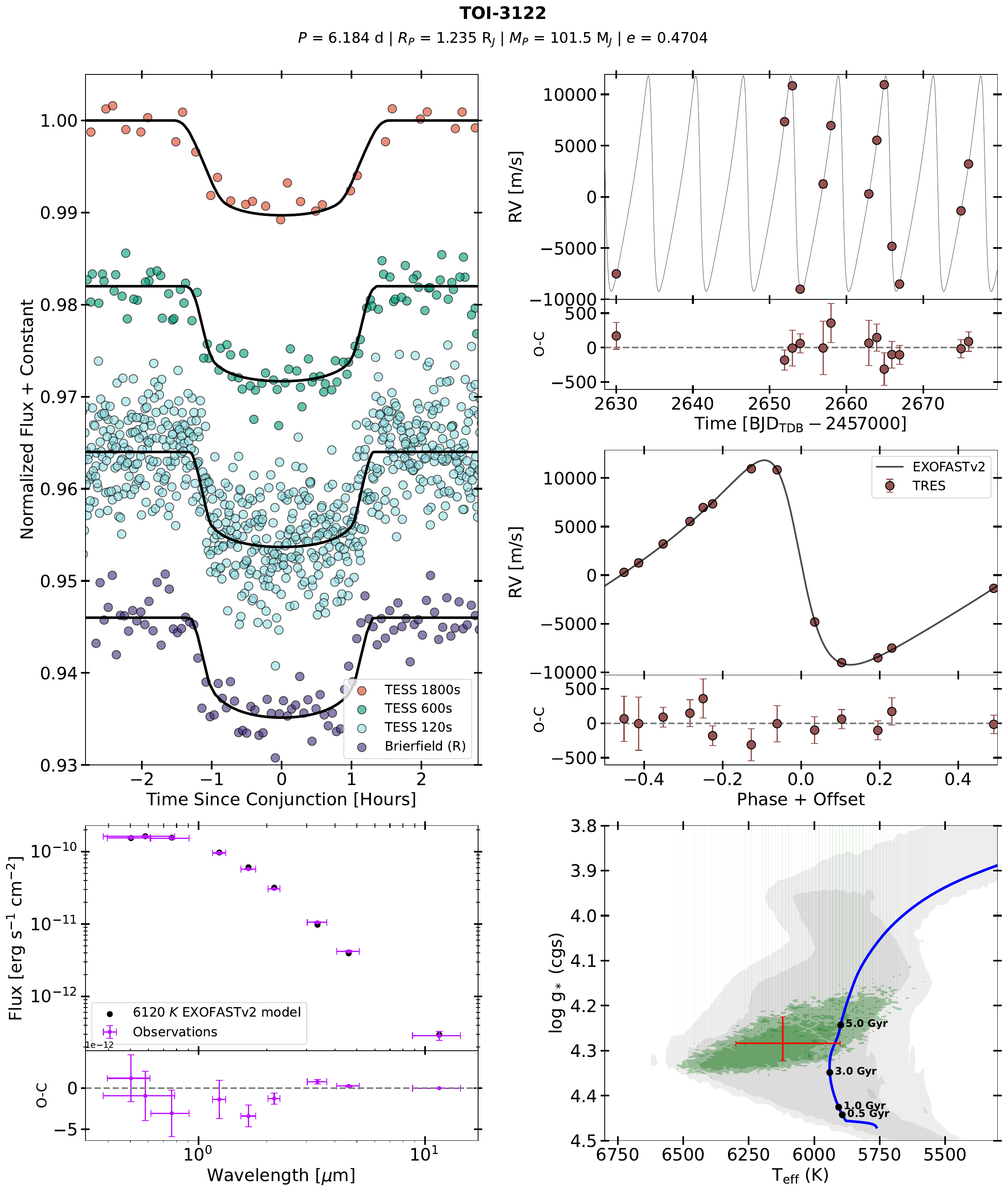}
    \caption{Same as Figure \ref{fig:2844} except for TOI-3122.}
    \label{fig:3122}
\end{figure*}

\begin{figure*}[]
    \includegraphics[width=\linewidth]{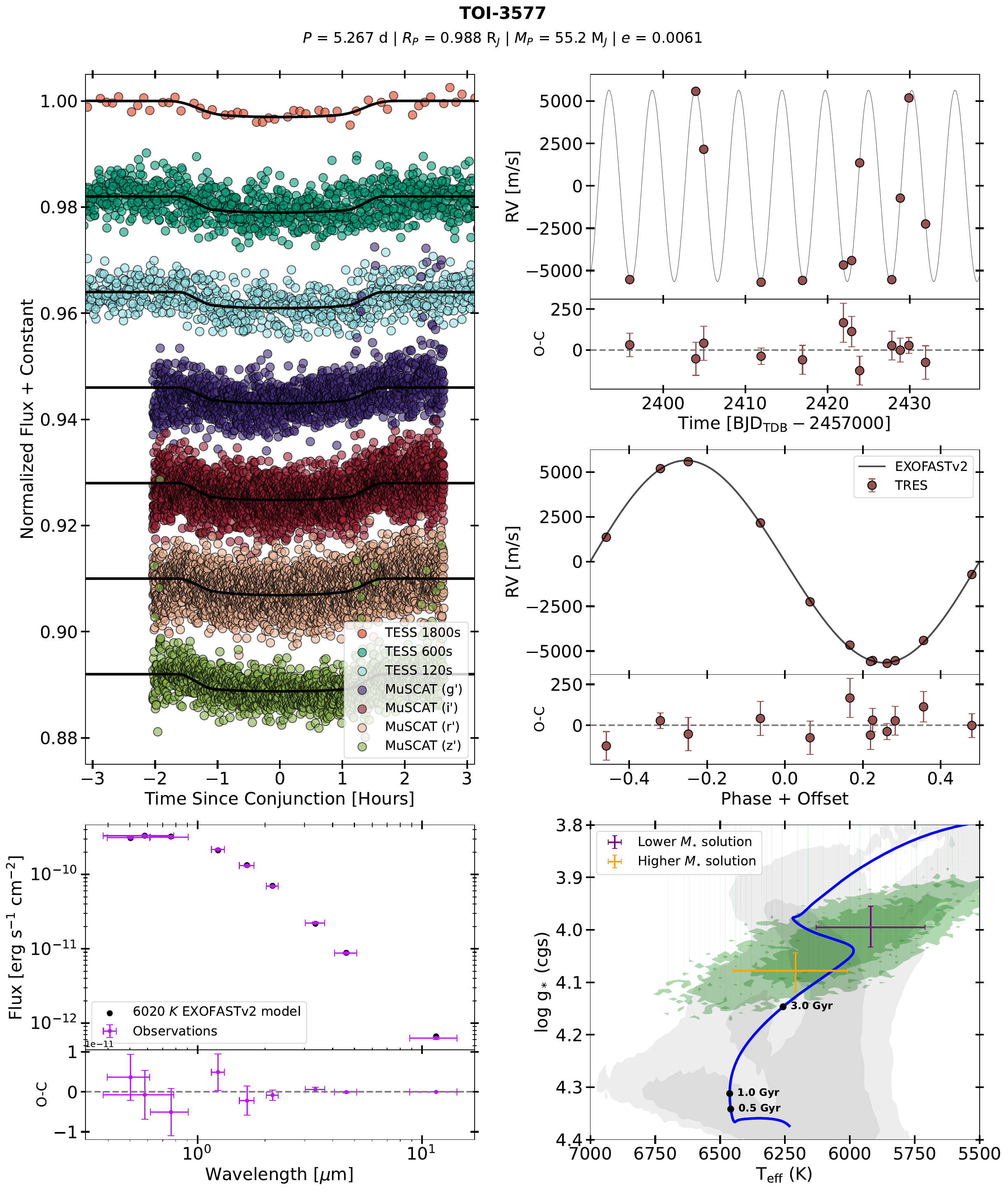}
    \caption{Same as Figure \ref{fig:2844} except for TOI-3577. TOI-3577's fit resulted in a bimodal solution. We characterized both solutions independently as described in \S\ref{sec:analysis}, and they are both are shown in the \textbf{bottom right} plot.}
    \label{fig:3577}
\end{figure*}

\begin{figure*}[]
    \includegraphics[width=\linewidth]{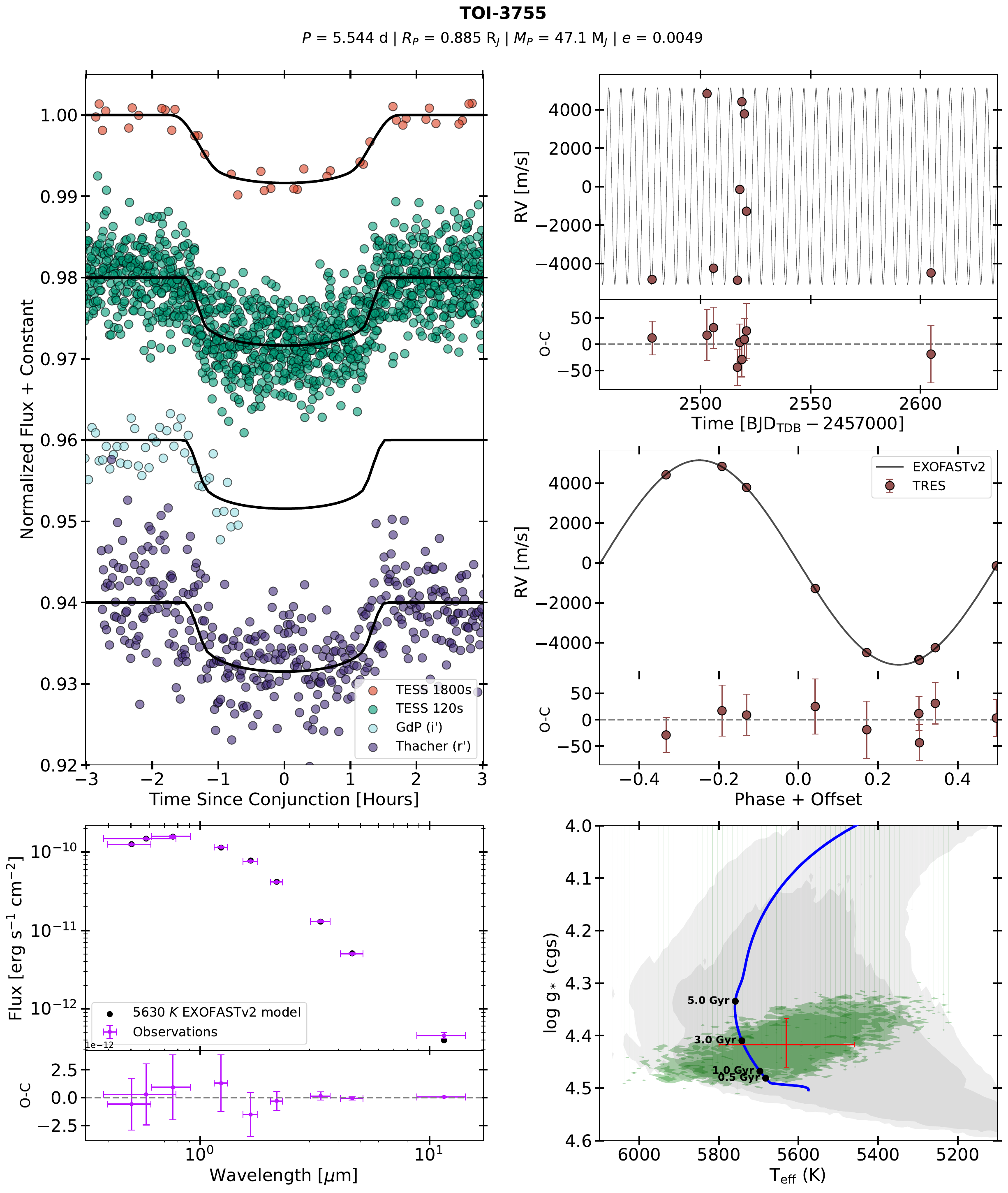}
    \caption{Same as Figure \ref{fig:2844} except for TOI-3755.}
    \label{fig:3755}
\end{figure*}

\begin{figure*}[]
    \includegraphics[width=\linewidth]{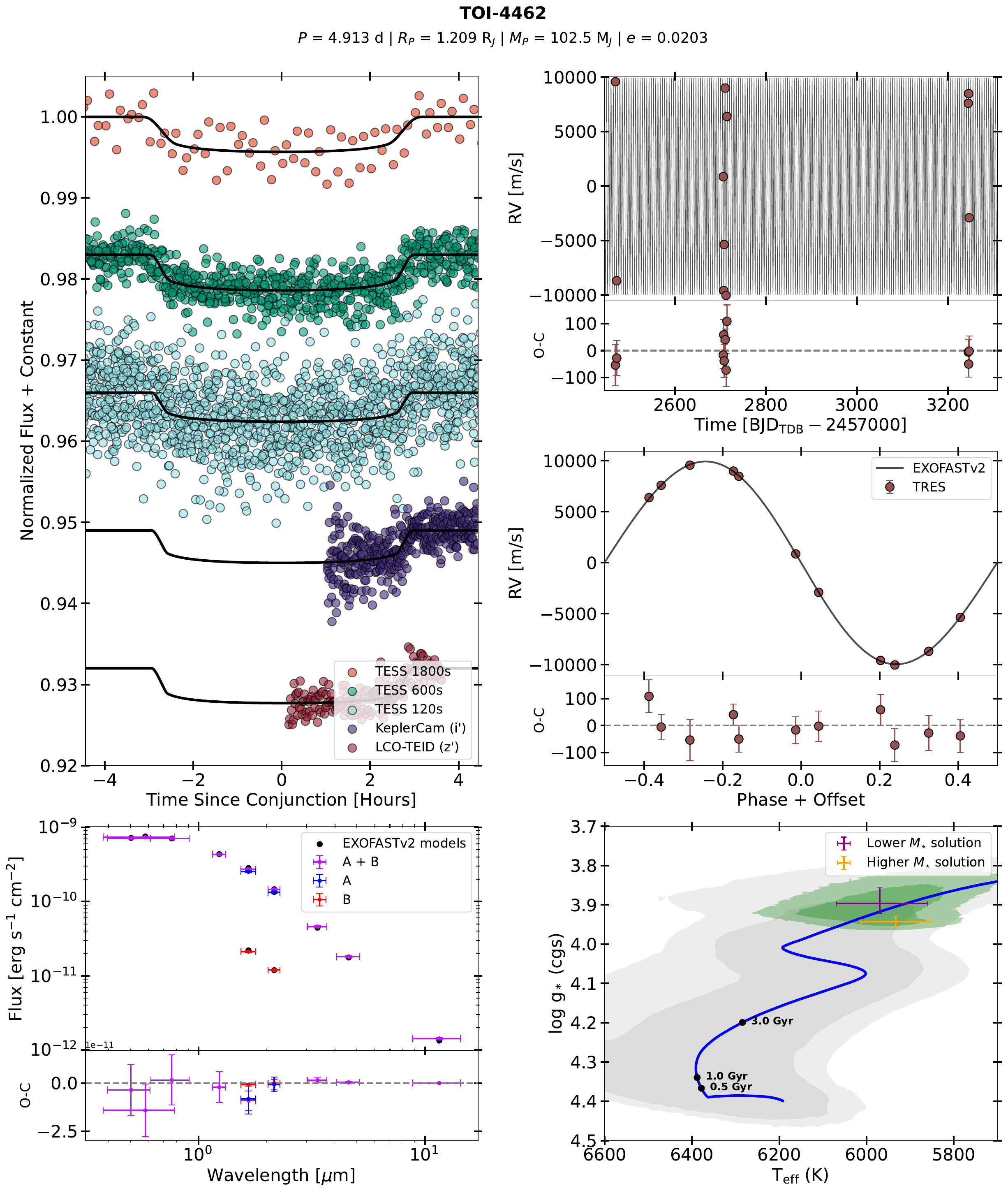}
    \caption{Same as Figure \ref{fig:3577} except for TOI-4462. Both TOI-4462 A and B are shown in the \textbf{bottom left}. The \exofast\ models for TOI-4462 A and B are 5970 $K$ and and 4660 $K$ respectively.}
    \label{fig:4462}
\end{figure*}

\begin{figure*}[]
    \includegraphics[width=\linewidth]{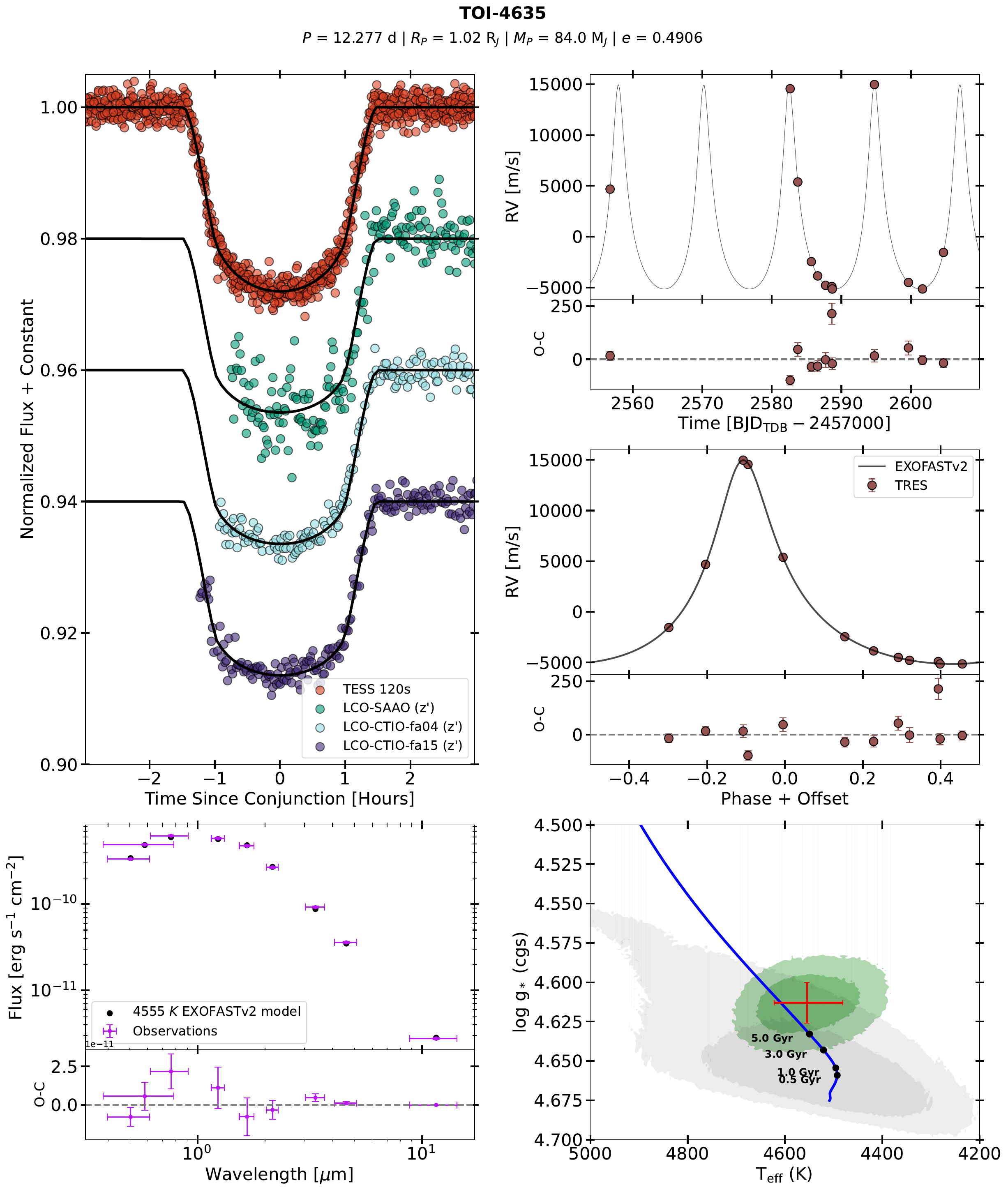}
    \caption{Same as Figure \ref{fig:2844} except for TOI-4635.}
    \label{fig:4635}
\end{figure*}

\begin{figure*}[]
    \includegraphics[width=\linewidth]{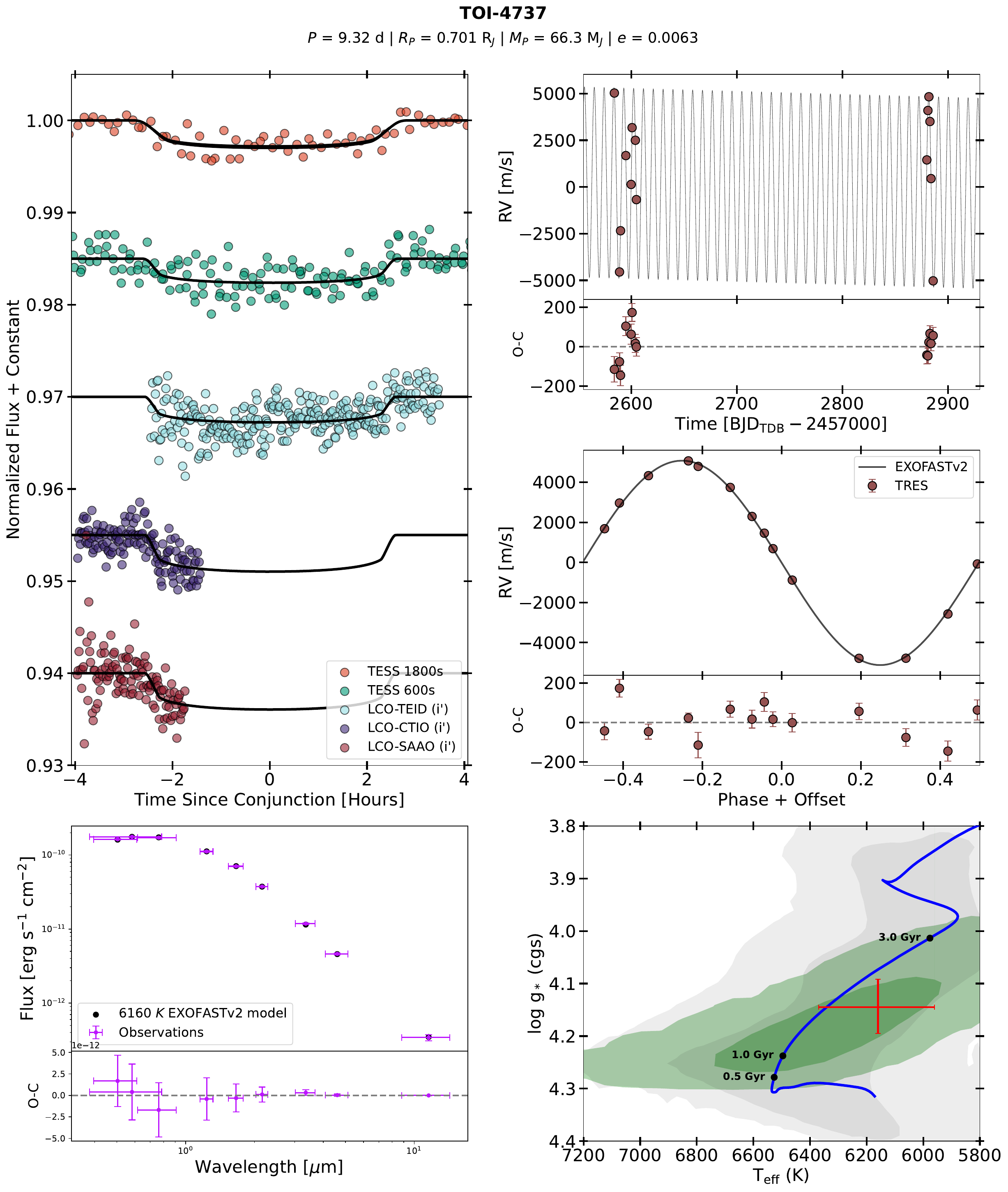}
    \caption{Same as Figure \ref{fig:2844} except for TOI-4737.}
    \label{fig:4737}
\end{figure*}

\begin{figure*}[]
    \includegraphics[width=\linewidth]{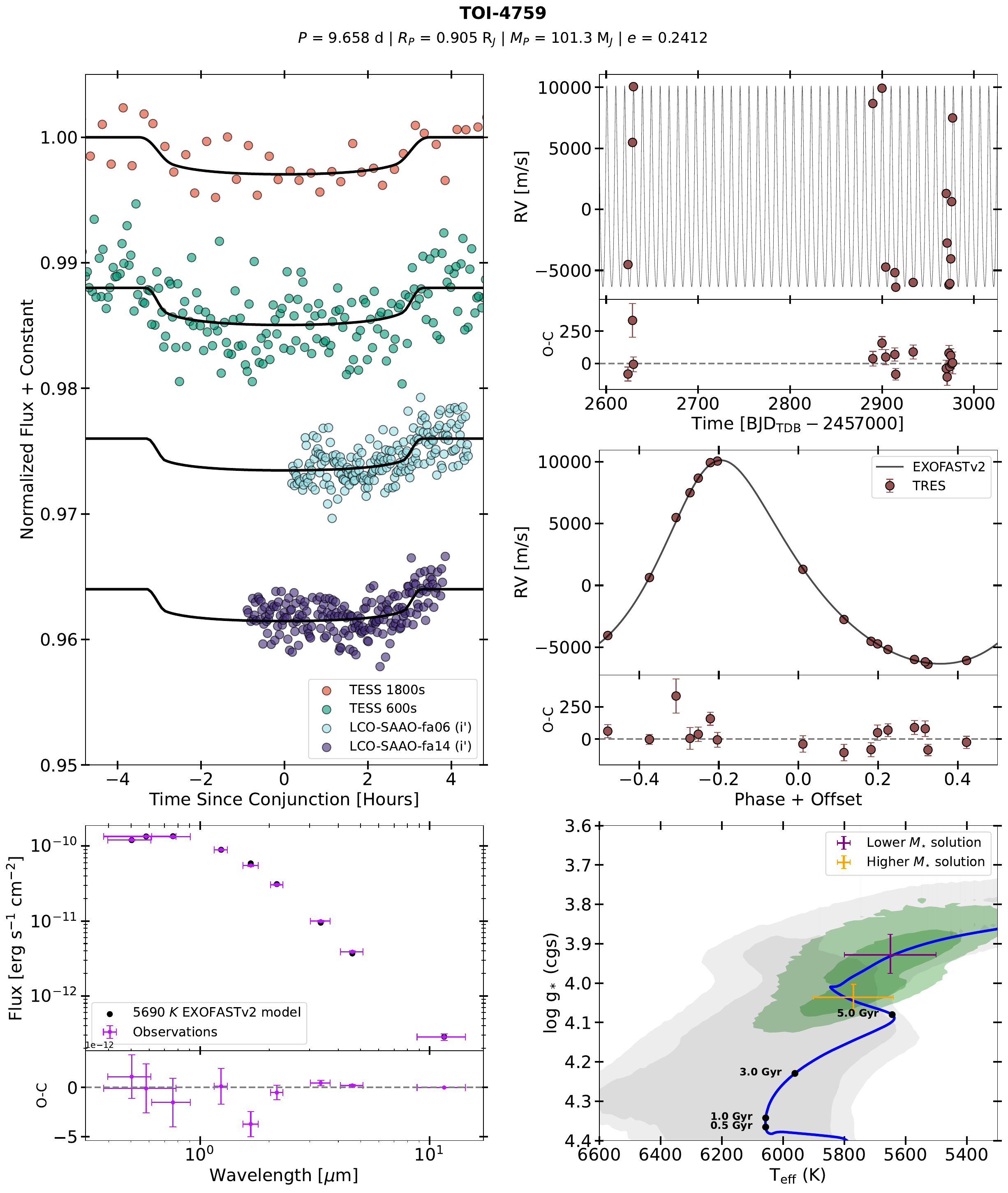}
    \caption{Same as Figure \ref{fig:3577} except for TOI-4759.}
    \label{fig:4759}
\end{figure*}

\begin{figure*}[]
    \includegraphics[width=\linewidth]{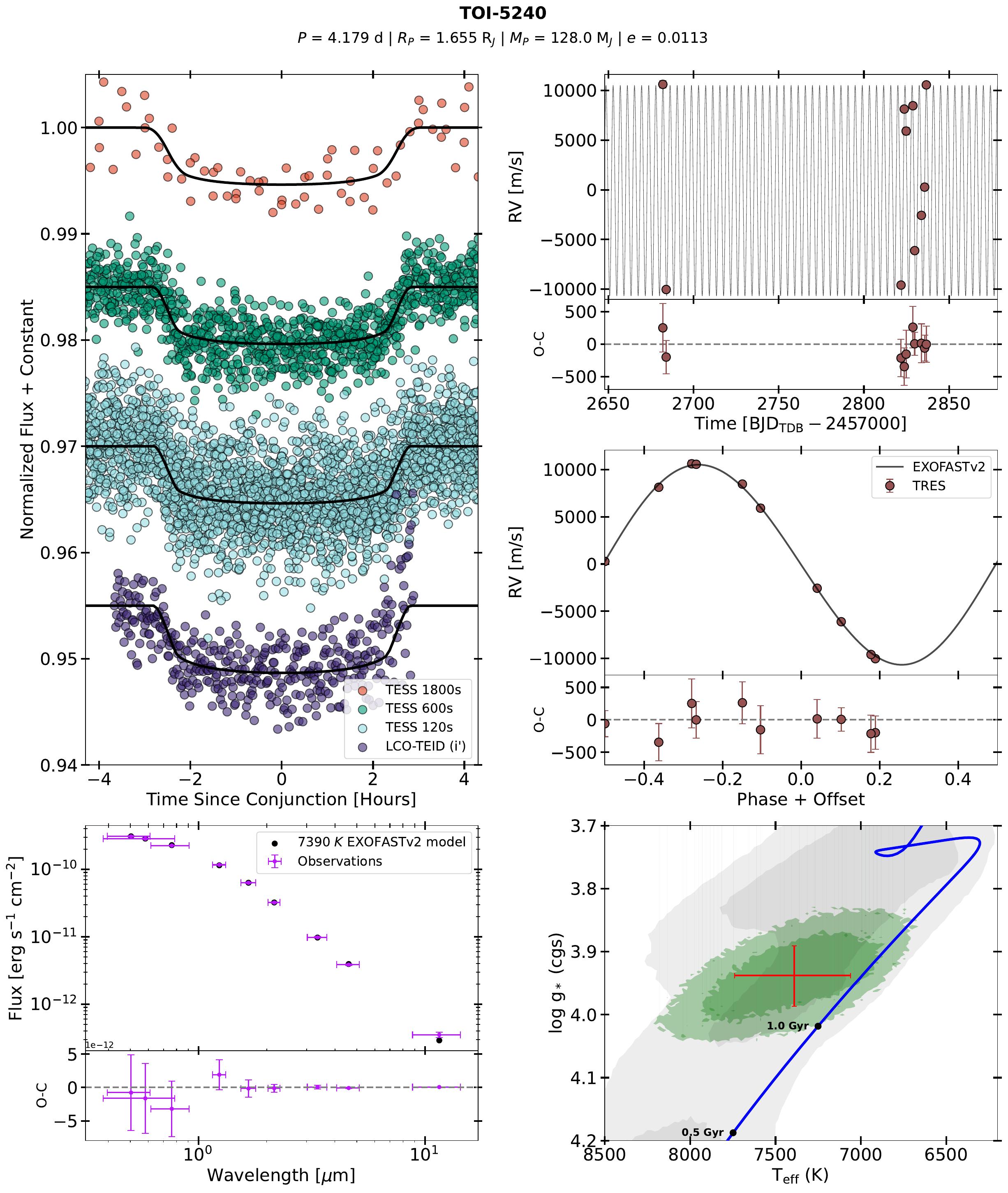}
    \caption{Same as Figure \ref{fig:2844} except for TOI-5240.}
    \label{fig:5240}
\end{figure*}

\begin{figure*}[]
    \includegraphics[width=\linewidth]{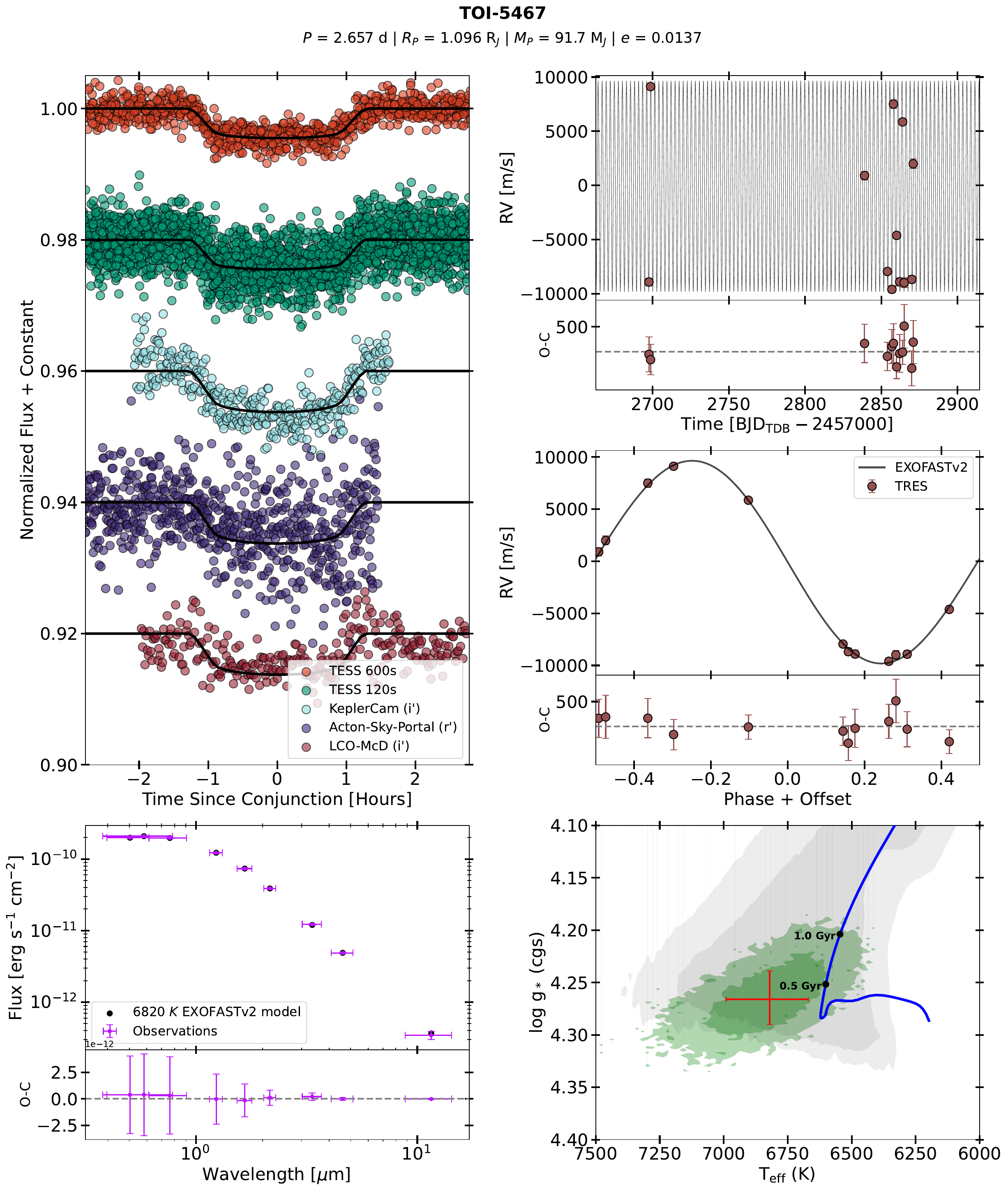}
    \caption{Same as Figure \ref{fig:2844} except for TOI-5467.}
    \label{fig:5467}
\end{figure*}

\begin{figure*}[]
    \includegraphics[width=\linewidth]{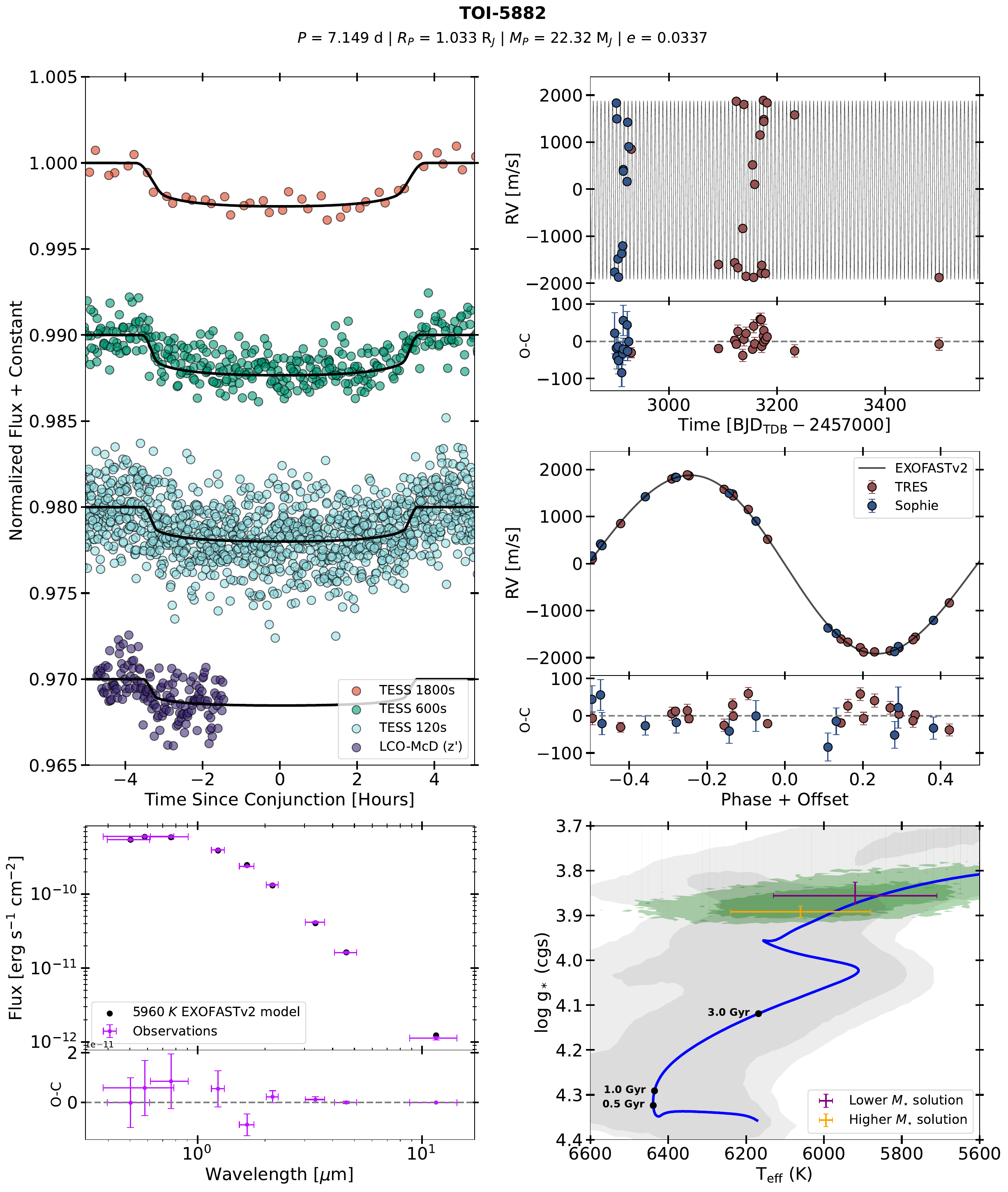}
    \caption{Same as Figure \ref{fig:3577}, except for TOI-5882.}
    \label{fig:5882}
\end{figure*}

Four of our fits resulted in bimodal posterior distributions. This typically arises when \exofast\ is unable to distinguish between a host star that is at the end of the main sequence versus the subgiant branch resulting in a high and low stellar mass solutions. Indeed, this was the case in all four bimodal systems presented here (TOI-3577, TOI-4462, TOI-4759, and TOI-5882). We characterized each solution independently by splitting the posterior distributions at the local minimum between the two solutions. We present both solutions for the sake of transparency, but in each case we adopt the higher probability solution as the preferred parameter set. See Table \ref{tab:medians_bimodal} for the priors used, and the median values determined for both solutions of these bimodal systems. Plots of the transit photometry, radial velocities, SED, and MIST evolutionary tracks for each system presented in this work are presented in Figures \ref{fig:2844}-\ref{fig:5882}.

\begin{deluxetable*}{l>{\centering}cccccc}
\tablecaption{Median Values and 68\% Confidence Intervals for Fitted Stellar and Planetary Parameters \label{tab:medians}}
\tabletypesize{\scriptsize}
\input{median_table_3}
 \footnotesize{\textbf{NOTES:}\\
 The priors listed at the top of the table are labeled as $\mathcal{G}$[mean, standard deviation] if they are Gaussian priors and $\mathcal{U}$[lower limit, upper limit] if they are uniform priors.\\
}
\end{deluxetable*}
\addtocounter{table}{-1}

\begin{deluxetable*}{l>{\centering}cccccc}
\tablecaption{\textit{(Continued)}}
\tabletypesize{\scriptsize}
\input{median_table_10}
\end{deluxetable*}

\begin{deluxetable*}{l>{\centering}cccccc}
\tablecaption{Median Values and 68\% Confidence Intervals for Fitted Stellar and Planetary Parameters for Bimodal Systems \label{tab:medians_bimodal}}
\tabletypesize{\scriptsize}
\input{median_table_7}
\end{deluxetable*}
\addtocounter{table}{-1}

\begin{deluxetable*}{l>{\centering}cccccc}
\tablecaption{\textit{(Continued)}}
\tabletypesize{\scriptsize}
\input{median_table_11}
\end{deluxetable*}

\subsection{Multi-star Fitting in \exofast}
\label{sec:toi4462}

As discussed in \S\ref{sec:ao}, a stellar companion to TOI-4462 was detected 0.4$\arcsec$ away, which was blended in all catalog photometry and a significant factor in the dilution of the transit light curves. Given that the probability of a chance alignment is low, and the high Gaia Re-normalized Unit Weight Error (RUWE) of 3.13, we assumed this companion is bound to the primary star. We undid the deblending that SPOC applies to the TESS lightcurves so that we could more accurately model it based on our multi-component SED model. We modeled both stars simultaneously, each with their own MIST evolutionary model, while assuming that the age, initial metallicity, distance, and extinction is the same for both stars. In addition, we modeled a spectral energy distribution for each star, constraining the sum of both stars with the catalog photometry of the unresolved TOI-4462 system, and the difference between the two stars with the AO photometry from PHARO, as shown in Figure \ref{fig:AO} (Left). We therefore fit for dilution terms that were then constrained by the multi-component SED model, integrated at the transit-observed bands assuming a 5\% floor in the theoretical dilution from the model atmospheres. That is, we applied an adaptive prior penalty of

\begin{equation}
    \ln{\mathscr{L}} = 0.5 \left(\frac{D_{\rm Step} - D_{\rm SED}}{0.02 D_{\rm Step}}\right)^2
\end{equation}

\noindent where $D_{\rm Step}$ is the modeled dilution at the current MCMC step and $D_{\rm SED}$ is the SED-derived dilution. This naturally propagates the uncertainty in the stellar properties, accounting for systematics in the theoretical atmospheres, to the light curve de-blending and transit depth.

\section{Discussion}
\label{sec:discussion}
The 5 BD-mass companions presented here increase the population of transiting brown dwarfs to over 50. While this number is expected to continue growing, it is worth analyzing the sizable population that has been put together thus far in the context of planet-like and star-like formation. We have also added 6 new transiting low-mass stars to the population $>80$ $M_J$.  Accurate mass and radius measurements are rare for these objects which will be vital for anchoring our understanding of stellar formation and evolution. We may find that early trends or features observed in the growing population from previous efforts have been reinforced, or lost significance (perhaps even disappearing) in the wake of new discoveries. One such feature of particular interest is the so-called "brown dwarf desert" and its potential role in dividing the brown dwarfs into distinct planet-like and star-like groups. We discuss these trends below in \S\ref{sec:population} and \S\ref{sec:feh}. We also note that one of the BD-hosting stars presented here, TOI-5882, has a significant absorption feature at 6708 \AA\ which we attribute to lithium. We discuss in \S\ref{sec:toi5882} the implications of this, as well as how it affects our determination of the system's age.

\subsection{The Transiting Brown Dwarf Desert}
\label{sec:population}
Perhaps the most discussed feature to emerge from the growing transiting BD population is that of the brown dwarf desert. The phrase was originally coined in the earliest days of exoplanet discovery \citep{Marcy1997, Latham1998} in reference to the lack of BD-mass companions discovered by RV surveys at that time. It has since evolved over time with studies like \citet{Grether&Lineweaver2006} which analyzed RV-detected companions with periods less than 5 years and found the "driest" part of the desert was between $13 - 56$ \mj. Then \citet{Ma&Ge2014} examined trends in the RV discovered population and refined the measurement to be between $35 < m\,\sin{i} < 55$ \mj\ and with $Period < 100$ days.

\begin{figure*}[]
    \includegraphics[trim = 0 0 0 0, width=\linewidth]{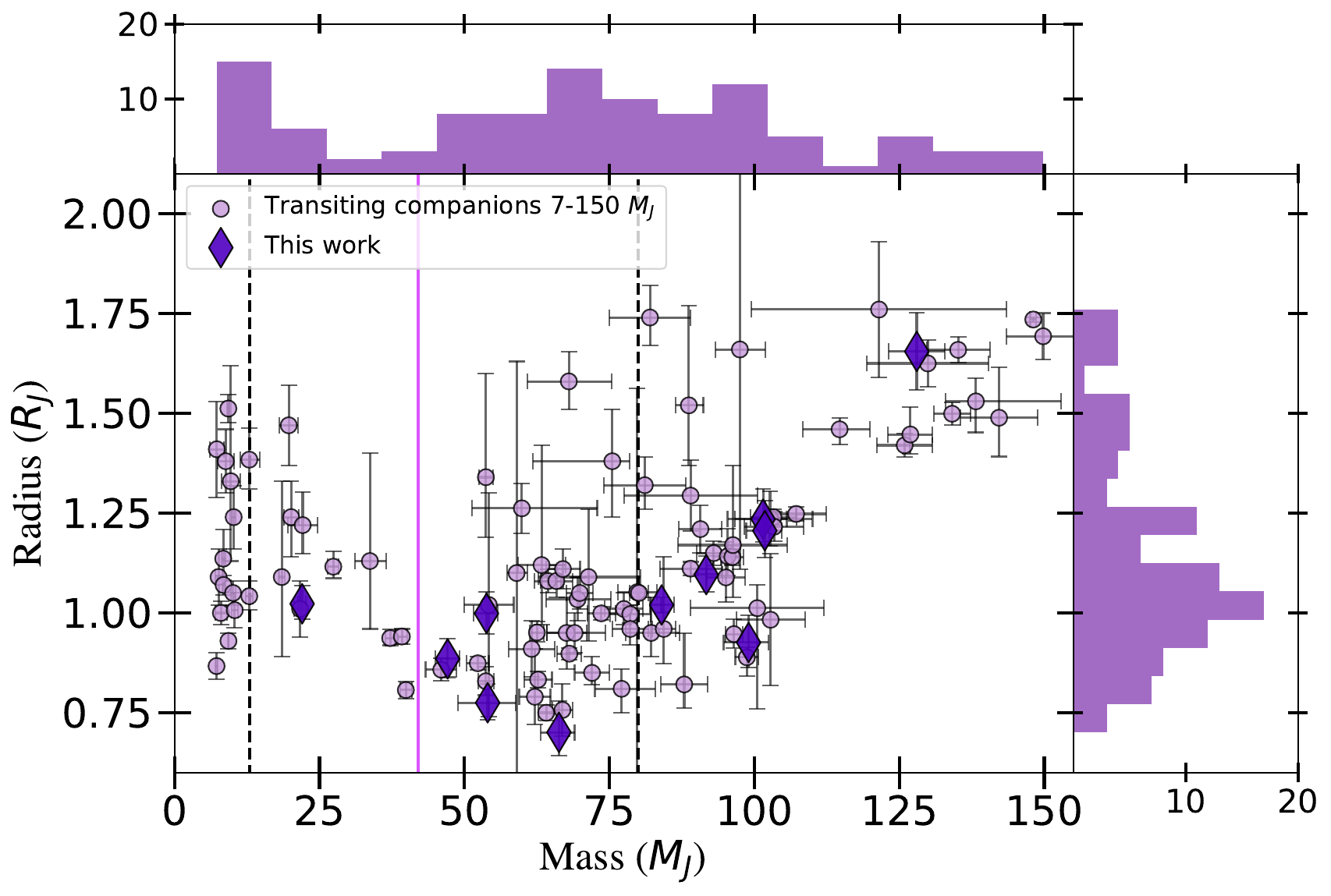}
    \caption{Radius versus mass for all transiting companions from 7 - 150 \mj. The black dashed lines depict the canonical 13 and 80 \mj\ BD boundaries. The solid purple line at 42 \mj\ shows the proposed \citet{Ma&Ge2014} boundary between planet and star-like BDs. \textbf{Note:} systems where the primary object is a white dwarf or brown dwarf are not shown. \textbf{References:} \citet{Henderson2024b} and references therein as well as \citet{Bakos:2010, Buchhave:2011, Tingley2011, Parviainen2014, Bonomo2015, Esteves2015, Stassun2017, Bento2018, Canas2018, Cooke2020, Cortez-Zuleta2020, El-Badry2023, Lambert2023, Schmidt2023, Dalba2024, Davis2024, Eberhardt2023, Henderson2024a, Swayne2024, Wang2024}}
    \label{fig:massradius}
\end{figure*}

We now have a more substantial population of transiting BD systems which have precisely measured radii and masses. It's worth exploring how well the trends found in the the RV discovered sample hold up in the transiting BD regime. Of course, because of the transit probability decreases with period, transiting systems tend to have much shorter orbital periods than their counterparts discovered through direct imaging and RV campaigns. As a result we are largely investigating a different, more limited parameter space than the previous studies of \citet{Ma&Ge2014} for example. We found that the most sparsely populated area of the transiting BD desert appears to be the entire low-mass BD regime ($<\sim$ 42\mj), and the work presented here contributes 1 new BD (TOI-5882) to this underpopulated region (see Figure \ref{fig:massradius}). We also note that the apparent drop off in systems with companions above the substellar limit (80 \mj) is likely unphysical and is more plausibly due to selection bias since most of these systems have been discovered via exoplanet discovery pipelines. In order to more accurately describe trends emerging near the hydrogen fusion boundary, a more unbiased sample will need to be produced. 

\begin{figure*}[]
    \includegraphics[trim = 0 0 0 0, width=\linewidth]{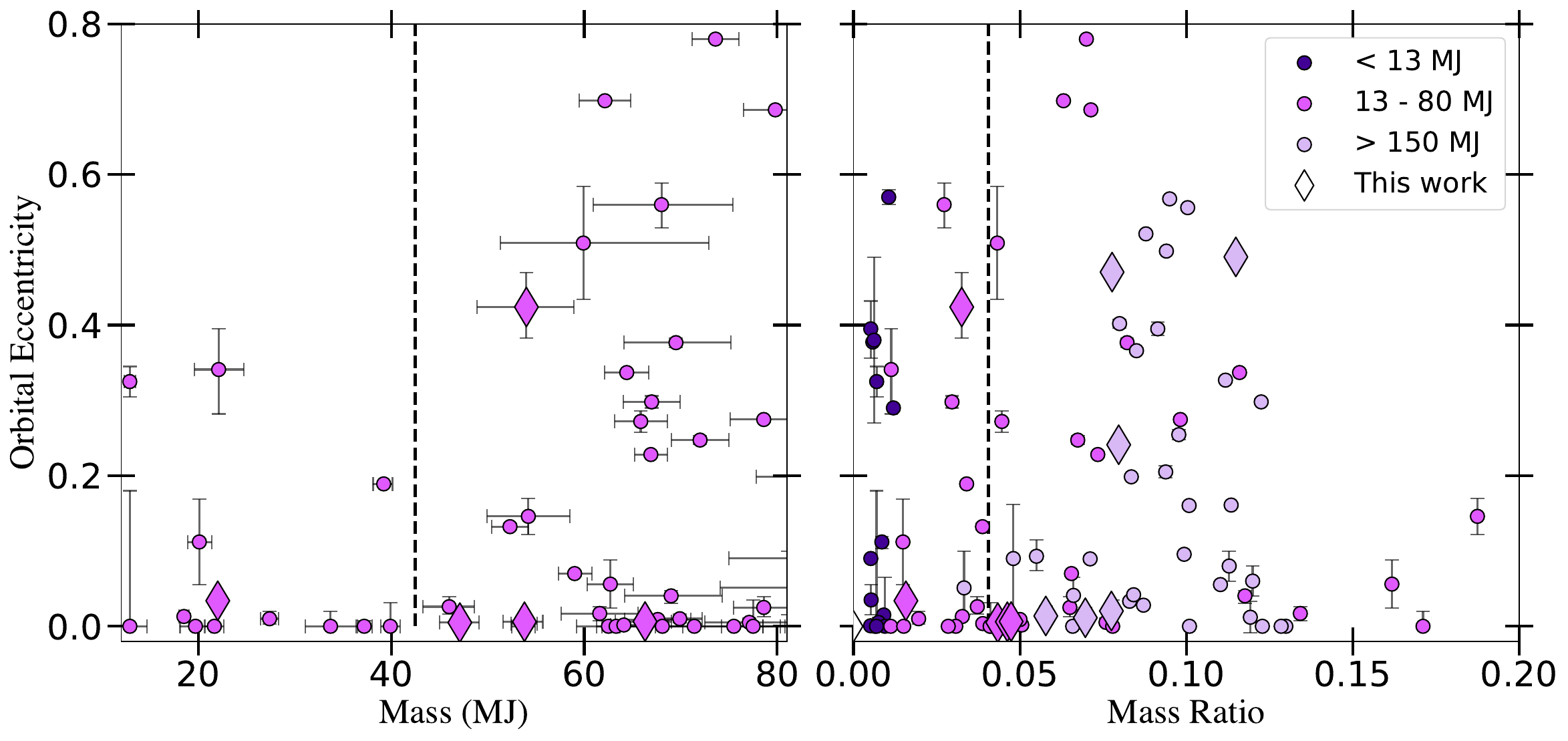}
    \caption{(Left) Eccentricity versus mass for transiting BDs with a dotted line at the proposed \citet{Ma&Ge2014} boundary between planet and star-like BDs. (Right) All transiting companions ranging from 7 - 150 \mj\ in eccentricity versus mass ratio with a dashed line at the same 42 \mj\ location assuming a 1.0 \msun\ host star. As discussed in \S\ref{sec:population}, the eccentricity dichotomy between low and high mass BDs does not hold up when plotted against mass ratio, suggesting that this feature may not represent the boundary between planet-like and star-like BDs.}
    \label{fig:massecc}
\end{figure*}

Another key trend that was first noted by the \citet{Ma&Ge2014} RV study is in the eccentricity vs $m\,\sin{i}$ distribution. They found that eccentricity decreases as companion mass increases up until $m\,\sin{i} \sim$ 42 \mj, right in the middle of the driest part of the brown dwarf desert. On the other hand, companions more massive than 42 \mj\ cover a much larger range in eccentricity and little to no correlation with the companion mass. The authors attributed this trend as evidence of a $\sim$42 \mj\ transition point between the planet and star formation mechanisms. As the transiting BD population has developed, several studies have drawn comparisons to these results, some finding evidence for the same trends \citep{Grieves2021, Henderson2024a}, while others have noted low-mass BDs with higher than expected eccentricities \citep{Page2024}. However, such claims have historically been subject to the small sample size and selection effects that accompany the transiting BD population. Now that this population exceeds 50 systems, we can at least start to alleviate the risks of small number statistics. Figure \ref{fig:massecc} (Left) shows the eccentricity vs. companion mass distribution for the transiting brown dwarfs, and it is clear that there are more systems above 42 \mj\ with high eccentricity than there are below. About 30\% of systems below 42 \mj\ have eccentricities $> 0.1$ compared to about 45\% for their higher-mass counterparts. However, we argue that this trend alone is not necessarily supportive of a 42 \mj\ transition between the planet and stellar formation mechanisms. If the low-mass transiting BDs are indeed dominated by the planet formation mechanism, then they should be subject to the same evolutionary pathways as the hot Jupiters. The hot Jupiter eccentricity distribution has been shown to be most consistent with high eccentricity migration mechanisms and thus are ultimately sculpted by tidal recircularization \citep{Rodriguez2022, Schulte2024}. We know that this process depends more fundamentally on the mass ratio of the system rather than just the companion mass as evidenced by the tidal recircularization time scale \citep[Equation 2 of][]{Adams:2006}. So, if the trend in eccentricity vs companion mass were indeed indicative of a separation between planet-like and star-like formation processes, then we should expect to see the same trends emerge in eccentricity vs mass ratio. Namely, low mass ratio systems should exhibit a much smaller range of eccentricities than their high mass ratio counterparts. However, as shown in Figure \ref{fig:massecc} (Right), we see the opposite. The eccentricity dichotomy between low and high mass companions seems to disappear when plotted against mass ratio. We therefore argue that the eccentricity vs companion mass distribution of transiting companions does not support a 42 \mj\ transition point. 

\subsection{Transiting Brown Dwarf Metallicities}
\label{sec:feh}
\begin{figure}[]
    \includegraphics[width=\linewidth]{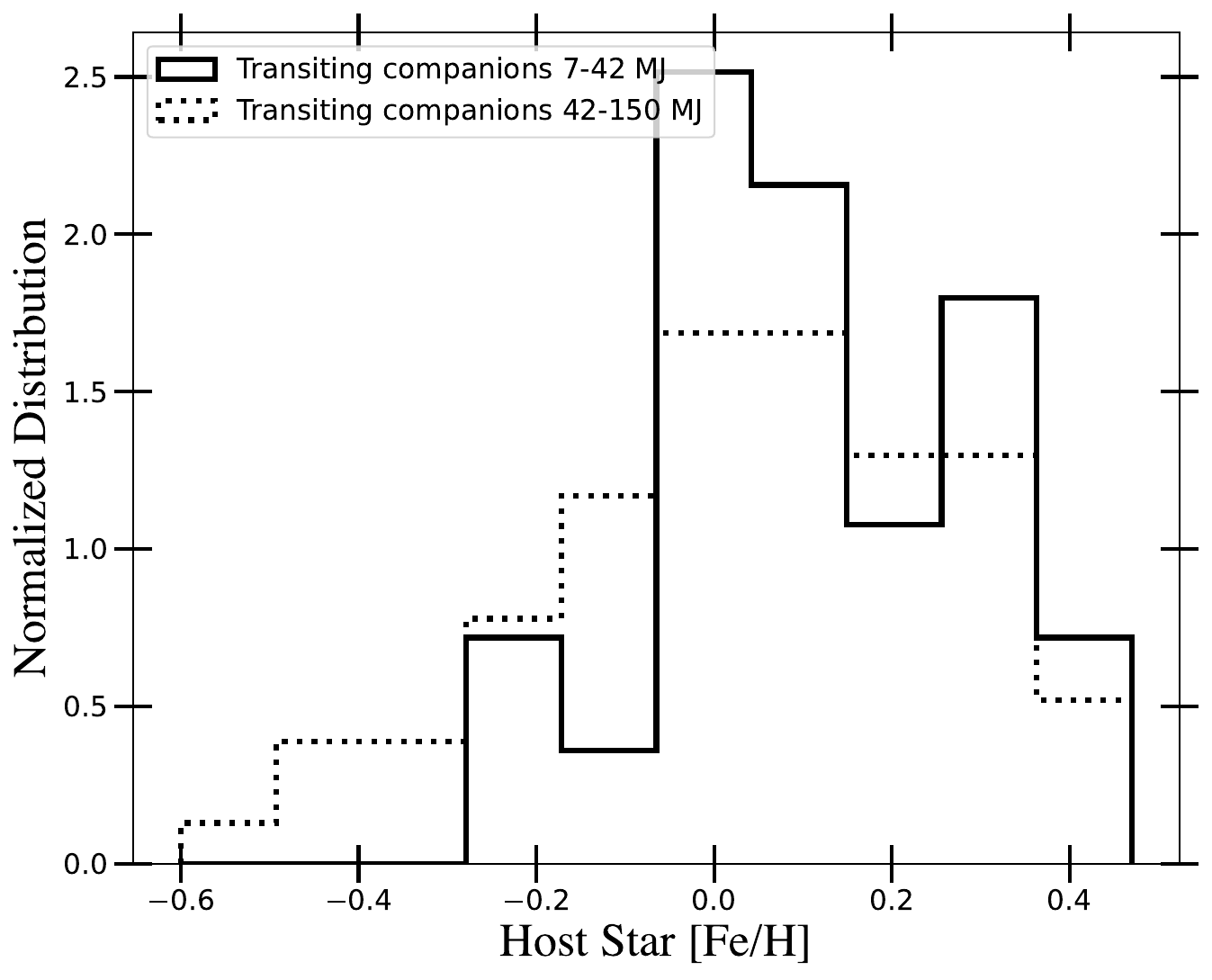}
    \caption{The solid-lined histogram depicts the metallicity distribution of transiting companions ranging from 7 - 42 \mj\ while the dotted-lined histogram depicts the metallicity distribution of transiting companions from 42 - 150 \mj. Both histograms are normalized such that their areas are equal to 1. There appears to be a slight trend towards higher metallicities for host stars with lower mass companions. \textbf{Note:} The metallicities shown here are the values cited by their original discovery papers, and hence represent a heterogeneous sample with a variety of different measurement techniques.}
    \label{fig:feh}
\end{figure}

Eccentricity is likely not the only parameter that could offer insight into at which critical companion mass the dominant formation mechanism turns over from planet-like to star-like. For nearly three decades we have known of the giant planet-metallicity correlation, in which hot Jupiter hosting stars tend to be more metal rich than their counterparts with no discovered planets \citep{Gonzalez1997, Santos2003, Fischer2005}. If the low-mass transiting BDs are predominantly forming in the same way as the hot Jupiters, then we should expect their host stars to exhibit the same metallicity enhancement when compared to the high-mass BD hosts. \citet{Schlaufman2018} tested this hypothesis using the metallicities of transiting companions in the range $ 0.1 - 300$ \mj\ to show that transition between core accretion and fragmentation may be as low as $4-10$ \mj. However, at the time of this study, there were only 27 transiting companions known between $13 - 300$ \mj\, limiting the ability to probe potential higher-mass transition points. Now that we have access to significantly more systems in this mass regime, we can better probe the same $\sim$42 \mj\ transition. In Figure \ref{fig:feh} we show a preliminary look at testing this hypothesis. Qualitatively, it appears that the lower-mass companions ($7 - 42$ \mj) preferentially orbit more metal-rich host stars. However, a two-sample Kolmogovorov-Smirnov test yields a $p$-value of 0.35, too high to reject the null hypothesis that the high mass and low mass samples are drawn from the same underlying distribution. 

We note also a few important caveats for interpreting the metallicity distribution. First, we chose a 42 \mj\ cutoff for historical reasons in order to compare to the original hypothesis presented by \citet{Ma&Ge2014} as well as the eccentricity distribution presented in \S\ref{sec:population}. It may be that a more appropriate boundary separating the population will be found after a more comprehensive analysis which may be in better agreement with the lower-mass \citet{Schlaufman2018} transition. With just 26 companions below 42 \mj\ in our $7 - 150$ \mj\ sample, we chose note to investigate possible lower-mass transitions. A more in-depth analysis including the population of giant planets will need to be done to more precisely probe a lower mass transition. An unbiased sample of the population of companions across the substellar limit will also be required to better understand the selection effects currently affecting this population. We note also that the metallicities presented here are the reported values from each system's discovery which have been measured using a variety of different techniques and therefore the underlying biases affecting each measurement are not explored here.

\subsection{Lithium Detected in TOI-5882}
\label{sec:toi5882}
\begin{figure}[]
    \includegraphics[width=\linewidth]{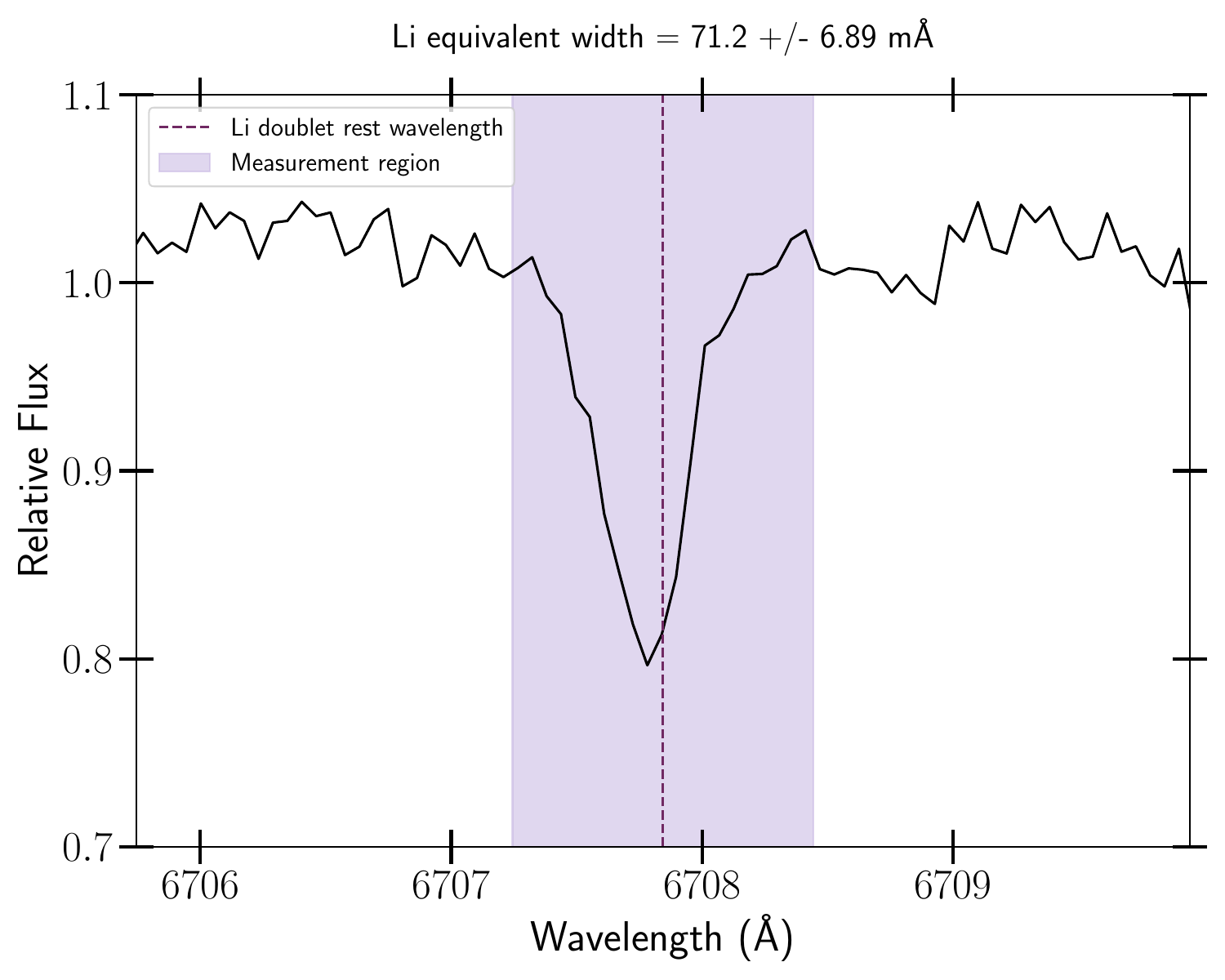}
    \caption{The co-added spectra of TOI-5882 zoomed in to the Li absorption line at 6708 \AA. The measurement region used to determine the equivalent width is shaded in purple with the dashed line at the center indicating the rest wavelength of the Li doublet, 6707.844 \AA.}
    \label{fig:lithium}
\end{figure}

During our analysis of TOI-5882's spectra, we found a significant absorption feature at 6708 \AA\ which we attribute to the lithium doublet. We measured the equivalent width of this feature using the \texttt{specutils} \citep{Earl:2022} package in Python (see Figure \ref{fig:lithium}). To perform this measurement, we first co-added all of the observed TRES spectra, after correcting each for the RV shift, to increase the signal-to-noise resolution (SNR) of the Li feature. The resulting co-added echelle order containing Li has an SNR of 56.8. Then, we defined a 1.4 \AA\ region centered on the rest wavelength of the Li doublet at 6707.844 \AA\ to measure the equivalent width. The resulting equivalent width is $71.2 \pm 6.89$ m\AA, where the uncertainty was estimated using Equation 6 in \cite{Cayrel:1988}.

The presence of lithium in stars is typically interpreted as an indicator of youth. This is due to the temperatures and pressures in the core being sufficiently high to destroy Li, which results in Li visible on the stellar surface slowly depleting as transport occurs between the core and the surface of the star \citep{Soderblom2014}. Despite this, we claim that TOI-5882 is likely not a young star, since we found no other signs of youth. To verify this, we performed a period search on each sector of TOI-5882's \tess\ light curves to characterize the rotation of the host star. Since young stars are typically born rapidly rotating and gradually spin down over time, an age can often be inferred from a star's rotation period if it is below the Kraft break \citep{Bouma2023}. We found a significant peak in the periodogram at 9.6 days, however, we are hesitant to adopt this as the true rotation period since periodicity beyond 1/3 of a \tess\ observing sector ($\sim$ 9 days) can be unreliable due to aliases induced by the \tess\ observing strategy and processing of light curves. Even if we were to believe that the 9.6 day periodicity is truly due to stellar rotation, it is still anomalous when compared to the observed rotation periods of young stars. T Tauri stars for example rarely exhibit rotation periods longer than 8 days \citep{Serna2021}, and gyrochronology shows that a 9.6 day period would be indicative of an age of approximately 1 Gyr given this star's effective temperature \citep{Bouma2023}. While a young age for TOI-5882 cannot be conclusively ruled out by its rotation, it is unlikely, especially combined with the lack of other youth indicators.

For example, we also searched for nearby comoving stars using \texttt{FriendFinder} \footnote{\url{https://github.com/adamkraus/comove}} \citep{Tofflemire2021} since their presence would indicate that TOI-5882 and its hypothetical nearby comovers have not yet dispersed from their birth location and hence would be young. \texttt{FriendFinder} identifies all nearby sources that fall within a selected search radius, and calculates the predicted tangential velocity $v_{\rm tan}$ for each source, assuming that they have Galactic velocity components ($U$,$V$,$W$) identical to TOI-5882. This predicted $v_{tan}$ is then compared to the true $v_{\rm tan}$ which is derived from \textit{Gaia} proper motions. Using a physical search radius of 30 pc around TOI-5882 and a difference between the predicted and measured $v_{\rm tan}$ of $<5\,$\kms\ we find no evidence that TOI-5882 is part of a comoving group. Furthermore, the nearest star-forming regions in Cygnus, where TOI-5882 is located, are much further away \citep[$> 1$ kpc;][]{Reipurth2008}.

Finally, we looked for an infrared excess as well as H$\alpha$ emission. Young stellar objects that retain a circumstellar disk show increased emission at infrared wavelengths \citep{Cotten2016}. We ruled out an infrared excess for TOI-5882 via our SED fitting in the global analysis where we see no significant infrared emission above the blackbody model in any of the WISE $W1$, $W2$, and $W3$ bandpasses. The presence of H$\alpha$ lines in emission is also characteristic of active young stars \citep{Briceno2019}, and we found no evidence of such emission. While the lack of these additional youth indicators do not completely rule out the possibility of a young host star, we believe it is more likely that TOI-5882 is a late subgiant star, as indicated by our most probable \exofast\ solution. This older age could then imply that the presence of Li is due to the infall of planetary material onto the host star. For a deeper dive into the origin of Li in TOI-5882, including its potential as system that has undergone a planetary engulfment, we alert the reader to Kotten et al. (in prep).

\section{Conclusions}
\label{sec:conclusions}
In this paper, we present the discovery of 11 new transiting companions from the \tess\ mission. We collected photometric time-series, spectroscopic, and high resolution imaging follow-up as a part of the \tess\ Follow-up Observing Program (TFOP) to rule our false positives and further characterize each system. Using \exofast, we performed a global fit on each system using the space and ground-based transits, spectroscopic RVs, and archival photometry to characterize both the host stars, and their transiting companions. We found that 5 of these systems are brown dwarfs ($13 < M_2 < 80$\mj) and 6 of them are very low mass stars from $ 80 < M_2 < 130$ \mj. This contribution to the transiting brown dwarf population increases it to 54 systems, a milestone that represents the population outgrowing the burdens of small sample statistics. 

Using this population that \tess\ has rapidly developed to a significant size, we offered some initial insight into the features that have started to appear. We revisit the idea of the "brown dwarf" desert for the short orbital periods probed by the transit method. We revisited the eccentricity-mass distribution that has been claimed as evidence of a 42 \mj\ transition between planet and star formation and showed that this trend does not seem to hold in eccentricity versus mass-ratio, calling into question whether eccentricity truly does offer insight into the formation mechanisms behind these rare objects. We also examined the metallicity distribution of transiting BD host stars for the first time, and find that a 42 \mj\ transition does not divide the population into two distinct populations with any statistical significance. 

Finally, we noted the presence of Li in the spectrum of TOI-5882, the host star of our lowest mass BD. We measured the equivalent width of the Li line and search for other signs of youth. Seeing no evidence of youth from any of the other indicators that we examined, we adopt the age provided by our global \exofast fit and do not interpret the presence of Li as a sign of youth. Instead, we noted that the Li may actually be a signature of engulfed planetary material, and that more work will be required to explore this hypothesis. 

\section*{Acknowledgements}
This paper includes data collected by the \tess\ mission that are publicly available from the Mikulski Archive for Space Telescopes \citep[MAST;][]{TICdoi}. Funding for the \tess\ mission is provided by NASA's Science Mission Directorate. We acknowledge the use of public \tess\ data from pipelines at the \tess\ Science Office and at the \tess\ Science Processing Operations Center. Resources supporting this work were provided by the NASA High-End Computing (HEC) Program through the NASA Advanced Supercomputing (NAS) Division at Ames Research Center for the production of the SPOC data products. This research has made use of the NASA Exoplanet Archive and the Exoplanet Follow-up Observation Program (ExoFOP; DOI: 10.26134/ExoFOP5) website, which is operated by the California Institute of Technology, under contract with the National Aeronautics and Space Administration under the Exoplanet Exploration Program. We acknowledge financial support from the Agencia Estatal de Investigaci\'on of the Ministerio de Ciencia e Innovaci\'on MCIN/AEI/10.13039/501100011033 and the ERDF “A way of making Europe” through project PID2021-125627OB-C32, and from the Centre of Excellence “Severo Ochoa” award to the Instituto de Astrofisica de Canarias. This work makes use of observations from the LCOGT network. Part of the LCOGT telescope time was granted by NOIRLab through the Mid-Scale Innovations Program (MSIP). MSIP is funded by NSF. This paper is based on observations made with the Las Cumbres Observatory’s education network telescopes that were upgraded through generous support from the Gordon and Betty Moore Foundation. Based on observations obtained at the Hale Telescope, Palomar Observatory, as part of a collaborative agreement between the Caltech Optical Observatories and the Jet Propulsion Laboratory operated by Caltech for NASA. DRC and CAC acknowledge partial support from NASA Grant 18-2XRP18\_2-0007. Based in part on observations obtained at the Southern Astrophysical Research (SOAR) telescope, which is a joint project of the Minist\'{e}rio da Ci\^{e}ncia, Tecnologia e Inova\c{c}\~{o}es (MCTI/LNA) do Brasil, the U.S. National Science Foundation NOIRLab, the University of North Carolina at Chapel Hill (UNC), and Michigan State University (MSU). This work is partly supported by JSPS KAKENHI Grant Number JP24H00017, JP24K00689 and JSPS Bilateral Program Number JPJSBP120249910. This paper is based on observations made with the MuSCAT2 instrument, developed by ABC, at Telescopio Carlos Sánchez operated on the island of Tenerife by the IAC in the Spanish Observatorio del Teide. We thank the staff of the Observatoire de Haute-Provence for their support at the 1.93 m telescope and on SOPHIE.

NV is supported by the NASA FINESST program. The postdoctoral fellowship of KB is funded by F.R.S.-FNRS grant T.0109.20 and by the Francqui Foundation. KAC and CNW acknowledge support from the TESS mission via subaward s3449 from MIT. I.A.S. acknowledges the support of M.V. Lomonosov Moscow State University Program of Development. F. M. acknowledge support from the Agencia Estatal de Investigaci\'{o}n del Ministerio de Ciencia, Innovaci\'{o}n y Universidades (MCIU/AEI) through grant PID2023-152906NA-I00. LM acknowledges financial contribution from PRIN MUR 2022 project 2022J4H55R. TF's work is supported by the French National Research Agency in the framework of the Investisse-ments d’Avenir program (ANR-15-IDEX-02), through the funding of the “Origin of Life” project of the Grenoble-Alpes University

\bibliographystyle{apj}
\bibliography{refs}

\end{document}

%% file: authors.tex
\correspondingauthor{Noah Vowell} 
\email{vowellno@msu.edu}

\author[0000-0002-0701-4005]{Noah Vowell}
\affiliation{\msu}
\affiliation{\cfa}

\author[0000-0001-8812-0565]{Joseph E. Rodriguez} 
\affiliation{\msu}

\author[0000-0001-9911-7388]{David W. Latham} 
\affiliation{\cfa}

\author[0000-0002-8964-8377]{Samuel N. Quinn} 
\affiliation{\cfa}

\author[0000-0002-7382-0160]{Jack Schulte} 
\affiliation{\msu}

\author[0000-0003-3773-5142]{Jason D. Eastman} 
\affiliation{\cfa}

\author[0000-0001-6637-5401]{Allyson Bieryla} 
\affiliation{\cfa}

\author[0000-0003-1464-9276]{Khalid Barkaoui} 
\affiliation{Astrobiology Research Unit, Universit\'e de Li\`ege, 19C All\'ee du 6 Ao\^ut, 4000 Li\`ege, Belgium}
\affiliation{Department of Earth, Atmospheric and Planetary Science, Massachusetts Institute of Technology, 77 Massachusetts Avenue, Cambridge, MA 02139, USA}
\affiliation{Instituto de Astrof\'isica de Canarias (IAC), Calle V\'ia L\'actea s/n, 38200, La Laguna, Tenerife, Spain}


\author[0000-0002-5741-3047]{David R. Ciardi} 
\affiliation{NASA Exoplanet Science Institute-Caltech/IPAC, Pasadena, CA 91125 USA}

\author[0000-0001-6588-9574]{Karen A.\ Collins} 
\affiliation{\cfa}



\author[0000-0002-5443-3640]{Eric Girardin} 
\affiliation{Grand Pra Observatory, 1984 Les Hauderes, Switzerland}

\author{Guillaume H\'ebrard}
\affiliation{Institut d'astrophysique de Paris, UMR7095 CNRS, Universit\'e Pierre \& Marie Curie, 98bis boulevard Arago, 75014 Paris, France}
\affiliation{Observatoire de Haute-Provence, CNRS, Universit\'e d'Aix-Marseille, 04870 Saint-Michel-l'Observatoire, France}

\author[0009-0001-2900-7834]{Elisabeth Heldridge} 
\affiliation{The Thacher School, 5025 Thacher Rd., Ojai, CA 93023, USA}

\author[0000-0001-8019-6661]{Marziye Jafariyazani}
\affiliation{SETI Institute, Mountain View, CA 94043 USA/NASA Ames Research Center, Moffett Field, CA 94035 USA}

\author[0009-0008-5864-9415]{Brooke Kotten} 
\affiliation{Department of Astronomy, University of Michigan, Ann Arbor, MI 48109, USA}

\author[0000-0002-9428-8732]{Luigi Mancini} 
\affiliation{Department of Physics, University of Rome ``Tor Vergata'', Via della Ricerca Scientifica 1, I-00133, Rome, Italy}
\affiliation{Max Planck Institute for Astronomy, K\"{o}nigstuhl 17, D-69117, Heidelberg, Germany}
\affiliation{INAF -- Osservatorio Astrofisico di Torino, via Osservatorio 20, I-10025, Pino Torinese, Italy}

\author[0000-0001-9087-1245]{Felipe Murgas} 
\affiliation{Instituto de Astrof\'isica de Canarias (IAC), E-38205 La Laguna, Tenerife, Spain}
\affiliation{Departamento de Astrof\'isica, Universidad de La Laguna (ULL), E-38206 La Laguna, Tenerife, Spain}

\author[0000-0001-8511-2981]{Norio Narita} 
\affiliation{Komaba Institute for Science, The University of Tokyo, 3-8-1 Komaba, Meguro, Tokyo 153-8902, Japan}
\affiliation{Astrobiology Center, 2-21-1 Osawa, Mitaka, Tokyo 181 8588, Japan}
\affiliation{Instituto de Astrof\'isica de Canarias (IAC), E-38205 La Laguna, Tenerife, Spain}

\author[0000-0002-3940-2360]{D. J. Radford} 
\affiliation{Brierfield Observatory, Bowral, NSW Australia}

\author[0009-0009-5132-9520]{Howard M. Relles} 
\affiliation{\cfa}

\author[0000-0002-1836-3120]{Avi Shporer} 
\affiliation{Department of Physics and Kavli Institute for Astrophysics and Space Research, Massachusetts Institute of Technology, Cambridge, MA 02139, USA}


\author[0000-0001-7493-7419]{Melinda Soares-Furtado} 
\affiliation{Department of Astronomy, University of Wisconsin-Madison, 475 N. Charter St., Madison, WI 53706, USA}

\author[0000-0003-0647-6133]{Ivan A. Strakhov} 
\affiliation{Sternberg Astronomical Institute, Lomonosov Moscow State University, Universitetsky pr. 13, Moscow 119234, Russia}

\author[0000-0002-0619-7639]{Carl Ziegler} 
\affiliation{Department of Physics, Engineering and Astronomy, Stephen F. Austin State University, 1936 North St, Nacogdoches, TX 75962, USA}


\author{Isabelle Boisse}
\affiliation{Laboratoire d'astrophysique de Marseille, Univ. de Provence, UMR6110 CNRS, 38 r. F. Joliot Curie, 13388 Marseille cedex 13, France}
\affiliation{Aix Marseille Univ, CNRS, CNES, LAM, Marseille, France}

\author[0000-0001-7124-4094]{C\'{e}sar Brice\~{n}o} 
\affiliation{SOAR Telescope/NSF NOIRLab, Casilla 603, La Serena, Chile}


\author{Michael L. Calkins} 
\affiliation{\cfa}

\author[0000-0002-2361-5812]{Catherine A. Clark} 
\affiliation{NASA Exoplanet Science Institute-Caltech/IPAC, Pasadena, CA 91125 USA}

\author[0000-0003-2781-3207]{Kevin I.\ Collins} 
\affiliation{George Mason University, 4400 University Drive, Fairfax, VA, 22030 USA}


\author{Jerome de Leon}
\affiliation{Komaba Institute for Science, The University of Tokyo, 3-8-1 Komaba, Meguro, Tokyo 153-8902, Japan}
\affiliation{Instituto de Astrof\'isica de Canarias (IAC), Calle V\'ia L\'actea s/n, 38200, La Laguna, Tenerife, Spain}

\author[0000-0002-9789-5474]{Gilbert A. Esquerdo} 
\affiliation{\cfa }

\author[0000-0001-9309-0102]{Sergio B. Fajardo-Acosta} 
\affiliation{Caltech/IPAC, Mail Code 100-22, Pasadena, CA 91125}

\author{Thierry Forveille}
\affiliation{Universit\'e Grenoble Alpes, CNRS, IPAG, 38000 Grenoble, France}

\author[0000-0001-8511-2981]{Akihiko Fukui} 
\affiliation{Komaba Institute for Science, The University of Tokyo, 3-8-1 Komaba, Meguro, Tokyo 153-8902, Japan}
\affiliation{Instituto de Astrof\'isica de Canarias (IAC), E-38205 La Laguna, Tenerife, Spain}


\author[0000-0001-8621-6731]{Cristilyn N.\ Watkins} 
\affiliation{Center for Astrophysics \textbar \ Harvard \& Smithsonian, 60 Garden Street, Cambridge, MA 02138, USA}

\author[0000-0002-5031-7853]{Ruixuan He} 
\affiliation{The Thacher School, 5025 Thacher Rd., Ojai, CA 93023, USA}

\author{Neda Heidari}
\affiliation{Institut d'astrophysique de Paris, UMR7095 CNRS, Universit\'e Pierre \& Marie Curie, 98bis boulevard Arago, 75014 Paris, France}


\author[0000-0003-1728-0304]{Keith Horne} 
\affiliation{SUPA Physics and Astronomy, University of St. Andrews, Fife, KY16 9SS Scotland, UK}

\author[0000-0002-4715-9460]{Jon M. Jenkins} 
\affiliation{\ames}



\author[0000-0003-3654-1602]{Andrew W. Mann} 
\affiliation{Department of Physics and Astronomy, The University of North Carolina at Chapel Hill, Chapel Hill, NC 27599-3255, USA}


\author[0000-0001-9390-0988]{Luca Naponiello} 
\affiliation{INAF -- Osservatorio Astrofisico di Torino, via Osservatorio 20, I-10025, Pino Torinese, Italy}

\author[0000-0003-0987-1593]{Enric Palle} 
\affiliation{Instituto de Astrof\'isica de Canarias (IAC), E-38205 La Laguna, Tenerife, Spain}
\affiliation{Departamento de Astrof\'isica, Universidad de La Laguna (ULL), E-38206 La Laguna, Tenerife, Spain}



\author[0000-0001-8227-1020]{Richard P. Schwarz} 
\affiliation{Center for Astrophysics \textbar \ Harvard \& Smithsonian, 60 Garden Street, Cambridge, MA 02138, USA}

\author[0000-0002-6892-6948]{S.~Seager} 
\affiliation{Department of Physics and Kavli Institute for Astrophysics and Space Research, Massachusetts Institute of Technology, Cambridge, MA 02139, USA}
\affiliation{Department of Earth, Atmospheric and Planetary Sciences, Massachusetts Institute of Technology, Cambridge, MA 02139, USA}
\affiliation{Department of Aeronautics and Astronautics, MIT, 77 Massachusetts Avenue, Cambridge, MA 02139, USA}


\author[0000-0002-3807-3198]{John Southworth} 
\affiliation{Astrophysics Group, Keele University, Staffordshire, ST5 5BG, UK}

\author{Gregor Srdoc} 
\affiliation{Kotizarovci Observatory, Sarsoni 90, 51216 Viskovo, Croatia}

\author[0000-0002-9486-818X]{Jonathan J. Swift} 
\affiliation{The Thacher School, 5025 Thacher Rd., Ojai, CA 93023, USA}

\author[0000-0002-4265-047X]{Joshua N.\ Winn} 
\affiliation{Department of Astrophysical Sciences, Princeton University, Princeton, NJ 08544, USA}

%% file: affiliations.tex
\newcommand{\cfa}{Center for Astrophysics \textbar \ Harvard \& Smithsonian, 60 Garden St, Cambridge, MA 02138, USA}
\newcommand{\msu}{Center for Data Intensive and Time Domain Astronomy, Department of Physics and Astronomy, Michigan State University, East Lansing, MI 48824, USA}
\newcommand{\umich}{Astronomy Department, University of Michigan, 1085 S University Avenue, Ann Arbor, MI 48109, USA}
\newcommand{\utaustin}{Department of Astronomy, The University of Texas at Austin, Austin, TX 78712, USA}
\newcommand{\MIT}{Department of Physics and Kavli Institute for Astrophysics and Space Research, Massachusetts Institute of Technology, Cambridge, MA 02139, USA}
\newcommand{\MITEPS}{Department of Earth, Atmospheric and Planetary Sciences, Massachusetts Institute of Technology,  Cambridge,  MA 02139, USA}
\newcommand{\uflorida}{Department of Astronomy, University of Florida, 211 Bryant Space Science Center, Gainesville, FL, 32611, USA}
\newcommand{\riverside}{Department of Earth Sciences, University of California, Riverside, CA 92521, USA}
\newcommand{\usq}{University of Southern Queensland, West St, Darling Heights QLD 4350, Australia}
\newcommand{\ames}{NASA Ames Research Center, Moffett Field, CA, 94035, USA}
\newcommand{\geneva}{Geneva Observatory, University of Geneva, Chemin des Mailettes 51, 1290 Versoix, Switzerland}
\newcommand{\uw}{Astronomy Department, University of Washington, Seattle, WA 98195 USA}
\newcommand{\warwick}{Deptartment of Physics, University of Warwick, Gibbet Hill Road, Coventry CV4 7AL, UK}
\newcommand{\warwickceh}{Centre for Exoplanets and Habitability, University of Warwick, Gibbet Hill Road, Coventry CV4 7AL, UK}
\newcommand{\princeton}{Department of Astrophysical Sciences, Princeton University, 4 Ivy Lane, Princeton, NJ, 08544, USA}
\newcommand{\liege}{Space Sciences, Technologies and Astrophysics Research (STAR) Institute, Universit\'e de Li\`ege, 19C All\'ee du 6 Ao\^ut, 4000 Li\`ege, Belgium}
\newcommand{\vanderbilt}{Department of Physics and Astronomy, Vanderbilt University, Nashville, TN 37235, USA}
\newcommand{\fisk}{Department of Physics, Fisk University, 1000 17th Avenue North, Nashville, TN 37208, USA}
\newcommand{\columbia}{Department of Astronomy, Columbia University, 550 West 120th Street, New York, NY 10027, USA}
\newcommand{\toronto}{Dunlap Institute for Astronomy and Astrophysics, University of Toronto, Ontario M5S 3H4, Canada}
\newcommand{\unc}{Department of Physics and Astronomy, University of North Carolina at Chapel Hill, Chapel Hill, NC 27599, USA}
\newcommand{\iac}{Instituto de Astrof\'isica de Canarias (IAC), E-38205 La Laguna, Tenerife, Spain}
\newcommand{\lalaguna}{Departamento de Astrof\'isica, Universidad de La Laguna (ULL), E-38206 La Laguna, Tenerife, Spain}
\newcommand{\louisville}{Department of Physics and Astronomy, University of Louisville, Louisville, KY 40292, USA}
\newcommand{\aavso}{American Association of Variable Star Observers, 49 Bay State Road, Cambridge, MA 02138, USA}
\newcommand{\utokyo}{The University of Tokyo, 7-3-1 Hongo, Bunky\={o}, Tokyo 113-8654, Japan}
\newcommand{\naoj}{National Astronomical Observatory of Japan, 2-21-1 Osawa, Mitaka, Tokyo 181-8588, Japan}
\newcommand{\jstpresto}{JST, PRESTO, 7-3-1 Hongo, Bunkyo-ku, Tokyo 113-0033, Japan}
\newcommand{\astrobiojapan}{Astrobiology Center, 2-21-1 Osawa, Mitaka, Tokyo 181-8588, Japan}
\newcommand{\ctio}{Cerro Tololo Inter-American Observatory, Casilla 603, La Serena, Chile}
\newcommand{\nexsci}{Caltech IPAC -- NASA Exoplanet Science Institute 1200 E. California Ave, Pasadena, CA 91125, USA}
\newcommand{\ucsc}{Department of Astronomy and Astrophysics, University of
California, Santa Cruz, CA 95064, USA}
\newcommand{\gsfc}{Exoplanets and Stellar Astrophysics Laboratory, Code 667, NASA Goddard Space Flight Center, Greenbelt, MD 20771, USA}
\newcommand{\sgtinc}{SGT, Inc./NASA AMES Research Center, Mailstop 269-3, Bldg T35C, P.O. Box 1, Moffett Field, CA 94035, USA}
\newcommand{\chile}{Center of Astro-Engineering UC, Pontificia Universidad Cat\'olica de Chile, Av. Vicu\~{n}a Mackenna 4860, 7820436 Macul, Santiago, Chile}
\newcommand{\Pontificia}{Instituto de Astrof\'isica, Pontificia Universidad Cat\'olica de Chile, Av.\ Vicu\~na Mackenna 4860, Macul, Santiago, Chile}
\newcommand{\Millennium}{Millennium Institute for Astrophysics, Chile}
\newcommand{\maxplank}{Max-Planck-Institut f\"ur Astronomie, K\"onigstuhl 17, Heidelberg 69117, Germany}
\newcommand{\utdallas}{Department of Physics, The University of Texas at Dallas, 800 West
Campbell Road, Richardson, TX 75080-3021 USA}
\newcommand{\MauryLewin}{Maury Lewin Astronomical Observatory, Glendora, CA 91741, USA}
\newcommand{\umbc}{University of Maryland, Baltimore County, 1000 Hilltop Circle, Baltimore, MD 21250, USA}
\newcommand{\osu}{Department of Astronomy, The Ohio State University, 140 West 18th Avenue, Columbus, OH 43210, USA}
\newcommand{\MITAA}{Department of Aeronautics and Astronautics, MIT, 77 Massachusetts Avenue, Cambridge, MA 02139, USA}
\newcommand{\openu}{School of Physical Sciences, The Open University, Milton Keynes MK7 6AA, UK}
\newcommand{\swarthmore}{Department of Physics and Astronomy, Swarthmore College, Swarthmore, PA 19081, USA}
\newcommand{\seti}{SETI Institute, Mountain View, CA 94043, USA}
\newcommand{\lehigh}{Department of Physics, Lehigh University, 16 Memorial Drive East, Bethlehem, PA 18015, USA}
\newcommand{\utah}{Department of Physics and Astronomy, University of Utah, 115 South 1400 East, Salt Lake City, UT 84112, USA}
\newcommand{\USNA}{Department of Physics, United States Naval Academy, 572C Holloway Rd., Annapolis, MD 21402, USA}
\newcommand{\DTM}{Department of Terrestrial Magnetism, Carnegie Institution for Science, 5241 Broad Branch Road, NW, Washington, DC 20015, USA}
\newcommand{\UPenn}{The University of Pennsylvania, Department of Physics and Astronomy, Philadelphia, PA, 19104, USA}
\newcommand{\montana}{Department of Physics and Astronomy, University of Montana, 32 Campus Drive, No. 1080, Missoula, MT 59812 USA}
\newcommand{\psu}{Department of Astronomy \& Astrophysics, The Pennsylvania State University, 525 Davey Lab, University Park, PA 16802, USA}
\newcommand{\psust}{Center for Exoplanets and Habitable Worlds, The Pennsylvania State University, 525 Davey Lab, University Park, PA 16802, USA}
\newcommand{\Kutztown}{Department of Physical Sciences, Kutztown University, Kutztown, PA 19530, USA}
\newcommand{\udel}{Department of Physics \& Astronomy, University of Delaware, Newark, DE 19716, USA}
\newcommand{\Westminster}{Department of Physics, Westminster College, New Wilmington, PA 16172}
\newcommand{\steward}{Department of Astronomy and Steward Observatory, University of Arizona, Tucson, AZ 85721, USA}
\newcommand{\saao}{South African Astronomical Observatory, PO Box 9, Observatory, 7935, Cape Town, South Africa}
\newcommand{\salt}{Southern African Large Telescope, PO Box 9, Observatory, 7935, Cape Town, South Africa}
\newcommand{\ssl}{Societ\`{a} Astronomica Lunae, Italy}
\newcommand{\spot}{Spot Observatory, Nashville, TN 37206, USA}
\newcommand{\txamGP}{George P.\ and Cynthia Woods Mitchell Institute for Fundamental Physics and Astronomy, Texas A\&M University, College Station, TX77843 USA}
\newcommand{\txam}{Department of Physics and Astronomy, Texas A\&M university, College Station, TX 77843 USA}
\newcommand{\wellesley}{Department of Astronomy, Wellesley College, Wellesley, MA 02481, USA}
\newcommand{\byu}{Department of Physics and Astronomy, Brigham Young University, Provo, UT 84602, USA}
\newcommand{\Hazelwood}{Hazelwood Observatory, Churchill, Victoria, Australia}
\newcommand{\pest}{Perth Exoplanet Survey Telescope}
\newcommand{\Winer}{Winer Observatory, PO Box 797, Sonoita, AZ 85637, USA}
\newcommand{\icpo}{Ivan Curtis Private Observatory}
\newcommand{\elsauce}{El Sauce Observatory, Chile}
\newcommand{\crow}{Atalaia Group \& CROW Observatory, Portalegre, Portugal}
\newcommand{\dfus}{Dipartimento di Fisica ``E.R.Caianiello'', Universit\`a di Salerno, Via Giovanni Paolo II 132, Fisciano 84084, Italy}
\newcommand{\indfn}{Istituto Nazionale di Fisica Nucleare, Napoli, Italy}
\newcommand{\sotes}{Gabriel Murawski Private Observatory (SOTES)}
\newcommand{\lco}{Las Cumbres Observatory Global Telescope, 6740 Cortona Dr., Suite 102, Goleta, CA 93111, USA}
\newcommand{\ucsb}{Department of Physics, University of California, Santa Barbara, CA 93106-9530, USA}
\newcommand{\yale}{Department of Astronomy, Yale University, 52 Hillhouse Avenue, New Haven, CT 06511, USA}
\newcommand{\eso}{European Southern Observatory, Alonso de C\'ordova 3107, Vitacura, Casilla 19001, Santiago, Chile}
\newcommand{\stsci}{Space Telescope Science Institute, Baltimore, MD 21218, USA}
\newcommand{\keele}{Astrophysics Group, Keele University, Staffordshire ST5 5BG, UK}
\newcommand{\gsfcsellers}{GSFC Sellers Exoplanet Environments Collaboration, NASA Goddard Space Flight Center, Greenbelt, MD 20771 }
\newcommand{\usno}{U.S. Naval Observatory, Washington, DC 20392, USA}
\newcommand{\kansas}{Department of Physics and Astronomy, University of Kansas, 1251 Wescoe Hall Dr., Lawrence, KS 66045, USA}
\newcommand{\gmu}{George Mason University, 4400 University Drive MS 3F3, Fairfax, VA 22030, USA}
\newcommand{\unsw}{Exoplanetary Science at UNSW, School of Physics, UNSW Sydney, NSW 2052, Australia}
\newcommand{\sifa}{School of Physics, Sydney Institute for Astronomy (SIfA), The University of Sydney, NSW 2006, Australia}
\newcommand{\nanjing}{School of Astronomy and Space Science, Key Laboratory of Modern Astronomy and Astrophysics in Ministry of Education, Nanjing University, Nanjing 210046, Jiangsu, China}
\newcommand{\jhuapl}{Johns Hopkins APL, 11100 Johns Hopkins Rd, Laurel, MD 20723, USA}
\newcommand{\torres}{\altaffiliation{Juan Carlos Torres Fellow}}
\newcommand{\sagan}{\altaffiliation{NASA Sagan Fellow}}
\newcommand{\bernoulli}{\altaffiliation{Bernoulli fellow}}
\newcommand{\gruber}{\altaffiliation{Gruber fellow}}
\newcommand{\kavli}{\altaffiliation{Kavli Fellow}}
\newcommand{\peg}{\altaffiliation{51 Pegasi b Fellow}}
\newcommand{\pappalardo}{\altaffiliation{Pappalardo Fellow}}
\newcommand{\hubble}{\altaffiliation{NASA Hubble Fellow}}
\newcommand{\nsf}{\altaffiliation{National Science Foundation Graduate Research Fellow}}

%% file: lit_table_4.tex
\providecommand{\bjdtdb}{\ensuremath{\rm {BJD_{TDB}}}}
\providecommand{\feh}{\ensuremath{\left[{\rm Fe}/{\rm H}\right]}}
\providecommand{\teff}{\ensuremath{T_{\rm eff}}}
\providecommand{\teq}{\ensuremath{T_{\rm eq}}}
\providecommand{\ecosw}{\ensuremath{e\cos{\omega_*}}}
\providecommand{\esinw}{\ensuremath{e\sin{\omega_*}}}
\providecommand{\msun}{\ensuremath{\,M_\Sun}}
\providecommand{\rsun}{\ensuremath{\,R_\Sun}}
\providecommand{\lsun}{\ensuremath{\,L_\Sun}}
\providecommand{\mj}{\ensuremath{\,M_{\rm J}}}
\providecommand{\rj}{\ensuremath{\,R_{\rm J}}}
\providecommand{\me}{\ensuremath{\,M_{\rm E}}}
\providecommand{\re}{\ensuremath{\,R_{\rm E}}}
\providecommand{\fave}{\langle F \rangle}
\providecommand{\fluxcgs}{10$^9$ erg s$^{-1}$ cm$^{-2}$}
\providecommand{\tess}{\textit{TESS}\xspace}
\tablecolumns{7}
\tablehead{ &  & \colhead{TOI-2844} & \colhead{TOI-3122} & \colhead{TOI-3577} & \colhead{TOI-3755} & \colhead{Source}}
\startdata
\multicolumn{7}{l}{\textbf{Other identifiers}:} \\
& \tess\ Input Catalog & TIC 387342052 & TIC 61117473 & TIC 396133015 & TIC 281196902\\
& TYCHO-2 & TYC 771-367-1 & TYC 6773-1-1 & TYC 3608-647-1 & ---\\
& 2MASS & J07204878+1301073 & J15074899-2809237 & J21482300+4820042 & J04385936+6640161\\
& Gaia DR3 & 3166196736096450816 & 6212565847439064192 & 1977894600881987328 & 483359575160953728\\
\hline
\multicolumn{7}{l}{\textbf{Astrometric Parameters}:} \\
$\alpha_{J2000}\ddagger$ & Right Ascension (h:m:s) & 07:20:48.78 & 15:07:48.99 & 21:48:23.01 & 04:38:59.37  & 1 \\
$\delta_{J2000}\ddagger$ & Declination (d:m:s) & 13:01:07.4 & -28:09:23.8 & 48:20:04.4 & 66:40:16.2  & 1 \\
$\mu_{\alpha}$ & Gaia DR3 proper motion in RA (mas yr$^{-1}$)& $-2.992 \pm 0.016$ & $-13.329 \pm 0.015$ & $5.052 \pm 0.011$ & $-11.614 \pm 0.007$  & 1 \\
$\mu_{\delta}$ & Gaia DR3 proper motion in Dec (mas yr$^{-1}$)& $-4.906 \pm 0.016$ & $-0.501 \pm 0.013$ & $-33.472 \pm 0.011$ & $11.407 \pm 0.009$  & 1 \\
$\pi$ & Gaia DR3 Parallax (mas) & $1.4262 \pm 0.0137$ & $1.9337 \pm 0.0147$ & $2.3184 \pm 0.011$ & $3.0749 \pm 0.0104$  & 1 \\
$v\sin{i_\star}$ & Projected rotational velocity (km s$^{-1}$) & $60.8 \pm 2.6$ & $23.2 \pm 5.1$ & $10.3 \pm 0.5$ & $5.1 \pm 0.5$  & 2 \\
\multicolumn{7}{l}{\textbf{Photometric Parameters}:} \\
${G}$ & Gaia $G$ mag. & $11.87 \pm 0.02$ & $12.52 \pm 0.02$ & $11.75 \pm 0.02$ & $12.62 \pm 0.02$  & 1 \\
$G_{\rm BP}$ & Gaia $G_{\rm BP}$ mag. & $12.08 \pm 0.02$ & $12.835 \pm 0.02$ & $12.07 \pm 0.02$ & $13.06 \pm 0.02$  & 1 \\
$G_{\rm RP}$ & Gaia $G_{\rm RP}$ mag. & $11.52 \pm 0.02$ & $12.05 \pm 0.02$ & $11.26 \pm 0.02$ & $12.01 \pm 0.02$  & 1 \\
${T}$ & TESS mag. & $11.583 \pm 0.007$ & $12.123 \pm 0.008$ & $11.321 \pm 0.006$ & $12.079 \pm 0.006$  & 3 \\
$J$ & 2MASS $J$ mag. & $11.168 \pm 0.022$ & $11.515 \pm 0.026$ & $10.638 \pm 0.023$ & $11.316 \pm 0.024$  & 4 \\
$H$ & 2MASS $H$ mag. & $10.967 \pm 0.027$ & $11.277 \pm 0.025$ & $10.381 \pm 0.03$ & $10.966 \pm 0.028$  & 4 \\
$K$ & 2MASS $K$ mag. & $10.927 \pm 0.021$ & $11.207 \pm 0.024$ & $10.318 \pm 0.020$ & $10.876 \pm 0.022$  & 4 \\
$W1$ & WISE $W1$ mag. & $10.90 \pm 0.03$ & $11.05 \pm 0.03$ & $10.24 \pm 0.03$ & $10.82 \pm 0.03$  & 5 \\
$W2$ & WISE $W2$ mag. & $10.93 \pm 0.03$ & $11.07 \pm 0.03$ & $10.27 \pm 0.03$ & $10.87 \pm 0.03$  & 5 \\
$W3$ & WISE $W3$ mag. & $10.987 \pm 0.143$ & $11.147 \pm 0.155$ & $10.298 \pm 0.046$ & $10.651 \pm 0.093$  & 5 \\
\enddata

%% file: lit_table_5.tex
\providecommand{\bjdtdb}{\ensuremath{\rm {BJD_{TDB}}}}
\providecommand{\feh}{\ensuremath{\left[{\rm Fe}/{\rm H}\right]}}
\providecommand{\teff}{\ensuremath{T_{\rm eff}}}
\providecommand{\teq}{\ensuremath{T_{\rm eq}}}
\providecommand{\ecosw}{\ensuremath{e\cos{\omega_*}}}
\providecommand{\esinw}{\ensuremath{e\sin{\omega_*}}}
\providecommand{\msun}{\ensuremath{\,M_\Sun}}
\providecommand{\rsun}{\ensuremath{\,R_\Sun}}
\providecommand{\lsun}{\ensuremath{\,L_\Sun}}
\providecommand{\mj}{\ensuremath{\,M_{\rm J}}}
\providecommand{\rj}{\ensuremath{\,R_{\rm J}}}
\providecommand{\me}{\ensuremath{\,M_{\rm E}}}
\providecommand{\re}{\ensuremath{\,R_{\rm E}}}
\providecommand{\fave}{\langle F \rangle}
\providecommand{\fluxcgs}{10$^9$ erg s$^{-1}$ cm$^{-2}$}
\providecommand{\tess}{\textit{TESS}\xspace}
\tablecolumns{7}
\tablehead{ &  & \colhead{TOI-4462} & \colhead{TOI-4635} & \colhead{TOI-4737} & \colhead{TOI-4759} & \colhead{Source}}
\startdata
\multicolumn{7}{l}{\textbf{Other identifiers}:} \\
& \tess\ Input Catalog & TIC 76420654 & TIC 337129672 & TIC 142532090 & TIC 49705089\\
& TYCHO-2 & TYC 2635-1030-1 & --- & --- & ---\\
& 2MASS & J18184078+3615175 & J02143112+0804481 & J06533851-1326106 & J06234422-2401288\\
& Gaia DR3 & 4605954852723545088 & 2521579495665163008 & 2949605211853441664 & 2936390357694302336\\
\hline
\multicolumn{7}{l}{\textbf{Astrometric Parameters}:} \\
$\alpha_{J2000}\ddagger$ & Right Ascension (h:m:s) & 18:18:40.78 & 02:14:31.26 & 06:53:38.51 & 06:23:44.23 & 1 \\
$\delta_{J2000}\ddagger$ & Declination (d:m:s) & 36:15:17.5 & 08:04:45.3 & -13:26:10.7 & -24:01:28.9  & 1 \\
$\mu_{\alpha}$ & Gaia DR3 proper motion in RA (mas yr$^{-1}$)& $6.959 \pm 0.038$ & $122.150 \pm 0.026$ & $11.332 \pm 0.013$ & $0.634 \pm 0.009$  & 1 \\
$\mu_{\delta}$ & Gaia DR3 proper motion in Dec (mas yr$^{-1}$)& $-3.105 \pm 0.045$ & $-203.829 \pm 0.019$ & $-20.658 \pm 0.014$ & $7.847 \pm 0.012$  & 1 \\
$\pi$ & Gaia DR3 Parallax (mas) & $2.5184 \pm 0.0365$ & $13.3018 \pm 0.0238$ & $1.7169 \pm 0.0132$ & $1.3243 \pm 0.0107$  & 1 \\
$v\sin{i_\star}$ & Projected rotational velocity (km s$^{-1}$) & $18.6 \pm 0.4$ & $3.5 \pm 1.1$ & $5.1 \pm 0.6$ & $13.4 \pm 0.6$  & 2 \\
\multicolumn{7}{l}{\textbf{Photometric Parameters}:} \\
${G}$ & Gaia $G$ mag. & $10.88 \pm 0.02$ & $11.32 \pm 0.02$ & $12.43 \pm 0.02$ & $12.73 \pm 0.02$  & 1 \\
$G_{\rm BP}$ & Gaia $G_{\rm BP}$ mag. & $11.17 \pm 0.02$ & $11.99 \pm 0.02$ & $12.78 \pm 0.02$ & $13.10 \pm 0.02$  & 1 \\
$G_{\rm RP}$ & Gaia $G_{\rm RP}$ mag. & $10.39 \pm 0.02$ & $10.53 \pm 0.02$ & $11.93 \pm 0.02$ & $12.20 \pm 0.02$  & 1 \\
${T}$ & TESS mag. & $10.445 \pm 0.006$ & $10.445 \pm 0.006$ & $11.992 \pm 0.006$ & $12.263 \pm 0.007$  & 3 \\
$J$ & 2MASS $J$ mag. & $9.882 \pm 0.020$ & $9.565 \pm 0.025$ & $11.355 \pm 0.024$ & $11.596 \pm 0.022$  & 4 \\
$H$ & 2MASS $H$ mag. & $9.585 \pm 0.020$ & $8.987 \pm 0.028$ & $11.057 \pm 0.025$ & $11.325 \pm 0.025$  & 4 \\
$K$ & 2MASS $K$ mag. & $9.513 \pm 0.020$ & $8.854 \pm 0.025$ & $10.988 \pm 0.025$ & $11.21 \pm 0.026$  & 4 \\
$W1$ & WISE $W1$ mag. & $9.46 \pm 0.03$ & $8.70 \pm 0.03$ & $10.93 \pm 0.03$ & $11.12 \pm 0.03$  & 5 \\
$W2$ & WISE $W2$ mag. & $9.49 \pm 0.03$ & $8.74 \pm 0.03$ & $10.97 \pm 0.03$ & $11.15 \pm 0.03$  & 5 \\
$W3$ & WISE $W3$ mag. & $9.410 \pm 0.033$ & $8.696 \pm 0.030$ & $10.960 \pm 0.107$ & $11.166 \pm 0.107$  & 5 \\
\enddata

%% file: lit_table_6.tex
\providecommand{\bjdtdb}{\ensuremath{\rm {BJD_{TDB}}}}
\providecommand{\feh}{\ensuremath{\left[{\rm Fe}/{\rm H}\right]}}
\providecommand{\teff}{\ensuremath{T_{\rm eff}}}
\providecommand{\teq}{\ensuremath{T_{\rm eq}}}
\providecommand{\ecosw}{\ensuremath{e\cos{\omega_*}}}
\providecommand{\esinw}{\ensuremath{e\sin{\omega_*}}}
\providecommand{\msun}{\ensuremath{\,M_\Sun}}
\providecommand{\rsun}{\ensuremath{\,R_\Sun}}
\providecommand{\lsun}{\ensuremath{\,L_\Sun}}
\providecommand{\mj}{\ensuremath{\,M_{\rm J}}}
\providecommand{\rj}{\ensuremath{\,R_{\rm J}}}
\providecommand{\me}{\ensuremath{\,M_{\rm E}}}
\providecommand{\re}{\ensuremath{\,R_{\rm E}}}
\providecommand{\fave}{\langle F \rangle}
\providecommand{\fluxcgs}{10$^9$ erg s$^{-1}$ cm$^{-2}$}
\providecommand{\tess}{\textit{TESS}\xspace}
\tablecolumns{6}
\tablehead{ &  & \colhead{TOI-5240} & \colhead{TOI-5467} & \colhead{TOI-5882} & \colhead{Source}}
\startdata
\multicolumn{6}{l}{\textbf{Other identifiers}:} \\
& \tess\ Input Catalog & TIC 40055053 & TIC 83275782 & TIC 232941965\\
& TYCHO-2 & TYC 2663-268-1 & --- & TYC 2695-1754-1\\
& 2MASS & J19322010+3456254 & J06173449+2826431 &  J20473329+3444151\\
& Gaia DR3 & 2046792606517797632 & 3433414139371114368 & 1869489729418662528\\
\hline
\multicolumn{6}{l}{\textbf{Astrometric Parameters}:} \\
$\alpha_{J2000}\ddagger$ & Right Ascension (h:m:s) & 19:32:20.11 & 06:17:34.49 &  20:47:33.29  & 1 \\
$\delta_{J2000}\ddagger$ & Declination (d:m:s) & 34:56:25.4 & 28:26:43.1 & 34:44:15.2 & 1 \\
$\mu_{\alpha}$ & Gaia DR3 proper motion in RA (mas yr$^{-1}$)& $-0.414 \pm 0.011$ & $0.599 \pm 0.016$ & $-14.084 \pm 0.014$  & 1 \\
$\mu_{\delta}$ & Gaia DR3 proper motion in Dec (mas yr$^{-1}$)& $0.502 \pm 0.012$ & $-13.726 \pm 0.012$ & $-17.246 \pm 0.017$  & 1 \\
$\pi$ & Gaia DR3 Parallax (mas) & $0.9894 \pm 0.0114$ & $1.7558 \pm 0.0138$ & $2.3859 \pm 0.0144$  & 1 \\
$v\sin{i_\star}$ & Projected rotational velocity (km s$^{-1}$) & $32.8 \pm 1.3$ & $31.2 \pm 0.4$ & $7.3 \pm 0.5$  & 2 \\
\multicolumn{6}{l}{\textbf{Photometric Parameters}:} \\
${G}$ & Gaia $G$ mag. & $11.92 \pm 0.02$ & $12.25 \pm 0.02$ & $11.11 \pm 0.02$  & 1 \\
$G_{\rm BP}$ & Gaia $G_{\rm BP}$ mag. & $12.09 \pm 0.02$ & $12.56 \pm 0.02$ & $11.47 \pm 0.02$  & 1 \\
$G_{\rm RP}$ & Gaia $G_{\rm RP}$ mag. & $11.63 \pm 0.02$ & $11.78 \pm 0.02$ & $10.58 \pm 0.02$  & 1 \\
${T}$ & TESS mag. & $11.692 \pm 0.009$ & $11.842 \pm 0.006$ & $10.634 \pm 0.006$  & 3 \\
$J$ & 2MASS $J$ mag. & $11.307 \pm 0.021$ & $11.255 \pm 0.021$ & $9.988 \pm 0.020$  & 4 \\
$H$ & 2MASS $H$ mag. & $11.176 \pm 0.022$ & $11.011 \pm 0.023$ & $9.736 \pm 0.020$  & 4 \\
$K$ & 2MASS $K$ mag. & $11.154 \pm 0.020$ & $10.947 \pm 0.020$ & $9.615 \pm 0.020$  & 4 \\
$W1$ & WISE $W1$ mag. & $11.14 \pm 0.03$ & $10.90 \pm 0.03$ & $9.57 \pm 0.03$  & 5 \\
$W2$ & WISE $W2$ mag. & $11.16 \pm 0.03$ & $10.91 \pm 0.03$ & $9.60 \pm 0.03$  & 5 \\
$W3$ & WISE $W3$ mag. & $10.928 \pm 0.096$ & $10.952 \pm 0.144$ & $9.662 \pm 0.055$  & 5 \\
\enddata

%% file: tess_table.tex
\startlongtable
\begin{deluxetable}{l l l l l}
\tabletypesize{\scriptsize}
\tablecaption{Summary of Observations from TESS \label{tab:tess}}
\tablewidth{0pt}
\tablehead{
\colhead{Target} & \colhead{TESS Sector} & \colhead{Cadence (s)} & \colhead{Pipeline}
}
\startdata
TOI-2844 & 7 & 1800 & TESS-SPOC \\
--- & 33 & 600 & TESS-SPOC \\
--- & 44 & 600 & TESS-SPOC \\
--- & 45 & 600 & TESS-SPOC \\
--- & 46 & 600 & TESS-SPOC \\
--- & 71 & 120 & SPOC \\
--- & 72 & 120 & SPOC \\
TOI-3122 & 11 & 1800 & QLP \\
--- & 38 & 600 & QLP \\
--- & 65 & 120 & SPOC \\
TOI-3577 & 8 & 1800 & QLP \\
--- & 56 & 120 & SPOC \\
--- & 76 & 120 & SPOC \\
TOI-3755 & 19 & 1800 & TESS-SPOC \\
--- & 59 & 120 & SPOC \\
--- & 73 & 120 & SPOC \\
TOI-4462 & 26 & 1800 & TESS-SPOC \\
--- & 40 & 600 & TESS-SPOC \\
--- & 53 & 600 & TESS-SPOC \\
--- & 54 & 600 & TESS-SPOC \\
--- & 74 & 120 & SPOC \\
TOI-4635 & 42 & 120 & SPOC \\
--- & 43 & 120 & SPOC \\
--- & 70 & 120 & SPOC \\
--- & 71 & 120 & SPOC \\
TOI-4737 & 6 & 1800 & TESS-SPOC \\
--- & 7 & 1800 & QLP \\
--- & 33 & 600 & TESS-SPOC \\
TOI-4759 & 6 & 1800 & QLP \\
--- & 33 & 600 & QLP \\
TOI-5240 & 14 & 1800 & QLP \\
--- & 40 & 600 & QLP \\
--- & 41 & 600 & TESS-SPOC \\
--- & 54 & 600 & QLP \\
--- & 55 & 600 & QLP \\
--- & 74 & 120 & SPOC \\
--- & 75 & 120 & SPOC \\
TOI-5467 & 43 & 600 & TESS-SPOC \\
--- & 44 & 600 & TESS-SPOC \\
--- & 45 & 600 & TESS-SPOC \\
--- & 71 & 120 & TESS-SPOC \\
--- & 72 & 120 & TESS-SPOC \\
TOI-5882 & 15 & 1800 & TESS-SPOC \\
--- & 41 & 600 & TESS-SPOC \\
--- & 55 & 600 & TESS-SPOC \\
--- & 75 & 120 & SPOC \\
\hline
\enddata
\end{deluxetable}

%% file: followup_table.tex
\begin{table*}[ht]
\movetableright=-0.9in
\centering
\caption{Follow-up observations}
\label{tab:followup}
\resizebox{0.95\textwidth}{!}{%
\begin{tabular}{llllllllll}
\hline \hline
TIC ID & TOI & Telescope & Camera & Observation Date (UT) & Telescope Size (m) & Filter & Pixel Scale (arcsec) & Exposure Time (s) & Detrend params \\
\hline
387342052 & 2844 & LCO-McD & QHY600 & 2023 April 5 & 0.35 & i' & 0.7 & 135 & Airmass \\
 &  & Zeiss Calar Alto & iKon-XL 230 & 2023 November 21 & 1.23 & R & 0.314 & 90 & None \\
 &  & LCO-TEID & Sinistro & 2023 November 29 & 1.0 & i' & 0.389 & 19 & None \\

61117473 & 3122 & Brierfield & Moravian 16803 & 2023 May 15 & 0.36 & R & 0.735 & 180 & Airmass \\

396133015 & 3577 & TCS & MuSCAT2 & 2023 July 16 & 1.52 & g' & 0.44 & 10 & None \\
 &  & TCS & MuSCAT2 & 2023 July 16 & 1.52 & r' & 0.44 & 5 & None \\
 &  & TCS & MuSCAT2 & 2023 July 16 & 1.52 & i' & 0.44 & 5 & None \\
 &  & TCS & MuSCAT2 & 2023 July 16 & 1.52 & z\_s & 0.44 & 10 & None \\

281196902 & 3755 & GdP & FLI4710 & 2022 March 3 & 0.4 & i' & 0.73 & 90 & None \\
&  & Thacher CDK-700 & Teledyne PIXIS
 & 2023 October 18 & 0.7 & r' & 0.608 & 40 & None \\

76420654 & 4462 & FLWO & KeplerCam & 2024 March 19 & 1.2 & i' & 0.672 & 6 & Airmass \\
&  & LCO-TEID & Sinistro & 2024 April 3 & 1.0 & i' & 0.389 & 38 & tot\_C\_cnts \\

337129672 & 4635 & LCO-SAAO & Sinistro & 2023 November 22 & 1.0 & z' & 0.389 & 37 & None \\
 &  & LCO-CTIO-fa04 & Sinistro & 2023 December 5 & 1.0 & z' & 0.389 & 37 & None \\
 &  & LCO-CTIO-fa15 & Sinistro & 2023 December 5 & 1.0 & z' & 0.389 & 37 & Airmass \\

142532090 & 4737 & LCO-TEID & Sinistro & 2023 November 26 & 1.0 & i' & 0.389 & 33 & None \\
 &  & LCO-CTIO & Sinistro & 2023 December 6 & 1.0 & i' & 0.389 & 33 & None \\
 &  & LCO-SAAO & Sinistro & 2023 December 24 & 1.0 & i' & 0.389 & 33 & Airmass \\

49705089 & 4759 & LCO-SAAO-fa06 & Sinistro & 2024 February 5 & 1.0 & i' & 0.389 & 44 & Airmass \\
 &  & LCO-SAAO-fa14 & Sinistro & 2024 February 5 & 1.0 & i' & 0.389 & 44 & None \\

40055053 & 5240 & LCO-TEID & Sinistro & 2023 August 2 & 1.0 & i' & 0.389 & 26 & None \\

83275782 & 5467 & FLWO & KeplerCam & 2023 March 4 & 1.2 & i' & 0.672 & 15 & Airmass \\
 &  & Acton Sky Portal & SBIG A4710 & 2023 March 20 & 0.36 & r' & 1.0 & 20 & Airmass \\
  &  & LCO-McD & Sinistro & 2023 October 13 & 1.0 & i' & 0.389 & 29 & tot\_C\_cnts \\

232941965 & 5882 & LCO-McD & Sinistro & 2023 June 16 & 1.0 & z' & 0.389 & 45 & None \\
\hline
\end{tabular}%
}
\begin{flushleft}
    \textbf{NOTE:} All lightcurves are available on ExoFOP\footnote{\url{https://exofop.ipac.caltech.edu}}.\\
\end{flushleft}
\end{table*}

%% file: rv_table.tex
\begin{deluxetable}{l l l l l}[bt]
\tabletypesize{\scriptsize}
\tablecaption{One Representative RV measurement for each system. \label{tab:rv}}
\tablewidth{0pt}
\tablehead{
\colhead{Target} & \colhead{\bjdtdb} & \colhead{RV (m s$^{-1}$)}  & \colhead{$\sigma_{RV}$ (m s$^{-1}$)} & \colhead{Spectrograph}
}
\startdata
TOI-2844 & 2459528.8676 & -6696 & 1791 & TRES \\
TOI-3122 & 2459651.9406 & 14841 & 145 & TRES \\
TOI-3577 & 2459395.9184 & 52 & 70 & TRES\\
TOI-3755 & 2459477.9662 & -8617 & 31 & TRES \\
TOI-4462 & 2459468.6623 & 944 & 76 & TRES \\
TOI-4635 & 2459556.7367 & 6170 & 19 & TRES \\
TOI-4737 & 2459583.9325 & 2529 & 65 & TRES \\
TOI-4759 & 2459623.7415 & -64 & 53 & TRES \\
TOI-5240 & 2459681.9612 & 20204 & 375 & TRES \\
TOI-5467 & 2459697.6358 & -237 & 352 & TRES \\
TOI-5882 & 2459899.6240 & 108 & 55 & TRES \\
TOI-5882 & 2459930.2705 & -22361.2 & 12.4 & SOPHIE \\ 
\hline
\enddata
\begin{flushleft}
    \textbf{NOTE:} The full table of RVs for each system is available in machine-readable form in the online journal.
\end{flushleft}
\end{deluxetable}

%% file: hri_table.tex
\begin{deluxetable*}{l l l l l l l l}[bt]
\tabletypesize{\scriptsize}
\tablecaption{Summary of High-Resolution Imaging Observations \label{tab:hri}}
\tablewidth{0pt}
\tablehead{
\colhead{Target} & \colhead{Telescope} & \colhead{Instrument} & \colhead{Image Type}  & \colhead{Filter} & \colhead{Contrast} & \colhead{Observation Date (UT)} & \colhead{Detection?$^\dagger$}
}
\startdata
TOI-2844 & SOAR (4.1~m) & HRCam & Speckle & $I_c$ & $\Delta$ 5.8 mag at 1$\arcsec$ & 2022 Apr 15 & No \\
--- & WIYN (3.5~m) & NESSI & Speckle & 832 nm & --- & 2022 Apr 18 & No \\
TOI-3122 & SOAR (4.1~m) & HRCam & Speckle & $I_c$ & $\Delta$ 5.0 mag at 1$\arcsec$ & 2022 Apr 25 & No \\
TOI-3577 & Palomar (5.0~m) & PHARO & AO & Br$\gamma$ & $\Delta$ 5.8 mag at 0.5$\arcsec$ & 2023 Jun 7 & No \\
TOI-3755 & SAI (2.5~m) & Speckle Polarimeter & Speckle & $I_c$ & $\Delta$ 5.7 mag at 1$\arcsec$ & 2021 Oct 29 & No \\
TOI-4462 & SAI (2.5~m) & Speckle Polarimeter & Speckle & $I_c$ & $\Delta$ 6.3 mag at 1$\arcsec$ & 2023 Jan 22 & Yes \\
--- & SAI (2.5~m) & Speckle Polarimeter & Speckle & $I_c$ & $\Delta$ 5.3 mag at 1$\arcsec$ & 2024 Feb 24 & Yes \\
--- & Palomar (5.0~m) & PHARO & AO & $H_{cont}$ & $\Delta$ 7.2 mag at 0.5$\arcsec$ & 2024 Apr 21 & Yes \\
--- & Palomar (5.0~m) & PHARO & AO & $K_{cont}$ & $\Delta$ 6.9 mag at 0.5$\arcsec$ & 2024 Apr 21 & Yes \\
TOI-4635 & Shane (3.0~m) & ShARCS & AO & $J$ & --- & 2021 Nov 21 & No \\
--- & Shane (3.0~m) & ShARCS & AO & $Ks$ & --- & 2021 Nov 21 & No \\
--- & SOAR (4.1~m) & HRCam & Speckle & $I_c$ & $\Delta$ 6.7 mag at 1$\arcsec$ & 2024 Jan 8 & No \\
TOI-4737 & Gemini (8.0~m) & Zorro & Speckle & $562$ nm & $\Delta$ 4.3 mag at 0.5$\arcsec$ & 2022 Mar 19 & No \\
--- & Gemini (8.0~m) & Zorro & Speckle & $832$ nm & $\Delta$ 6.0 mag at 0.5$\arcsec$ & 2022 Mar 19 & No \\
--- & SOAR (4.1~m) & HRCam & Speckle & $I_c$ & $\Delta$ 5.6 mag at 1$\arcsec$ & 2022 Apr 15 & No \\
TOI-4759 & SOAR (4.1~m) & HRCam & Speckle & $I_c$ & $\Delta$ 6.2 mag at 1$\arcsec$ & 2022 Apr 15 & No \\
TOI-5240 & Palomar (5.0~m) & PHARO & AO & Br$\gamma$ & $\Delta$ 6.7 mag at 0.5$\arcsec$ & 2023 Jun 6 & Yes \\
--- & SAI (2.5~m) & Speckle Polarimeter & Speckle & $I_c$ & $\Delta$ 6.3 mag at 1$\arcsec$ & 2023 Sep 1 & No \\
TOI-5467 & SAI (2.5~m) & Speckle Polarimeter & Speckle & $I_c$ & $\Delta$ 6.2 mag at 1$\arcsec$ & 2022 Dec 12 & No \\
TOI-5882 & Palomar (5.0~m) & PHARO & AO & Br$\gamma$ & $\Delta$ 6.8 mag at 0.5$\arcsec$ & 2023 Jun 6 & No \\
--- & SAI (2.5~m) & Speckle Polarimeter & Speckle & $I_c$ & $\Delta$ 7.4 mag at 1$\arcsec$ & 2023 Aug 28 & No \\
\hline
\enddata
\begin{flushleft}
    \textbf{NOTE:} All images and contrast curves are available on ExoFOP.\\
    $\dagger$ Detection refers to a positive detection of a star within the field of view of the AO or speckle instrument, subject to the maximum contrast possible with the instrument in question.
\end{flushleft}
\end{deluxetable*}

%% file: median_table_3.tex
\providecommand{\bjdtdb}{\ensuremath{\rm {BJD_{TDB}}}}
\providecommand{\feh}{\ensuremath{\left[{\rm Fe}/{\rm H}\right]}}
\providecommand{\teff}{\ensuremath{T_{\rm eff}}}
\providecommand{\teq}{\ensuremath{T_{\rm eq}}}
\providecommand{\ecosw}{\ensuremath{e\cos{\omega_*}}}
\providecommand{\esinw}{\ensuremath{e\sin{\omega_*}}}
\providecommand{\msun}{\ensuremath{\,M_\Sun}}
\providecommand{\rsun}{\ensuremath{\,R_\Sun}}
\providecommand{\lsun}{\ensuremath{\,L_\Sun}}
\providecommand{\mj}{\ensuremath{\,M_{\rm J}}}
\providecommand{\rj}{\ensuremath{\,R_{\rm J}}}
\providecommand{\me}{\ensuremath{\,M_{\rm E}}}
\providecommand{\re}{\ensuremath{\,R_{\rm E}}}
\providecommand{\fave}{\langle F \rangle}
\providecommand{\fluxcgs}{10$^9$ erg s$^{-1}$ cm$^{-2}$}
\providecommand{\tess}{\textit{TESS}\xspace}
\tablecolumns{6}
\tablehead{&  & \colhead{TOI-2844} & \colhead{TOI-3122} & \colhead{TOI-3755} & \colhead{TOI-4635}}
\startdata
\multicolumn{6}{l}{\textbf{Priors}:} \\
$\pi$ & Gaia Parallax (mas)& $\mathcal{G}$[1.4759, 0.01696] & $\mathcal{G}$[1.962, 0.01778] & $\mathcal{G}$[3.0924, 0.01443] & $\mathcal{G}$[13.33, 0.02582] \\
$[{\rm Fe/H}]$ & Metallicity (dex)& $\mathcal{G}$[0.025, 0.198] & $\mathcal{G}$[0.3193, 0.1188] & $\mathcal{G}$[0.3156, 0.1007] & $\mathcal{G}$[-0.1783, 0.1608] \\
$A_V$ & V-band extinction (mag)& $\mathcal{U}$[0, 0.2725] & $\mathcal{U}$[0, 0.4675] & $\mathcal{U}$[0, 0.9867] & $\mathcal{U}$[0, 0.3959] \\
$D_T$ & Dilution in \tess& $\mathcal{G}$[0, 0.008335] & $\mathcal{G}$[0, 0.027729] & $\mathcal{G}$[0, 0.012054] & $\mathcal{G}$[0, 0.002608] \\
\hline
\multicolumn{6}{l}{\textbf{Primary Star Parameters}:} \\
$M_*$ & Mass (\msun) & $1.585^{+0.071}_{-0.072}$ & $1.247^{+0.074}_{-0.091}$ & $1.037^{+0.066}_{-0.071}$ & $0.698^{+0.027}_{-0.025}$ \\
$R_*$ & Radius (\rsun) & $1.784^{+0.085}_{-0.08}$ & $1.336^{+0.062}_{-0.045}$ & $1.044^{+0.042}_{-0.038}$ & $0.683\pm 0.011$ \\
$L_*$ & Luminosity (\lsun) & $6.51^{+0.57}_{-0.48}$ & $2.27^{+0.24}_{-0.27}$ & $0.99^{+0.12}_{-0.11}$ & $0.182^{+0.011}_{-0.013}$ \\
$\rho_*$ & Density (cgs) & $0.394^{+0.065}_{-0.059}$ & $0.742^{+0.09}_{-0.12}$ & $1.29^{+0.18}_{-0.17}$ & $3.09\pm 0.12$ \\
$\log{g}$ & Surface gravity (cgs) & $4.135^{+0.048}_{-0.051}$ & $4.284^{+0.038}_{-0.059}$ & $4.417^{+0.043}_{-0.049}$ & $4.613\pm 0.013$ \\
$T_{\rm eff}$ & Effective temperature (K) & $6910.0\pm 210$ & $6120.0^{+180}_{-220}$ & $5630.0\pm 170$ & $4555.0^{+67}_{-74}$ \\
$[{\rm Fe/H}]$ & Metallicity (dex) & $0.06^{+0.12}_{-0.089}$ & $0.29\pm 0.11$ & $0.334^{+0.092}_{-0.098}$ & $-0.091^{+0.039}_{-0.033}$ \\
$[{\rm Fe/H}]_{0}$ & Initial metallicity & $0.22^{+0.1}_{-0.088}$ & $0.298^{+0.094}_{-0.093}$ & $0.317^{+0.085}_{-0.093}$ & $-0.077^{+0.055}_{-0.051}$ \\
Age & Age (Gyr) & $1.08^{+0.52}_{-0.42}$ & $2.5^{+2.6}_{-1.6}$ & $4.9^{+4.9}_{-3.5}$ & $7.5^{+4.2}_{-4.5}$ \\
EEP & Equal evolutionary phase & $342.5^{+9.5}_{-14}$ & $351.0^{+59}_{-32}$ & $362.0^{+44}_{-40}$ & $332.0^{+11}_{-24}$ \\
$A_V$ & V-band extinction (mag) & $0.132^{+0.084}_{-0.082}$ & $0.3^{+0.12}_{-0.17}$ & $0.5^{+0.15}_{-0.16}$ & $0.26^{+0.1}_{-0.15}$ \\
$d$ & Distance (pc) & $677.0^{+7.9}_{-7.6}$ & $510.0\pm 4.6$ & $323.5\pm 1.5$ & $75.01^{+0.15}_{-0.14}$ \\
\multicolumn{6}{l}{\textbf{Companion Parameters}:} \\
$P$ & Period (days) & $3.5524204\pm 0.000003$ & $6.1836025\pm 0.0000063$ & $5.543744^{+0.0000062}_{-0.0000061}$ & $12.2769349\pm 0.0000033$ \\
$R_P$ & Radius (\rj) & $0.775^{+0.047}_{-0.043}$ & $1.235^{+0.075}_{-0.057}$ & $0.885^{+0.051}_{-0.046}$ & $1.02\pm 0.019$ \\
$M_P$ & Mass (\mj) & $54.0^{+4.9}_{-5.1}$ & $101.5^{+4.1}_{-4.8}$ & $47.1^{+2}_{-2.1}$ & $84.0^{+2.1}_{-2}$ \\
$T_C$ & Time of conjunction (\bjdtdb) & $2459574.31396^{+0.00094}_{-0.00099}$ & $2459356.52122^{+0.00067}_{-0.00065}$ & $2459914.24483^{+0.00057}_{-0.00058}$ & $2459448.74844\pm 0.00018$ \\
$T_0$ & Optimal conjunction time (\bjdtdb) & $2459940.21326^{+0.00088}_{-0.00093}$ & $2459727.53737^{+0.00054}_{-0.00052}$ & $2459775.65123^{+0.00054}_{-0.00057}$ & $2460013.48744\pm 0.0001$ \\
$a$ & Semi-major axis (AU) & $0.0537^{+0.00079}_{-0.00082}$ & $0.0728^{+0.0014}_{-0.0018}$ & $0.0629^{+0.0013}_{-0.0015}$ & $0.0958^{+0.0012}_{-0.0011}$ \\
$i$ & Inclination (Degrees) & $83.7^{+2}_{-1.3}$ & $87.3^{+1.6}_{-1.5}$ & $87.51^{+0.35}_{-0.36}$ & $88.791^{+0.057}_{-0.056}$ \\
$e$ & Eccentricity & $0.424^{+0.046}_{-0.041}$ & $0.4704^{+0.008}_{-0.0077}$ & $0.0049^{+0.0031}_{-0.0026}$ & $0.4906\pm 0.0015$ \\
$\omega_*$ & Argument of periastron (Degrees) & $159.0\pm 11$ & $75.55^{+0.98}_{-0.91}$ & $21.0^{+44}_{-57}$ & $-5.99^{+0.78}_{-0.74}$ \\
$T_{eq}$ & Equilibrium temperature (K) & $1919.0^{+37}_{-33}$ & $1267.0^{+27}_{-32}$ & $1106.0\pm 28$ & $586.6^{+8.2}_{-9.5}$ \\
$\tau_{\rm circ}$ & Tidal circularization timescale (Gyr) & $27.0^{+17}_{-12}$ & $33.0^{+10}_{-9.7}$ & $500.0^{+170}_{-130}$ & $827.0^{+71}_{-65}$ \\
$K$ & RV semi-amplitude (m/s) & $5700.0^{+450}_{-470}$ & $10450.0\pm 110$ & $5127.0\pm 22$ & $10032.0^{+41}_{-43}$ \\
$R_P/R_*$ & Radius of planet in stellar radii  & $0.0447\pm 0.0013$ & $0.0952^{+0.0026}_{-0.0025}$ & $0.0872\pm 0.0025$ & $0.15339^{+0.00081}_{-0.00082}$ \\
$a/R_*$ & Semi-major axis in stellar radii  & $6.47\pm 0.34$ & $11.74^{+0.45}_{-0.68}$ & $12.96^{+0.57}_{-0.61}$ & $30.17\pm 0.38$ \\
Depth & \tess\ flux decrement at mid-transit & $0.00212\pm 0.00011$ & $0.01026^{+0.00056}_{-0.00055}$ & $0.0085\pm 0.00046$ & $0.02871^{+0.00036}_{-0.00035}$ \\
$\tau$ & Ingress/egress transit duration (days) & $0.0073^{+0.0016}_{-0.0015}$ & $0.01015^{+0.0015}_{-0.00082}$ & $0.0144^{+0.0017}_{-0.0014}$ & $0.02135\pm 0.00058$ \\
$T_{14}$ & Total transit duration (days) & $0.1258^{+0.0019}_{-0.0018}$ & $0.1075^{+0.0019}_{-0.0016}$ & $0.1267^{+0.0021}_{-0.0019}$ & $0.12316^{+0.00051}_{-0.0005}$ \\
$b$ & Transit impact parameter & $0.52^{+0.11}_{-0.19}$ & $0.3^{+0.14}_{-0.17}$ & $0.564^{+0.052}_{-0.058}$ & $0.509^{+0.017}_{-0.019}$ \\
$\rho_P$ & Density (cgs) & $143.0^{+31}_{-27}$ & $67.0^{+10}_{-12}$ & $84.0^{+16}_{-14}$ & $98.3^{+4.9}_{-4.6}$ \\
$\log{g_P}$ & Surface gravity  & $5.346^{+0.065}_{-0.069}$ & $5.218^{+0.044}_{-0.062}$ & $5.173^{+0.052}_{-0.055}$ & $5.302\pm 0.014$ \\
$\Theta$ & Safronov number & $4.71^{+0.49}_{-0.48}$ & $9.58^{+0.47}_{-0.54}$ & $6.45^{+0.36}_{-0.35}$ & $22.57^{+0.47}_{-0.46}$ \\
$T_S$ & Time of eclipse (\bjdtdb) & $2459575.217^{+0.086}_{-0.084}$ & $2459353.947^{+0.033}_{-0.034}$ & $2459911.4849^{+0.0084}_{-0.0091}$ & $2459446.27^{+0.011}_{-0.01}$ \\
$T_{S,14}$ & Total eclipse duration (days) & $0.142^{+0.035}_{-0.011}$ & $0.205^{+0.069}_{-0.2}$ & $0.1269^{+0.0021}_{-0.002}$ & $0.1137\pm 0.0013$ \\
$e\cos{\omega_*}$ & & $-0.391^{+0.042}_{-0.04}$ & $0.1173^{+0.0076}_{-0.0078}$ & $0.0034^{+0.0024}_{-0.0026}$ & $0.4878\pm 0.0016$ \\
$e\sin{\omega_*}$ & & $0.148^{+0.088}_{-0.078}$ & $0.4555^{+0.0082}_{-0.0078}$ & $0.001^{+0.0039}_{-0.0032}$ & $-0.0511^{+0.0066}_{-0.0063}$ \\
$M_P/M_*$ & Mass ratio  & $0.0326^{+0.0028}_{-0.003}$ & $0.0778^{+0.0023}_{-0.0018}$ & $0.04337^{+0.0011}_{-0.00093}$ & $0.1148\pm 0.0016$ \\
$d/R_*$ & Separation at mid-transit  & $4.62^{+0.56}_{-0.58}$ & $6.26^{+0.28}_{-0.36}$ & $12.95^{+0.58}_{-0.6}$ & $24.14\pm 0.41$ \\
\enddata

%% file: median_table_10.tex
\providecommand{\bjdtdb}{\ensuremath{\rm {BJD_{TDB}}}}
\providecommand{\feh}{\ensuremath{\left[{\rm Fe}/{\rm H}\right]}}
\providecommand{\teff}{\ensuremath{T_{\rm eff}}}
\providecommand{\teq}{\ensuremath{T_{\rm eq}}}
\providecommand{\ecosw}{\ensuremath{e\cos{\omega_*}}}
\providecommand{\esinw}{\ensuremath{e\sin{\omega_*}}}
\providecommand{\msun}{\ensuremath{\,M_\Sun}}
\providecommand{\rsun}{\ensuremath{\,R_\Sun}}
\providecommand{\lsun}{\ensuremath{\,L_\Sun}}
\providecommand{\mj}{\ensuremath{\,M_{\rm J}}}
\providecommand{\rj}{\ensuremath{\,R_{\rm J}}}
\providecommand{\me}{\ensuremath{\,M_{\rm E}}}
\providecommand{\re}{\ensuremath{\,R_{\rm E}}}
\providecommand{\fave}{\langle F \rangle}
\providecommand{\fluxcgs}{10$^9$ erg s$^{-1}$ cm$^{-2}$}
\providecommand{\tess}{\textit{TESS}\xspace}
\tablecolumns{5}
\tablehead{&  & \colhead{TOI-4737} & \colhead{TOI-5240} & \colhead{TOI-5467}}
\startdata
\multicolumn{5}{l}{\textbf{Priors}:} \\
$\pi$ & Gaia Parallax (mas)& $\mathcal{G}$[1.7399, 0.01656] & $\mathcal{G}$[1.0355, 0.01516] & $\mathcal{G}$[1.7819, 0.01704]\\
$[{\rm Fe/H}]$ & Metallicity (dex)& $\mathcal{G}$[0.3267, 0.1499] & $\mathcal{G}$[-0.1571, 0.2089] & $\mathcal{G}$[0.3257, 0.1645] \\
$A_V$ & V-band extinction (mag)& $\mathcal{U}$[0, 1.8386] & $\mathcal{U}$[0, 0.4879] & $\mathcal{U}$[0, 1.6687] \\
$D_T$ & Dilution in \tess& $\mathcal{G}$[0, 0.005883] & $\mathcal{G}$[0, 0.024878] & $\mathcal{G}$[0, 0.007941]\\
\hline
\multicolumn{5}{l}{\textbf{Primary Star Parameters}:} \\
$M_*$ & Mass (\msun) & $1.336^{+0.08}_{-0.09}$ & $1.754^{+0.094}_{-0.092}$ & $1.515^{+0.059}_{-0.058}$ \\
$R_*$ & Radius (\rsun) & $1.618^{+0.068}_{-0.066}$ & $2.35\pm 0.11$ & $1.503^{+0.046}_{-0.045}$ \\
$L_*$ & Luminosity (\lsun) & $3.39^{+0.34}_{-0.31}$ & $14.9^{+2.4}_{-2.1}$ & $4.4^{+0.41}_{-0.33}$ \\
$\rho_*$ & Density (cgs) & $0.443^{+0.073}_{-0.065}$ & $0.189^{+0.032}_{-0.026}$ & $0.63^{+0.054}_{-0.052}$ \\
$\log{g}$ & Surface gravity (cgs) & $4.145^{+0.05}_{-0.053}$ & $3.938^{+0.049}_{-0.047}$ & $4.266^{+0.024}_{-0.027}$ \\
$T_{\rm eff}$ & Effective temperature (K) & $6160.0^{+210}_{-200}$ & $7390.0^{+350}_{-330}$ & $6820.0^{+170}_{-150}$ \\
$[{\rm Fe/H}]$ & Metallicity (dex) & $0.24^{+0.1}_{-0.11}$ & $-0.12\pm 0.19$ & $0.265^{+0.098}_{-0.11}$ \\
$[{\rm Fe/H}]_{0}$ & Initial metallicity & $0.297^{+0.093}_{-0.096}$ & $-0.03^{+0.18}_{-0.19}$ & $0.316^{+0.083}_{-0.093}$ \\
Age & Age (Gyr) & $3.0^{+1.6}_{-1.3}$ & $1.19^{+0.23}_{-0.2}$ & $0.29^{+0.42}_{-0.2}$ \\
EEP & Equal evolutionary phase & $377.0^{+38}_{-32}$ & $374.0^{+13}_{-14}$ & $294.0^{+27}_{-37}$ \\
$A_V$ & V-band extinction (mag) & $0.4\pm 0.11$ & $0.23^{+0.15}_{-0.14}$ & $0.567^{+0.092}_{-0.08}$ \\
$d$ & Distance (pc) & $574.5^{+5.5}_{-5.4}$ & $966.0\pm 14$ & $560.5^{+5.4}_{-5.2}$ \\
\multicolumn{5}{l}{\textbf{Companion Parameters}:} \\
$P$ & Period (days) & $9.320278^{+0.000018}_{-0.000017}$ & $4.1793241\pm 0.0000058$ & $2.6570963^{+0.0000027}_{-0.0000028}$ \\
$R_P$ & Radius (\rj) & $0.701^{+0.079}_{-0.059}$ & $1.655^{+0.097}_{-0.096}$ & $1.096^{+0.046}_{-0.043}$ \\
$M_P$ & Mass (\mj) & $66.3^{+2.7}_{-3.1}$ & $128.0^{+4.9}_{-4.8}$ & $91.7^{+2.8}_{-2.7}$ \\
$T_C$ & Time of conjunction (\bjdtdb) & $2459222.4062^{+0.0017}_{-0.0018}$ & $2459444.90408^{+0.0009}_{-0.00091}$ & $2459545.35471^{+0.00052}_{-0.00053}$ \\
$T_0$ & Optimal conjunction time (\bjdtdb) & $2459949.3884\pm 0.0011$ & $2459950.60229\pm 0.00057$ & $2459927.97657\pm 0.00035$ \\
$a$ & Semi-major axis (AU) & $0.097^{+0.0019}_{-0.0022}$ & $0.0626\pm 0.0011$ & $0.04394^{+0.00056}_{-0.00055}$ \\
$i$ & Inclination (Degrees) & $87.82^{+0.51}_{-0.46}$ & $85.7^{+1.4}_{-1.1}$ & $83.21^{+0.35}_{-0.37}$ \\
$e$ & Eccentricity & $0.0063^{+0.007}_{-0.0042}$ & $0.0113^{+0.014}_{-0.0078}$ & $0.0137^{+0.013}_{-0.0083}$ \\
$\omega_*$ & Argument of periastron (Degrees) & $-120.0^{+40}_{-89}$ & $-115.0^{+28}_{-83}$ & $121.0^{+57}_{-28}$ \\
$T_{eq}$ & Equilibrium temperature (K) & $1214.0^{+26}_{-25}$ & $2187.0^{+80}_{-81}$ & $1924.0^{+39}_{-34}$ \\
$\tau_{\rm circ}$ & Tidal circularization timescale (Gyr) & $25000.0^{+15000}_{-11000}$ & $25.7^{+9.5}_{-6.6}$ & $18.1^{+3.8}_{-3.3}$ \\
$K$ & RV semi-amplitude (m/s) & $5118.0\pm 49$ & $10580.0^{+160}_{-150}$ & $9760.0^{+130}_{-140}$ \\
$R_P/R_*$ & Radius of planet in stellar radii  & $0.0443^{+0.0047}_{-0.003}$ & $0.0723\pm 0.0019$ & $0.075\pm 0.0017$ \\
$a/R_*$ & Semi-major axis in stellar radii  & $12.86^{+0.67}_{-0.66}$ & $5.72^{+0.31}_{-0.27}$ & $6.29^{+0.17}_{-0.18}$ \\
Depth & \tess\ flux decrement at mid-transit & $0.00216^{+0.00048}_{-0.00028}$ & $0.00563\pm 0.0003$ & $0.00569\pm 0.00025$ \\
$\tau$ & Ingress/egress transit duration (days) & $0.012^{+0.0019}_{-0.0015}$ & $0.0192^{+0.0023}_{-0.0022}$ & $0.015^{+0.0011}_{-0.001}$ \\
$T_{14}$ & Total transit duration (days) & $0.214^{+0.0029}_{-0.0027}$ & $0.232^{+0.0023}_{-0.0022}$ & $0.1053\pm 0.0012$ \\
$b$ & Transit impact parameter & $0.491^{+0.074}_{-0.095}$ & $0.436^{+0.084}_{-0.13}$ & $0.735^{+0.018}_{-0.02}$ \\
$\rho_P$ & Density (cgs) & $236.0^{+76}_{-67}$ & $35.0^{+7.2}_{-5.7}$ & $86.5^{+11}_{-9.8}$ \\
$\log{g_P}$ & Surface gravity  & $5.521^{+0.083}_{-0.098}$ & $5.063^{+0.056}_{-0.053}$ & $5.278^{+0.034}_{-0.035}$ \\
$\Theta$ & Safronov number & $13.7^{+1.3}_{-1.4}$ & $5.52^{+0.34}_{-0.31}$ & $4.85^{+0.22}_{-0.21}$ \\
$T_S$ & Time of eclipse (\bjdtdb) & $2459227.054^{+0.016}_{-0.022}$ & $2459446.983^{+0.012}_{-0.014}$ & $2459546.6736^{+0.0091}_{-0.011}$ \\
$T_{S,14}$ & Total eclipse duration (days) & $0.213^{+0.0035}_{-0.0036}$ & $0.2294^{+0.0042}_{-0.006}$ & $0.1054\pm 0.0011$ \\
$e\cos{\omega_*}$ & & $-0.002^{+0.0027}_{-0.0037}$ & $-0.0039^{+0.0043}_{-0.0053}$ & $-0.0057^{+0.0054}_{-0.0067}$ \\
$e\sin{\omega_*}$ & & $-0.0026^{+0.0046}_{-0.0089}$ & $-0.0065^{+0.0084}_{-0.017}$ & $0.009^{+0.015}_{-0.0093}$ \\
$M_P/M_*$ & Mass ratio  & $0.0474^{+0.0013}_{-0.0011}$ & $0.0696^{+0.0018}_{-0.0017}$ & $0.0578\pm 0.0012$ \\
$d/R_*$ & Separation at mid-transit  & $12.92^{+0.68}_{-0.67}$ & $5.77^{+0.32}_{-0.29}$ & $6.22^{+0.2}_{-0.21}$ \\
\enddata

%% file: median_table_7.tex
\providecommand{\bjdtdb}{\ensuremath{\rm {BJD_{TDB}}}}
\providecommand{\feh}{\ensuremath{\left[{\rm Fe}/{\rm H}\right]}}
\providecommand{\teff}{\ensuremath{T_{\rm eff}}}
\providecommand{\teq}{\ensuremath{T_{\rm eq}}}
\providecommand{\ecosw}{\ensuremath{e\cos{\omega_*}}}
\providecommand{\esinw}{\ensuremath{e\sin{\omega_*}}}
\providecommand{\msun}{\ensuremath{\,M_\Sun}}
\providecommand{\rsun}{\ensuremath{\,R_\Sun}}
\providecommand{\lsun}{\ensuremath{\,L_\Sun}}
\providecommand{\mj}{\ensuremath{\,M_{\rm J}}}
\providecommand{\rj}{\ensuremath{\,R_{\rm J}}}
\providecommand{\me}{\ensuremath{\,M_{\rm E}}}
\providecommand{\re}{\ensuremath{\,R_{\rm E}}}
\providecommand{\fave}{\langle F \rangle}
\providecommand{\fluxcgs}{10$^9$ erg s$^{-1}$ cm$^{-2}$}
\providecommand{\tess}{\textit{TESS}\xspace}
\tablecolumns{6}
\tablehead{& & \multicolumn{2}{c}{\textbf{TOI-3577}} & \multicolumn{2}{c}{\textbf{TOI-4462}} \\&  & \colhead{Low-mass solution} & \colhead{High-mass solution} & \colhead{Low-mass solution} & \colhead{High-mass solution} \\ & & (63.8\% probability) & (36.2\% probability) & (90.7\% probability) & (9.3\% probability)}
\startdata
\multicolumn{6}{l}{\textbf{Priors}:} \\
$\pi$ & Gaia Parallax (mas) & \multicolumn{2}{c}{$\mathcal{G}$[2.36440, 0.01487]} & \multicolumn{2}{c}{$\mathcal{G}$[2.53984, 0.03785]}\\
$[{\rm Fe/H}]$ & Metallicity (dex) & \multicolumn{2}{c}{$\mathcal{G}$[-0.0486, 0.0889]} & \multicolumn{2}{c}{$\mathcal{G}$[0.0873, 0.1781]} \\
$A_V$ & V-band extinction (mag) & \multicolumn{2}{c}{$\mathcal{U}$[0, 2.2165]} & \multicolumn{2}{c}{$\mathcal{U}$[0, 0.0949]} \\
$D_T$ & Dilution in \tess\ & \multicolumn{2}{c}{$\mathcal{G}$[0, 0.032559]} & \multicolumn{2}{c}{$\mathcal{G}$[0, 0.057591]} \\
\hline
\multicolumn{6}{l}{\textbf{Primary Star Parameters}:} \\
$M_*$ & Mass (\msun) & $1.111^{+0.057}_{-0.067}$ & $1.31^{+0.073}_{-0.056}$ & $1.252^{+0.053}_{-0.061}$ & $1.452^{+0.049}_{-0.047}$ \\
$R_*$ & Radius (\rsun) & $1.753^{+0.068}_{-0.067}$ & $1.733^{+0.069}_{-0.064}$ & $2.084^{+0.084}_{-0.059}$ & $2.128^{+0.046}_{-0.036}$ \\
$L_*$ & Luminosity (\lsun) & $3.38^{+0.5}_{-0.42}$ & $4.06^{+0.59}_{-0.52}$ & $4.98^{+0.28}_{-0.24}$ & $5.07^{+0.24}_{-0.22}$ \\
$\rho_*$ & Density (cgs) & $0.29^{+0.037}_{-0.034}$ & $0.355^{+0.047}_{-0.038}$ & $0.195^{+0.017}_{-0.023}$ & $0.2132^{+0.0075}_{-0.011}$ \\
$\log{g}$ & Surface gravity (cgs) & $3.995^{+0.038}_{-0.04}$ & $4.078^{+0.04}_{-0.034}$ & $3.898^{+0.026}_{-0.04}$ & $3.945^{+0.011}_{-0.015}$ \\
$T_{\rm eff}$ & Effective temperature (K) & $5920.0\pm 210$ & $6210.0^{+240}_{-200}$ & $5970.0\pm 110$ & $5930.0^{+84}_{-82}$ \\
$[{\rm Fe/H}]$ & Metallicity (dex) & $-0.031^{+0.071}_{-0.069}$ & $-0.026^{+0.081}_{-0.073}$ & $0.05^{+0.15}_{-0.13}$ & $0.19^{+0.13}_{-0.14}$ \\
$[{\rm Fe/H}]_{0}$ & Initial metallicity & $0.021^{+0.066}_{-0.064}$ & $0.074^{+0.076}_{-0.068}$ & $0.09\pm 0.12$ & $0.22^{+0.11}_{-0.12}$ \\
Age & Age (Gyr) & $6.8^{+1.8}_{-1.3}$ & $3.29^{+0.68}_{-0.89}$ & $4.73^{+0.75}_{-0.58}$ & $2.88^{+0.36}_{-0.33}$ \\
EEP & Equal evolutionary phase & $453.5^{+4.5}_{-7.7}$ & $398.0^{+15}_{-28}$ & $454.5^{+3.9}_{-5.4}$ & $407.3^{+6.4}_{-7.4}$ \\
$A_V$ & V-band extinction (mag) & $0.31^{+0.16}_{-0.17}$ & $0.53\pm 0.15$ & $0.054^{+0.029}_{-0.035}$ & $0.063^{+0.023}_{-0.036}$ \\
$d$ & Distance (pc) & $423.2\pm 2.7$ & $423.4\pm 2.7$ & $392.6^{+5.8}_{-5.6}$ & $395.2^{+5.7}_{-5.5}$ \\
\multicolumn{6}{l}{\textbf{Companion Parameters}:} \\
$P$ & Period (days) & $5.266759\pm 0.000013$ & $5.266759^{+0.000014}_{-0.000013}$ & $4.9132987^{+0.0000088}_{-0.0000089}$ & $4.9132998^{+0.0000088}_{-0.0000089}$ \\
$R_P$ & Radius (\rj) & $0.999^{+0.053}_{-0.051}$ & $0.967^{+0.053}_{-0.048}$ & $1.141^{+0.081}_{-0.078}$ & $1.158\pm 0.075$ \\
$M_P$ & Mass (\mj) & $53.8^{+1.9}_{-2.2}$ & $60.0^{+2.2}_{-1.7}$ & $101.7^{+2.8}_{-3.2}$ & $111.9^{+2.5}_{-2.4}$ \\
$T_C$ & Time of conjunction (\bjdtdb) & $2459847.67307^{+0.00091}_{-0.00092}$ & $2459847.673^{+0.00093}_{-0.0009}$ & $2459789.41867\pm 0.00079$ & $2459789.4187^{+0.00078}_{-0.00079}$ \\
$T_0$ & Optimal conjunction time (\bjdtdb) & $2460105.74424^{+0.00064}_{-0.00063}$ & $2460105.74423^{+0.0006}_{-0.00062}$ & $2459882.77168\pm 0.00077$ & $2459882.77175^{+0.00076}_{-0.00077}$ \\
$a$ & Semi-major axis (AU) & $0.0623^{+0.001}_{-0.0013}$ & $0.06576^{+0.0012}_{-0.00093}$ & $0.06251^{+0.00085}_{-0.001}$ & $0.06558^{+0.00072}_{-0.0007}$ \\
$i$ & Inclination (Degrees) & $83.56^{+0.36}_{-0.38}$ & $84.14\pm 0.34$ & $87.7^{+1.4}_{-1.3}$ & $89.02^{+0.69}_{-0.97}$ \\
$e$ & Eccentricity & $0.006^{+0.0081}_{-0.0042}$ & $0.0066^{+0.0087}_{-0.0047}$ & $0.0203^{+0.0034}_{-0.0038}$ & $0.0199^{+0.0034}_{-0.004}$ \\
$\omega_*$ & Argument of periastron (Degrees) & $-77.0^{+86}_{-47}$ & $-78.0^{+80}_{-37}$ & $96.5^{+5.9}_{-5.8}$ & $96.5^{+6.1}_{-6}$ \\
$T_{eq}$ & Equilibrium temperature (K) & $1512.0^{+47}_{-43}$ & $1540.0\pm 47$ & $1663.0^{+27}_{-23}$ & $1630.0^{+20}_{-19}$ \\
$\tau_{\rm circ}$ & Tidal circularization timescale (Gyr) & $258.0^{+80}_{-62}$ & $378.0^{+110}_{-87}$ & $214.0^{+90}_{-63}$ & $241.0^{+95}_{-64}$ \\
$K$ & RV semi-amplitude (m/s) & $5656.0^{+40}_{-48}$ & $5655.0^{+40}_{-46}$ & $9954.0^{+31}_{-33}$ & $9954.0^{+32}_{-34}$ \\
$R_P/R_*$ & Radius of planet in stellar radii  & $0.0586\pm 0.0012$ & $0.0574\pm 0.0011$ & $0.0561^{+0.0033}_{-0.0034}$ & $0.0558^{+0.0034}_{-0.0035}$ \\
$a/R_*$ & Semi-major axis in stellar radii  & $7.63\pm 0.31$ & $8.16^{+0.34}_{-0.31}$ & $6.45^{+0.18}_{-0.27}$ & $6.635^{+0.077}_{-0.11}$ \\
Depth & \tess\ flux decrement at mid-transit & $0.00317\pm 0.0001$ & $0.003137^{+0.000099}_{-0.000098}$ & $0.00359^{+0.00044}_{-0.00042}$ & $0.00361^{+0.00045}_{-0.00044}$ \\
$\tau$ & Ingress/egress transit duration (days) & $0.0262^{+0.0028}_{-0.0026}$ & $0.0222^{+0.0023}_{-0.0021}$ & $0.014^{+0.0015}_{-0.0012}$ & $0.01323^{+0.00092}_{-0.00088}$ \\
$T_{14}$ & Total transit duration (days) & $0.1372^{+0.0027}_{-0.0025}$ & $0.1339^{+0.0023}_{-0.0022}$ & $0.2447^{+0.0024}_{-0.0023}$ & $0.2433^{+0.0022}_{-0.0021}$ \\
$b$ & Transit impact parameter & $0.86^{+0.012}_{-0.014}$ & $0.838^{+0.015}_{-0.016}$ & $0.26^{+0.12}_{-0.15}$ & $0.112^{+0.11}_{-0.078}$ \\
$\rho_P$ & Density (cgs) & $66.7^{+12}_{-9.9}$ & $82.0^{+14}_{-12}$ & $84.0^{+20}_{-16}$ & $89.0^{+20}_{-15}$ \\
$\log{g_P}$ & Surface gravity  & $5.125\pm 0.048$ & $5.201^{+0.047}_{-0.045}$ & $5.286^{+0.061}_{-0.06}$ & $5.316^{+0.058}_{-0.054}$ \\
$\Theta$ & Safronov number & $6.03^{+0.33}_{-0.3}$ & $6.21\pm 0.32$ & $8.89^{+0.65}_{-0.59}$ & $8.72^{+0.61}_{-0.53}$ \\
$T_S$ & Time of eclipse (\bjdtdb) & $2459845.0427^{+0.011}_{-0.0087}$ & $2459845.0428^{+0.011}_{-0.0091}$ & $2459791.8682^{+0.0062}_{-0.0065}$ & $2459791.8684^{+0.0064}_{-0.0065}$ \\
$T_{S,14}$ & Total eclipse duration (days) & $0.1381^{+0.0036}_{-0.0031}$ & $0.1347^{+0.003}_{-0.0027}$ & $0.2539^{+0.0028}_{-0.0029}$ & $0.2528^{+0.0029}_{-0.003}$ \\
$e\cos{\omega_*}$ & & $0.0009^{+0.0034}_{-0.0026}$ & $0.0009^{+0.0034}_{-0.0028}$ & $-0.0023^{+0.002}_{-0.0021}$ & $-0.0022\pm 0.0021$ \\
$e\sin{\omega_*}$ & & $-0.0031^{+0.0044}_{-0.0097}$ & $-0.0041^{+0.005}_{-0.01}$ & $0.0201^{+0.0034}_{-0.0038}$ & $0.0197^{+0.0034}_{-0.0041}$ \\
$M_P/M_*$ & Mass ratio  & $0.04626^{+0.0011}_{-0.00087}$ & $0.04365^{+0.00075}_{-0.00091}$ & $0.0776^{+0.0014}_{-0.0012}$ & $0.07359\pm 0.0009$ \\
$d/R_*$ & Separation at mid-transit  & $7.67^{+0.33}_{-0.32}$ & $8.21^{+0.36}_{-0.32}$ & $6.32^{+0.18}_{-0.26}$ & $6.503^{+0.09}_{-0.12}$ \\
\enddata

%% file: median_table_11.tex
\providecommand{\bjdtdb}{\ensuremath{\rm {BJD_{TDB}}}}
\providecommand{\feh}{\ensuremath{\left[{\rm Fe}/{\rm H}\right]}}
\providecommand{\teff}{\ensuremath{T_{\rm eff}}}
\providecommand{\teq}{\ensuremath{T_{\rm eq}}}
\providecommand{\ecosw}{\ensuremath{e\cos{\omega_*}}}
\providecommand{\esinw}{\ensuremath{e\sin{\omega_*}}}
\providecommand{\msun}{\ensuremath{\,M_\Sun}}
\providecommand{\rsun}{\ensuremath{\,R_\Sun}}
\providecommand{\lsun}{\ensuremath{\,L_\Sun}}
\providecommand{\mj}{\ensuremath{\,M_{\rm J}}}
\providecommand{\rj}{\ensuremath{\,R_{\rm J}}}
\providecommand{\me}{\ensuremath{\,M_{\rm E}}}
\providecommand{\re}{\ensuremath{\,R_{\rm E}}}
\providecommand{\fave}{\langle F \rangle}
\providecommand{\fluxcgs}{10$^9$ erg s$^{-1}$ cm$^{-2}$}
\providecommand{\tess}{\textit{TESS}\xspace}
\tablecolumns{6}
\tablehead{& & \multicolumn{2}{c}{\textbf{TOI-4759}} & \multicolumn{2}{c}{\textbf{TOI-5882}} \\&  & \colhead{Low-mass solution} & \colhead{High-mass solution} & \colhead{Low-mass solution} & \colhead{High-mass solution} \\ & & (68.2\% probability) & (31.8\% probability) & (71.1\% probability) & (28.9\% probability)}
\startdata
\multicolumn{6}{l}{\textbf{Priors}:} \\
$\pi$ & Gaia Parallax (mas) & \multicolumn{2}{c}{$\mathcal{G}$[1.34392, 0.01465]} & \multicolumn{2}{c}{$\mathcal{G}$[2.42207, 0.01753]}\\
$[{\rm Fe/H}]$ & Metallicity (dex) & \multicolumn{2}{c}{$\mathcal{G}$[0.2723, 0.4141]} & \multicolumn{2}{c}{$\mathcal{G}$[0.1400, 0.1900]} \\
$A_V$ & V-band extinction (mag) & \multicolumn{2}{c}{$\mathcal{U}$[0, 0.1507]} & \multicolumn{2}{c}{$\mathcal{U}$[0, 0.8438]} \\
$D_T$ & Dilution in \tess\ & \multicolumn{2}{c}{$\mathcal{G}$[0, 0.038128]} & \multicolumn{2}{c}{$\mathcal{G}$[0, 0.008565]} \\
\hline
\hline
\multicolumn{6}{l}{\textbf{Stellar Parameters}:} \\
$M_*$ & Mass (\msun) & $1.186^{+0.061}_{-0.079}$ & $1.384^{+0.049}_{-0.048}$ & $1.334^{+0.055}_{-0.065}$ & $1.549^{+0.055}_{-0.053}$ \\
$R_*$ & Radius (\rsun) & $1.953^{+0.1}_{-0.094}$ & $1.868^{+0.083}_{-0.077}$ & $2.26^{+0.072}_{-0.052}$ & $2.334^{+0.047}_{-0.041}$ \\
$L_*$ & Luminosity (\lsun) & $3.5^{+0.2}_{-0.18}$ & $3.49^{+0.15}_{-0.16}$ & $5.67^{+0.84}_{-0.73}$ & $6.65^{+0.84}_{-0.73}$ \\
$\rho_*$ & Density (cgs) & $0.223^{+0.037}_{-0.034}$ & $0.299^{+0.038}_{-0.033}$ & $0.1635^{+0.0094}_{-0.015}$ & $0.1722^{+0.0065}_{-0.0078}$ \\
$\log{g}$ & Surface gravity (cgs) & $3.928^{+0.047}_{-0.052}$ & $4.036^{+0.035}_{-0.033}$ & $3.856^{+0.018}_{-0.03}$ & $3.892^{+0.012}_{-0.013}$ \\
$T_{\rm eff}$ & Effective temperature (K) & $5650.0\pm 150$ & $5770.0\pm 130$ & $5920.0\pm 210$ & $6060.0\pm 180$ \\
$[{\rm Fe/H}]$ & Metallicity (dex) & $0.17^{+0.21}_{-0.22}$ & $0.37^{+0.11}_{-0.15}$ & $0.18^{+0.16}_{-0.15}$ & $0.25\pm 0.15$ \\
$[{\rm Fe/H}]_{0}$ & Initial metallicity & $0.18^{+0.18}_{-0.2}$ & $0.376^{+0.088}_{-0.13}$ & $0.19^{+0.14}_{-0.13}$ & $0.28^{+0.12}_{-0.13}$ \\
$Age$ & Age (Gyr) & $6.25^{+1.1}_{-0.86}$ & $3.48^{+0.53}_{-0.54}$ & $4.11^{+0.66}_{-0.52}$ & $2.44^{+0.34}_{-0.33}$ \\
$EEP$ & Equal evolutionary phase & $457.5^{+5.7}_{-6.2}$ & $405.3^{+8.6}_{-14}$ & $455.4^{+4.4}_{-5.9}$ & $405.5^{+7.2}_{-9.2}$ \\
$A_V$ & V-band extinction (mag) & $0.094^{+0.041}_{-0.057}$ & $0.109^{+0.031}_{-0.054}$ & $0.33^{+0.17}_{-0.18}$ & $0.53^{+0.13}_{-0.14}$ \\
$d$ & Distance (pc) & $743.7^{+8.2}_{-8}$ & $743.8^{+8.2}_{-7.9}$ & $413.0^{+3}_{-2.9}$ & $413.8^{+3}_{-2.9}$ \\
\multicolumn{6}{l}{\textbf{Planetary Parameters}:} \\
$P$ & Period (days) & $9.657846^{+0.000036}_{-0.000039}$ & $9.657845^{+0.000036}_{-0.000039}$ & $7.148972\pm 0.000014$ & $7.148973\pm 0.000015$ \\
$R_P$ & Radius (\rj) & $0.926^{+0.069}_{-0.063}$ & $0.867^{+0.056}_{-0.052}$ & $1.023^{+0.045}_{-0.038}$ & $1.056^{+0.037}_{-0.034}$ \\
$M_P$ & Mass (\mj) & $99.0^{+3.4}_{-4.3}$ & $109.4^{+2.6}_{-2.5}$ & $22.01^{+0.61}_{-0.72}$ & $24.29^{+0.58}_{-0.56}$ \\
$T_C$ & Time of conjunction (\bjdtdb) & $2459226.0099\pm 0.0035$ & $2459226.0089^{+0.0035}_{-0.0034}$ & $2459818.9045^{+0.0013}_{-0.0012}$ & $2459818.9044^{+0.0013}_{-0.0012}$ \\
$T_0$ & Optimal conjunction time (\bjdtdb) & $2459370.8776^{+0.0035}_{-0.0034}$ & $2459370.8766\pm 0.0034$ & $2459768.8621^{+0.0013}_{-0.0012}$ & $2459768.8621^{+0.0013}_{-0.0012}$ \\
$a$ & Semi-major axis (AU) & $0.0964^{+0.0016}_{-0.0021}$ & $0.1013^{+0.0011}_{-0.0012}$ & $0.0804^{+0.0011}_{-0.0013}$ & $0.08445^{+0.00099}_{-0.00098}$ \\
$i$ & Inclination (Degrees) & $86.97^{+0.58}_{-0.53}$ & $87.98^{+0.75}_{-0.53}$ & $88.56^{+0.97}_{-1.1}$ & $89.25^{+0.52}_{-0.76}$ \\
$e$ & Eccentricity & $0.2411^{+0.0024}_{-0.0026}$ & $0.2413^{+0.0024}_{-0.0026}$ & $0.0339\pm 0.0041$ & $0.0332^{+0.004}_{-0.0042}$ \\
$\omega_*$ & Argument of periastron (Degrees) & $-19.2\pm 1.5$ & $-19.1\pm 1.5$ & $104.7^{+4.9}_{-4.5}$ & $105.0^{+5.1}_{-4.6}$ \\
$T_{eq}$ & Equilibrium temperature (K) & $1226.0^{+24}_{-18}$ & $1194.0\pm 14$ & $1515.0^{+50}_{-46}$ & $1537.0^{+45}_{-43}$ \\
$\tau_{\rm circ}$ & Tidal circularization timescale (Gyr) & $5400.0^{+2400}_{-1700}$ & $9200.0^{+3300}_{-2400}$ & $369.0^{+74}_{-70}$ & $385.0^{+66}_{-57}$ \\
$K$ & RV semi-amplitude (m/s) & $8247.0\pm 44$ & $8247.0^{+44}_{-45}$ & $1895.9\pm 7$ & $1895.9^{+7.2}_{-6.9}$ \\
$R_P/R_*$ & Radius of planet in stellar radii  & $0.0487\pm 0.0019$ & $0.0477\pm 0.0018$ & $0.0465\pm 0.0013$ & $0.0465^{+0.0012}_{-0.0013}$ \\
$a/R_*$ & Semi-major axis in stellar radii  & $10.59^{+0.55}_{-0.57}$ & $11.66^{+0.48}_{-0.45}$ & $7.66^{+0.14}_{-0.24}$ & $7.786^{+0.097}_{-0.12}$ \\
$Depth$ & \tess\ flux decrement at mid-transit & $0.00263\pm 0.00019$ & $0.00261^{+0.0002}_{-0.00019}$ & $0.00251\pm 0.00014$ & $0.0025\pm 0.00014$ \\
$\tau$ & Ingress/egress transit duration (days) & $0.0183^{+0.0027}_{-0.0022}$ & $0.0146^{+0.0017}_{-0.0015}$ & $0.01376^{+0.0011}_{-0.00061}$ & $0.01337^{+0.00052}_{-0.00044}$ \\
$T_{14}$ & Total transit duration (days) & $0.2698^{+0.0087}_{-0.0083}$ & $0.267^{+0.0086}_{-0.0082}$ & $0.2969\pm 0.0029$ & $0.2951\pm 0.0029$ \\
$b$ & Transit impact parameter & $0.573^{+0.069}_{-0.089}$ & $0.42^{+0.095}_{-0.15}$ & $0.19^{+0.13}_{-0.12}$ & $0.098^{+0.097}_{-0.068}$ \\
$\rho_P$ & Density (cgs) & $154.0^{+37}_{-30}$ & $208.0^{+42}_{-35}$ & $25.4^{+2.9}_{-3}$ & $25.6^{+2.5}_{-2.4}$ \\
$logg_P$ & Surface gravity  & $5.455^{+0.064}_{-0.065}$ & $5.557\pm 0.053$ & $4.716^{+0.032}_{-0.036}$ & $4.732^{+0.027}_{-0.028}$ \\
$\Theta$ & Safronov number & $17.4^{+1.3}_{-1.2}$ & $18.5^{+1.2}_{-1.1}$ & $2.59^{+0.1}_{-0.11}$ & $2.505^{+0.084}_{-0.085}$ \\
$T_S$ & Time of eclipse (\bjdtdb) & $2459222.573^{+0.014}_{-0.015}$ & $2459222.573^{+0.014}_{-0.015}$ & $2459822.44\pm 0.012$ & $2459822.44\pm 0.012$ \\
$T_{S,14}$ & Total eclipse duration (days) & $0.2433^{+0.008}_{-0.0075}$ & $0.2339^{+0.0068}_{-0.0065}$ & $0.3158\pm 0.004$ & $0.3141^{+0.004}_{-0.0039}$ \\
$e\cos{\omega_*}$ & & $0.2277^{+0.0024}_{-0.0026}$ & $0.2279^{+0.0025}_{-0.0027}$ & $-0.0086\pm 0.0026$ & $-0.0085^{+0.0025}_{-0.0027}$ \\
$e\sin{\omega_*}$ & & $-0.0791^{+0.0061}_{-0.0062}$ & $-0.0791\pm 0.0062$ & $0.0327^{+0.0041}_{-0.0042}$ & $0.032^{+0.0041}_{-0.0043}$ \\
$M_P/M_*$ & Mass ratio  & $0.0797^{+0.002}_{-0.0015}$ & $0.0755\pm 0.001$ & $0.01574^{+0.00027}_{-0.00022}$ & $0.01497^{+0.00018}_{-0.00019}$ \\
$d/R_*$ & Separation at mid-transit  & $10.83^{+0.57}_{-0.58}$ & $11.92^{+0.5}_{-0.46}$ & $7.4^{+0.15}_{-0.23}$ & $7.53^{+0.11}_{-0.13}$ \\
\enddata